\documentclass[longauth]{aa}  

\usepackage{graphicx}
\usepackage{rotating}
\usepackage{siunitx}
\usepackage{txfonts}

\usepackage{hyperref}
\hypersetup{
    colorlinks=true,
    linkcolor=blue,
    citecolor=blue,
    }
\usepackage[normalem]{ulem}    
\usepackage{xcolor}

\makeatletter
\renewcommand*\aa@pageof{, page \thepage{} of \pageref*{LastPage}}
\makeatother

\newcommand{\fuse}{FUSE}
\newcommand{\stis}{STIS}
\newcommand{\cosp}{COS}
\newcommand{\hst}{HST}

\newcommand{\xshooter}{X-shooter}

\newcommand{\teff}{$T_\text{eff}$}
\newcommand{\logg}{$\log g$}

\newcommand{\vmac}{$\varv_{\rm mac}$}
\newcommand{\vsini}{$\varv\sin\,i$}
\newcommand{\vrad}{$\varv_{\rm rad}$}

\newcommand{\zsun}{${\rm Z}_{\odot}$}

\newcommand{\cmfgen}{\textsc{CMFGEN}}

\begin{document}

\title{X-Shooting ULLYSES: Massive stars at low metallicity }
\subtitle{IV. Spectral analysis methods and exemplary results for O stars%
\thanks{Based on observations collected at the European Southern Observatory under ESO program 106.211Z.001.} }

\titlerunning{X-Shooting ULLYSES -- IV. Methods and exemplary results for O stars}

\author{
      {A.\,A.\,C. Sander\inst{\ref{ARI}}}
    \and
      {J.-C. Bouret\inst{\ref{LAM}}}
    \and
      {M. Bernini-Peron\inst{\ref{ARI}}}
    \and
      {J. Puls\inst{\ref{LMU}}}
    \and
      {F. Backs\inst{\ref{API}}}
    \and
      {S. R. Berlanas\inst{\ref{IAC}, \ref{ULL}}}
    \and
      {J.\,M. Bestenlehner\inst{\ref{UoS}}}
    \and
      {S.\,A. Brands\inst{\ref{API}}}
    \and
      {A. Herrero\inst{\ref{IAC}, \ref{ULL}}}
    \and 
      {F. Martins\inst{\ref{LUPM}}}
    \and 
      {O. Maryeva\inst{\ref{OND}}}
    \and
      {D. Pauli\inst{\ref{UP}}}  
    \and
      {V. Ramachandran\inst{\ref{ARI}}}
    \and
      {P.\,A. Crowther\inst{\ref{UoS}}}
    \and 
      {V.\,M.\,A. G\'omez-Gonz\'alez\inst{\ref{UP}}}
    \and
      {A. C. Gormaz-Matamala\inst{\ref{CAMK},\ref{UAIVina},\ref{PUC}}}
    \and
       {W.-R. Hamann\inst{\ref{UP}}} 
    \and
      {D. J. Hillier\inst{\ref{PITT}}}
    \and
      {R.~Kuiper\inst{\ref{UDE}}}
    \and 
      {C.\,J.\,K. Larkin\inst{\ref{ARI},\ref{MPIK},\ref{MPIA}}}     
    \and 
      {R.\,R. Lefever\inst{\ref{ARI}}}     
    \and 
      {A. Mehner\inst{\ref{ESO}}}
    \and
       {F. Najarro\inst{\ref{CAB}}}
    \and
       {L.\,M. Oskinova\inst{\ref{UP}}} 
    \and
      {E.\,C. Sch{\"o}sser\inst{\ref{ARI}}}
    \and 
      {T. Shenar\inst{\ref{TAU}}}
    \and 
      {H. Todt\inst{\ref{UP}}}  
    \and
      {A. ud-Doula\inst{\ref{PENN}}}
    \and 
      {J.\,S. Vink\inst{\ref{AOP}}}
}

\institute{
   {Zentrum f{\"u}r Astronomie der Universit{\"a}t Heidelberg, Astronomisches Rechen-Institut, M{\"o}nchhofstr. 12-14, 69120 Heidelberg, Germany \label{ARI}}
   \and 
   {Aix Marseille Univ, CNRS, CNES, LAM, Marseille, France\label{LAM}}
   \and
   {LMU M\"unchen, Universit\"atssternwarte, Scheinerstr. 1, D-81679 M\"unchen, Germany \label{LMU}}
   \and 
   {Anton Pannekoek Institute for Astronomy, Universiteit van Amsterdam, Science Park 904, 1098 XH Amsterdam, The Netherlands\label{API}}   
   \and
   {Instituto de Astrof{\'i}sica de Canarias, 38200, La Laguna, Tenerife, Spain\label{IAC}}
   \and
   {Departamento de Astrof{\'i}sica, Universidad de La Laguna, 38205, La Laguna, Tenerife, Spain\label{ULL}}
   \and
   {Department of Physics \& Astronomy, University of Sheffield, Hicks Building, Hounsfield Road, Sheffield S3 7RH, United Kingdom\label{UoS}}
   \and
   {LUPM, Universit\'e de Montpellier, CNRS, Place Eug\`ene Bataillon, F-34095 Montpellier, France \label{LUPM}} 
   \and
   {Astronomical Institute of the Czech Academy of Sciences, Fri\v{c}ova 298, 25165 Ond\v{r}ejov, Czech Republic \label{OND}}
   \and
   {Institut f{\"u}r Physik und Astronomie, Universit{\"a}t Potsdam, Karl-Liebknecht-Str. 24/25, 14476 Potsdam, Germany\label{UP}}
   \and
   {Nicolaus Copernicus Astronomical Center, Polish Academy of Sciences, Bartycka 18, 00-716 Warsaw, Poland\label{CAMK}}
   \and
   {Departamento de Ciencias, Facultad de Artes Liberales, Universidad Adolfo Ib\'a\~nez, Vi\~na del Mar, Chile\label{UAIVina}}
   \and
   {Instituto de Astrof\'isica, Facultad de F\'isica, Pontificia Universidad Cat\'olica de Chile, 782-0436 Santiago, Chile\label{PUC}}
   \and
   {Department of Physics and Astronomy \& Pittsburgh Particle Physics, Astrophysics and Cosmology Center (PITT PACC), University of Pittsburgh, 3941 O'Hara Street, Pittsburgh, PA 15260, USA\label{PITT}}
   \and 
   {Faculty of Physics, University of Duisburg-Essen, Lotharstraße 1, D-47057 Duisburg, Germany \label{UDE}}
   \and
   {Max-Planck-Institut f\"ur Kernphysik, Saupfercheckweg 1, D-69117 Heidelberg, Germany \label{MPIK}}
   \and
   {Max-Planck-Institut f\"ur Astronomie, K\"onigstuhl 17, D-69117 Heidelberg, Germany \label{MPIA}}
     \and
    {ESO - European Organisation for Astronomical Research in the Southern Hemisphere, Alonso de Cordova 3107, Vitacura, Santiago de Chile, Chile\label{ESO}}
   \and
   {Departamento de Astrof\'{\i}sica, Centro de Astrobiolog\'{\i}a, (CSIC-INTA), Ctra. Torrej\'on a Ajalvir, km 4,  28850 Torrej\'on de Ardoz, Madrid, Spain\label{CAB}}
   \and
   {The School of Physics and Astronomy, Tel Aviv University, Tel Aviv 6997801, Israel\label{TAU}}
   \and
    {Penn State Scranton, 120 Ridge View Drive, Dunmore, PA 18512, USA\label{PENN}}
   \and
     {Armagh Observatory and Planetarium, College Hill, BT61 9DG Armagh, Northern Ireland \label{AOP}}
}

   \date{Received 1 March 2024 / Accepted 29 July 2024}

 
  \abstract
   {The spectral analysis of hot, massive stars is a fundamental astrophysical method of determining their intrinsic properties and feedback. With their inherent, radiation-driven winds, the quantitative spectroscopy for hot, massive stars requires detailed numerical modeling of the atmosphere and an iterative treatment in order to obtain the best solution within a given framework.}
   {We present an overview of different techniques for the quantitative spectroscopy of hot stars employed within the X-Shooting ULLYSES collaboration, ranging from grid-based approaches to tailored spectral fits. By performing a blind test for selected targets, we gain an overview of the similarities and differences between the resulting stellar and wind parameters. Our study is not a systematic benchmark between different codes or methods; our aim is to provide an overview of the parameter spread caused by different approaches.}
   {For three different stars from the XShooting ULLYSES sample (SMC O5 star AzV\,377, LMC O7 star Sk\,-69$^{\circ}$ 50, and LMC O9 star Sk\,-66$^{\circ}$ 171), we employ different stellar atmosphere codes (CMFGEN, \textsc{Fastwind}, PoWR) and different strategies to determine their best-fitting model solutions. For our analyses, UV and optical spectroscopy are used to derive the stellar and wind properties with some methods relying purely on optical data for comparison. To determine the overall spectral energy distribution, we further employ additional photometry from the literature.}
   {The effective temperatures found for each of the three different sample stars agree within $3\,$kK, while the differences in $\log g$ can be up to $0.2\,$dex. Luminosity differences of up to $0.1\,$dex result from different reddening assumptions, which seem to be systematically larger for the methods employing a genetic algorithm. All sample stars are found to be enriched in nitrogen. The terminal wind velocities are surprisingly similar and do not strictly follow the $\varv_\infty$--$T_\text{eff}$ relation.}
   {We find reasonable agreement in terms of the derived stellar and wind parameters between the different methods. Tailored fitting methods tend to be able to minimize or avoid discrepancies obtained with coarser or increasingly automatized treatments. The inclusion of UV spectral data is essential for the determination of realistic wind parameters. For one target (Sk\,-69$^{\circ}$ 50), we find clear indications of an evolved status.}

   \keywords{Stars: early-type - Stars: massive - Stars: evolution - Stars: winds, outflows - Stars: abundances - Stars: fundamental parameters}

   \maketitle
%

\section{Introduction}

The study of metal-poor massive O-type stars has received renewed interest in recent years. They dominate the rest-frame ultraviolet (UV) spectroscopic appearance of high-redshift ($z$) galaxies \citep{Rix2004} and are the source of the ionizing radiation responsible for their rest-frame optical and UV nebular properties \citep{Steidel2014,Lecroq2024}. Early release observations with JWST \citep{Pontoppidan2022} have already identified metal-poor, star-forming galaxies at $z \geq 7$ \citep[e.g.,][]{Schaerer2022, Arellano-Cordova2022, ArrabalHaro2023, Curti2023, Robertson2023, Trussler2023}, highlighting the vital role of massive stars in the first gigayear and the demand for accurate knowledge of massive stars in metal-poor environments. In addition, metal-poor massive binaries are considered to be progenitors of black hole and neutron star mergers \citep[e.g.,][]{Neijssel2019, Boco2021, Stevance2023}, which are increasingly frequently detected by gravitational wave observatories \citep[e.g.,][]{Abbott2021}.

High-quality optical spectroscopy of O stars in the Magellanic Clouds -- our closest metal-poor star-forming galaxies -- was scarce until the advent of multiobject and integral field spectrographs on large ground-based telescopes greatly improved the samples \citep[e.g.,][]{Evans2004a,Evans2006,Evans2011,Castro2018,Ramachandran2019}. In the near future, the 1001MC survey performed with 4MOST will provide a further order-of-magnitude increase in sample size \citep{Cioni2019}. In contrast, far-ultraviolet (FUV) spectroscopy of Magellanic Cloud OB stars --directly sampling P Cygni wind diagnostic lines -- remains exceptionally scarce, and has often been limited to low-spectral-resolution observations \citep[e.g.,][]{Walborn1995, Walborn2000, Crowther2016, Rickard2022}. 

A significant new Hubble Space Telescope (HST) initiative, the Ultraviolet Legacy Library of Young Stars as Essential Standards \citep[ULLYSES,][]{Roman-Duval2020}, seeks to address this deficiency through the acquisition of high-quality UV spectroscopy of several hundred OB stars in the Magellanic Clouds, each of which has also been observed in the optical range with the \xshooter\ spectrograph at the Very Large Telescope (VLT) in the framework of the XShooting ULLYSES (``XShootU'') initiative \citep{Vink2023}.
Historically, detailed studies of individual O stars in the Magellanic Clouds involved application of plane-parallel model atmospheres not assuming local thermodynamic equilibrium (``non-LTE'') to optical spectroscopy \citep{Lanz1996, Heap2006}, with stellar winds handled separately \citep{Puls1996}. Spherical, non-LTE model atmosphere codes incorporating the effects of metal line blanketing were subsequently developed, namely CMFGEN \citep{Hillier1998}, WM-BASIC \citep{Pauldrach2001}, PoWR \citep{Graefener2002}, and \textsc{Fastwind} \citep{Puls2005}. Of these, all are capable of analyzing UV and optical spectroscopy, with the exception of WM-BASIC, whose focus is on UV spectroscopic studies \citep[e.g.,][]{Garcia2004}.

These sophisticated model atmosphere codes have been applied to optical spectroscopic samples of OB stars in the Magellanic Clouds; see, for example, \citet{Bestenlehner2014}, \citet{Ramachandran2018}, and \citet{Massey2009} and \citet{RiveroGonzalez2012} using CMFGEN, PoWR, and \textsc{Fastwind}, respectively. \citet{Massey2013} made a rare comparison of two particular codes, finding broad agreement in the temperatures of O stars using CMFGEN and \textsc{Fastwind}, although systematically lower gravities were obtained with the latter.
Combined UV and optical spectroscopic studies of OB stars in the Clouds were rare until recently \citep{Ramachandran2018Of, Bouret2021, Hawcroft2021, Brands2022, Rickard2022}, albeit with notable exceptions \citep{Crowther2002, Hillier2003, Evans2004b, Bouret2013}. 

The era of ULLYSES/XShootU permits combined UV and optical studies of large samples of OB stars in the Magellanic Clouds, but it is critical to quantify any systematic differences between the model atmosphere codes and the various techniques employed by individual groups. This is the focus of the present study, where we analyze ULLYSES/XShootU spectroscopy of representative O stars in the Magellanic Clouds ---showing prominent stellar winds--- with different methods and provide a detailed comparison of the results. The paper is structured as follows: In Sect.\,\ref{sec:sample}, we present a summary of the UV and optical spectroscopic datasets. Section\,\ref{analysis} outlines the various analysis techniques, with the comparison of results being presented in Sect.\,\ref{sec:results}. A discussion of the implications of the derived stellar and wind parameters follows in Sect.\,\ref{sec:discussion} before we draw conclusions in Sect.\,\ref{sec:perspectives}. In the Appendix, we provide detailed information about the different codes and methods as well as large parameter and atomic data comparison tables.

\section{Sample}
\label{sec:sample}

To compare different analysis techniques, we selected three stars from the ULLYSES database that sample spectral types from early to late O-type, and luminosity classes from dwarfs to supergiants, with no a priori indication for binarity: We selected one O5 dwarf in the SMC, namely \object{AzV\,377}, which has a fine-classification as O5 V((f)) \citep{Evans2004a} following the additional ``((f))'' notation from \citet{Walborn1971}. This means that the \ion{He}{II}\,4686\AA\ line is in absorption while the \ion{N}{iii} line complex at 4634$-$4640$-$4642\,\AA\ is seen in emission. The two other sample stars are located in the LMC and consist of the O9 supergiant Sk\,-66$^{\circ}$ 171 \citep[classified as O9Ia by][]{Fitzpatrick1988} and the O7 star Sk\,-69$^{\circ}$ 50, which does not formally have a luminosity class and belongs ---with its O7(n)(f)p fine classification \citep{Walborn2010}--- to the group of Ofnp stars \citep{Walborn1973}. This group is marked as peculiar (``p'') and is characterized by broadened absorption lines (``n'') as well as the above-mentioned ``f''-character, albeit with the involved \ion{N}{iii} lines portraying stronger emission in the case of Sk\,-69$^{\circ}$ 50 compared to AzV\,377, hence the fine classification with single parenthesis compared to the double parenthesis designation for \object{AzV\,377}.
The SMC star (\object{AzV\,377}) is the only one to have been analyzed previously with quantitative spectroscopy \citep{Massey2004}.

Optical and near-infrared (NIR) photometry is gathered from the ULLYSES project for each star, and is summarized in Table \ref{tab:sample}.
As introduced in the first paper of the XShootU series \citep{Vink2023}, we adopt $d_\text{SMC}= 62.44\,$kpc, corresponding to DM(SMC) = 18.98 mag \citep{Graczyk2020}, and $d_\text{LMC} = 49.59\,$kpc, corresponding to DM(LMC) = 18.48 mag \citep{Pietrzynski2019}.

\subsection{ULLYSES ultraviolet spectroscopy}

The three stars have been observed with the Far Ultraviolet Spectroscopic Explorer \citep[\fuse,][]{Moos2000}, providing spectroscopic coverage of $\lambda\lambda$905--1187~\AA\ \citep[$R\sim 20\,000$; for an OB atlas see][]{Walborn2002}. \object{AzV 377} has been observed with \hst\ in the FUV with the Cosmic Origins Spectrograph \citep[\cosp,][]{Green2012} using the G130M/1291 ($\lambda\lambda$1132--1433~\AA, $R\sim 14\,000$) and G160M/1611 ($\lambda\lambda$1419--1790~\AA, $R\sim 14\,000$) gratings, while the Space Telescope Imaging Spectrograph \citep[\stis,][]{Kimble1998} was used for \object{Sk\,-69$^{\circ}$ 50} (added later to the ULLYSES dataset) and \object{Sk\,-66$^{\circ}$ 171} with the E140M grating ($\lambda\lambda$1143--1710~\AA, $R\sim 46\,000$). For the latter star, the spectral coverage extends into the near-UV (NUV) due to an additional observation with \stis\ using the E230M/1978 grating ($\lambda\lambda$1607--2366~\AA, $R\sim 30\,000$). Only the observations of \object{Sk\,-66$^{\circ}$ 171}, performed on January 28, 2022 (GO/DD 16365, PI Roman-Duval), were obtained within the DDT provided for the ULLYSES project, while the UV observations for \object{AzV 377} were part of GO 15837 (PI Oskinova), performed on June 25, 2020, and the observations of \object{Sk\,-69$^{\circ}$ 50} date back to October 11, 2011 and were part of GO 12218 (PI Massa).

\subsection{XShootU optical spectroscopy}

Optical, normalized \xshooter\ \citep{Vernet2011} spectroscopy of each star from VLT/\xshooter\  was reduced and processed according to \citet[eDR1]{Sana2024}. In this work, we use the reduced data for the UBV ($\lambda\lambda$3100-5600~\AA, $R\sim 6\,700$) and VIS ($\lambda\lambda$5600--10240~\AA, $R\sim11\,400$) arms. The data reduction for the NIR arm requires additional work and is therefore not yet available. For the purpose of our  O star analysis, the broad wavelength coverage from the combined UV and optical spectra contains a sufficient number of spectral lines from different elements and ionization stages in order to avoid any deficiencies due to the absence of NIR data.

 \begin{table*}[]
    \centering
    \caption{Photometry of the sample stars.}
    \label{tab:sample}
    \begin{tabular}{lccccccccc}\hline
    \rule{0cm}{2.2ex}
    object      &  RA/Dec\tablefootmark{(a)} and host galaxy    & sp.\ type                             & U               & B             & V             & J             & H             & K$_s$ \\ \hline 
    \rule{0cm}{2.2ex}
    AzV 377     & 01 05 07.38 -72 48 18.71 (\textsc{smc})               &O5\,V((f))\tablefootmark{(b)}\! & 13.19\tablefootmark{(c)}      & 14.25\tablefootmark{(c)}      & 14.45\tablefootmark{(c)}      & 15.21\tablefootmark{(d)}        & 15.18\tablefootmark{(e)}\!    & 15.33\tablefootmark{(d)}   \\
    \rule{0cm}{2.2ex}
    Sk\,-69$^{\circ}$ 50        &  04 57 15.09 -69 20 19.95 (\textsc{lmc})              &O7(n)(f)p\tablefootmark{(f)}\! & 12.16\tablefootmark{(c)}        & 13.15\tablefootmark{(c)}      & 13.31\tablefootmark{(c)}      & 13.60\tablefootmark{(d)}        & 13.67\tablefootmark{(g)}\! & 13.66\tablefootmark{(d)}  \\ 
    \rule{0cm}{2.2ex}
    Sk\,-66$^{\circ}$ 171 & 05 37 02.42 -66 38 37.03 (\textsc{lmc})             &O9\,Ia\tablefootmark{(h)}\!    & 11.02\tablefootmark{(i)}        & 12.04\tablefootmark{(i)}      & 12.19\tablefootmark{(i)}      & 12.58\tablefootmark{(d)}        & 12.57\tablefootmark{(g)}\!  & 12.62\tablefootmark{(d)}  \\ \hline
\end{tabular}
    \tablefoot{
     \tablefoottext{a}{Coordinates from \citet{GAIA2023}, given in ICRS (epoch=2000)}
     \tablefoottext{b}{\citet{Evans2004a}}
     \tablefoottext{c}{\citet{Massey2002}}
     \tablefoottext{d}{\citet{Cioni2011}}
     \tablefoottext{e}{\citet{Cutri2012}}
     \tablefoottext{f}{\citet{Walborn2010}}
     \tablefoottext{g}{\citet{Cutri+2003}}
     \tablefoottext{h}{\citet{Fitzpatrick1988}}
     \tablefoottext{i}{\citet{Isserstedt1975}}
    }
\end{table*}

\section{Analysis methods}\label{analysis}

The three targets in the present work are limited to the regime of O stars. All selected stars have noticeable winds that leave an imprint (i.e., diagnostic) in the spectrum and therefore mark ideal targets for our method comparison. Thus, we only use the expanding atmosphere codes CMFGEN, PoWR, and FASTWIND. For O dwarfs with weak winds, plane-parallel model atmosphere codes, such as TLUSTY \citep{Lanz2003, Lanz2007}, would be suitable as well. Investigations of ULLYSES B-type stars will be presented in subsequent papers.

\subsection{Common aspects}
  \label{sec:anacommon}

The three atmosphere codes utilized in this work ---\textsc{CMFGEN}, \textsc{Fastwind}, and \textsc{PoWR}--- are 1D codes assuming spherical symmetry and a stationary outflow. Targeting hot stars, they account for an expanding, non-LTE environment by iteratively solving the equations of statistical equilibrium for a large set of levels of individual elements and ions together with the solution of the radiative transfer. For \textsc{PoWR} and \textsc{CMFGEN}, the radiative transfer is completely solved in the co-moving frame (CMF). In the \textsc{Fastwind} version \citep{Sundqvist2018} applied in this work, only a few ``explicit'' elements (cf.\, Sect.\,\ref{sec:fastwind}), plus the most important lines from the other elements, are treated in the CMF. For all other  lines, a Sobolev approach with a pseudo-continuum irradiation ---accounting for the combined line-opacity/emissivity from the metallic background--- is used. Both approximations enable comparatively short computation times. After discussing common aspects and tools in this subsection, the following subsections introduce the different model atmosphere codes and their specific application methods. In the Appendix, we provide a more in-depth review of the physical treatments for all expanding atmosphere codes employed in this work (Appendix\,\ref{sec:codedetails}). Detailed method descriptions sorted by the different aspects necessary (including aspects such as the determination of the projected rotational velocity or the bolometric luminosity) for quantitative spectral analysis are given in Appendix\,\ref{sec:mdetail}. In the following, each individual method is denoted by a letter and a number with the letter denoting the initial of the employed atmosphere code.

\subsubsection{Velocity and density structure}

The models computed in this work are not dynamically consistent, but assume a prescribed velocity $\varv(r)$ in the form of a so-called $\beta$-law
\begin{equation}
  \label{eq:betalaw}
  \varv(r) = \varv_\infty \left( 1 - \frac{R_0}{r} \right)^{\beta},
\end{equation}
where $\varv_\infty$ is the velocity for $r \rightarrow \infty$, $\beta$ is a free parameter, and $R_0$ is a reference radius. In methods where a pre-calculated grid of models is used, $\beta$ is often fixed to a specific value; for example, $\beta=0.8$ motivated by extensions of the CAK \citep*[named after][]{Castor1975} theory \citep[e.g.,][]{Pauldrach1986}. When individual models are calculated, the value of $\beta$ can instead be determined from combining constraints from UV and optical lines that are affected by the stellar wind.
In the subsonic part, the models aim to obtain a (quasi-)hydrostatic stratification, with the detailed techniques and the connection to the supersonic $\beta$-law differing between the codes. The solution techniques vary between the different codes. A notable difference with respect to the derived values of the surface gravity $\log g$ can arise due to different assumptions in the solution of the hydrostatic equation
\begin{equation}
  \label{eq:hydrostat}
  \frac{\mathrm{d}P}{\mathrm{d}r} = - \rho(r) \left[ g(r) - g_\text{rad}(r) \right]
,\end{equation}
with $P$ denoting the pressure and $\rho$ the matter density.
In addition to the required knowledge of the radiative acceleration $g_\text{rad}(r)$, the solution of Eq.\,\eqref{eq:hydrostat} demands an equation of state. In all model codes used in this work, this is the ideal gas equation, which we can write as $P(r) = \rho(r) a^2(r)$. For the speed $a$, there is the opportunity to not only include the (thermal) speed of sound, but also a turbulence term, such that
\begin{equation}
   \label{eq:a2def}
   a^2(r) = \frac{k_\text{B} T(r)}{\mu(r) m_\text{H}} + \frac{1}{2} \xi^2(r),
\end{equation}
with $T(r)$ denoting the (electron) temperature, $\mu(r)$ the mean particle mass (including electrons), $k_\text{B}$ Boltzmann's constant, and $m_\text{H}$ the mass of a hydrogen atom. From the codes used in this work, only PoWR has the option to include a nonzero term for a turbulent pressure described by a velocity $\xi(r)$ when solving the hydrostatic equation \citep{Sander2015}. The value of  $\xi(r)$ in Eq.\,\eqref{eq:a2def} can ---but does not have to--- be chosen similar to the microturbulence entering the formal integral. The use of $\xi > 0$ in the solution of the hydrostatic equation will lead to larger values for the determined $\log g$. The difference can be estimated via
\begin{align}
  \label{eq:loggturb}
  \Delta\left(\log g\right) &= \log \left(1 + \frac{\xi^2 \mu m_\text{H}}{ 2 k_\text{B} T } \right).
\end{align}

\subsubsection{Wind inhomogenieties}\label{sec:clumping}

All atmosphere codes take wind inhomogenieties (``clumping'') into account. Most of the applied methods only make use of the so-called ``microclumping'' approximation, whereby it is assumed that clumping is limited to small scales and the clumps themselves are optically thin. The medium between the clumps is assumed to be void. Defining an average density via the (stationary) equation of continuity
\begin{equation}
   \langle\rho\rangle = \frac{\dot{M}}{4 \pi r^2 \varv(r)}
,\end{equation}
with $\dot{M}$ denoting the mass-loss rate, one can define a ``clumping factor'':
\begin{equation}
   f_\text{cl} = \frac{\langle\rho^2\rangle}{\langle\rho\rangle^2}\text{.}
\end{equation}
Inside the clumps, the over-density relative to a smooth wind can be described by a factor $D = \rho_\text{cl} / \rho_\text{smooth}$ with $\rho_\text{cl}$ denoting the density inside the clumps and $\rho_\text{smooth}$ denoting the density of a smooth wind with the same mass-loss rate. The mean density can further be expressed as $\langle\rho\rangle = f_\text{V} \rho_\text{cl}$, with $f_\text{V}$ describing the volume filling factor of the clumps.

In practice, the different atmosphere codes employ different quantities as their free parameter: \textsc{Fastwind} uses $f_\text{cl}$, while CMFGEN requires $f_\text{V}$ to be given, and PoWR has $D$ as its free parameter. Fortunately, these values can easily be converted into each other for a void interclump  medium and optically thin clumps, namely
\begin{equation}
  f_\text{cl} \equiv D \equiv f_\text{V}^{-1}\text{.}
\end{equation}
This relation does not hold for optically thick clumps or a nonvoid interclump   medium, which is used in one of the employed methods (F3). The more detailed clumping implementations in the different codes, including the optically thick clumping approach in \textsc{Fastwind}, are described in Appendix\,\ref{sec:clumpingdetails}. For $f_\text{cl} = 1$, a smooth (``unclumped'') wind situation is recovered in all cases. 

\subsubsection{Abundance notations}

The input and output format for abundances differ between the atmosphere codes. While for example \textsc{PoWR} expects either mass fractions or absolute number fractions to be given, \textsc{Fastwind} requires number ratios and \textsc{CMFGEN} can handle a mixture of number ratios and mass fractions. Fortunately, these quantities can easily be converted as long as information on all elements with major abundances is provided. From a given set of either absolute number fractions $N(i)$ or number ratios relative to an element such as hydrogen $N(i)/N(\text{H})$, the (absolute) mass fraction $X_{i}$ for an arbitrary element $i$ can be determined via
\begin{equation}
   \label{eq:converttomassfraction}
   X_{i} = \frac{\mathcal{A}_i \frac{N(i)}{N(\text{H})}}{\sum\limits_{j=1}^{n_\text{elem}} \mathcal{A}_j \frac{N(j)}{N(\text{H})}} 
   = \frac{\mathcal{A}_i N(\text{i})}{\sum\limits_{j=1}^{n_\text{elem}} \mathcal{A}_j N(\text{j})},
\end{equation}
with $\mathcal{A}_i$ denoting the atomic weight of element $i$. If absolute mass fractions are provided, the absolute number fractions are given by
\begin{equation}
   \label{eq:converttonumberfraction}
    N(i) = \frac{X_{i}}{\mathcal{A}_i \sum\limits_{j=1}^{n_\text{elem}} \frac{ X_{j} }{ \mathcal{A}_j }}.
\end{equation}
In \textsc{Fastwind} and \textsc{CMFGEN}, the \ion{He}{}/\ion{H}{} number ratio
\begin{equation}
   y_\text{He} = \frac{ N(\text{He}) }{ N(\text{H}) }
\end{equation}
is an input quantity, and is commonly also simply  denoted $Y$ in the literature. We refrain from using the latter notation and instead use $y_\text{He}$ in this work to avoid any confusion with the common ($X$, $Y$, $Z$) notation, which refers to the mass fractions of hydrogen, helium, and all other elements, respectively, with these latter being commonly referred to as ``metals'' in astrophysics. This fraction of metals $Z$ is referred to as metallicity.

A further common astrophysical abundance notation is 
\begin{equation}
   \epsilon(x) = \log \frac{x}{H} + 12 \equiv \log \frac{N(x)}{N(\text{H})} + 12,
\end{equation}
with the first expression referring to the typical literature standard and the second being the equivalent in our notation accounting for the different format specifications. In the extragalactic community, $Z$ is sometimes also used as a label for $\left[\text{O}/\text{H}\right] = \epsilon(\text{O}) - \epsilon(\text{O})_{\odot}$ or $\left[\text{Fe}/\text{H}\right]$. When gas-phase abundances are measured from nebular lines, $\epsilon(\text{O})$ is often treated as a proxy for $Z$ or is even termed ``metallicity''. We only use the term in its original meaning, referring to the total mass fraction of all elements beyond helium. 

\subsubsection{Rotation, macroturbulence, and IACOB-Broad}

To determine the projected rotational and macroturbulent velocities, which broaden the spectral lines, a couple of methods use the \texttt{iacob-broad}\footnote{http://research.iac.es/proyecto/iacob/pages/en/useful-tools.php} package described in \citet[][see also references therein]{SimonDiazHerrero2014}. \texttt{iacob-broad} combines the Fourier transform method (based on the presence of zeros introduced by the transform function of the rotational profile) and the ``goodness of fit'' method (based on the best fit of a combination of rotational and radial-tangential macroturbulent profiles) to the observed spectral lines. In their package, \citet{SimonDiazHerrero2014} make three major assumptions: the stellar surface is spherical, the rotational and macroturbulent profiles are convolved with the emergent flux profiles (and not with the intensity profiles), and other broadening mechanisms (e.g., collisional and instrumental) are comparatively small. The last approximation implies that \ion{H}{} and \ion{He}{} lines should be avoided if possible, as they will be broadened by the linear Stark effect\footnote{Some \ion{He}{i} lines are affected only by the quadratic Stark effect, meaning that they can serve as an alternative, e.g., if no resolved metal lines are available.}. The selection of lines is decided by the user and can be adjusted for each star depending on the available spectra and the strength of the individual lines therein.

The results obtained for the projected rotational velocity $\varv \sin i$ and the macroturbulence $\varv_\text{mac}$ can be degenerate, depending on the spectral resolution, the available lines, and ---if a Fourier transform method such as \texttt{iacob-broad} is used--- the selection of the correct zero in Fourier space \citep[see, e.g., the discussions][for more details]{SimonDiaz2007,SimonDiazHerrero2014}. Therefore, some of the methods employed in this work choose to fix $\varv_\text{mac}$, while others keep it as a free parameter. The origin of macroturbulence in massive stars and the interpretation of the derived $\varv_\text{mac}$ values are a topic of  active research \citep[e.g.,][]{Aerts2009,Sundqvist2013,Grassitelli2016,Debnath2024}.

\subsection{FASTWIND (F methods)}\label{sec:fastwind}

Developed with the intent to provide a computationally fast non-LTE scheme \citep{SantolayaRey1997}, \textsc{Fastwind} focuses on providing models and synthetic spectra for OBA stars with winds that are not significantly optically thick in the (optical) continuum. The initial efforts of the code are documented in \citet{SantolayaRey1997} with subsequent improvements and extensions described in \citet{Puls2005,RiveroGonzalez2012,Carneiro2016,Sundqvist2018}. Unlike the other codes applied in this work, \textsc{Fastwind} distinguishes between line and continuum transfer as well as ``explicit'' and ``background'' elements. Only the explicit elements have a flexible, user-supplied model atom\footnote{The background elements are described with the WM-basic atomic database, see \citet{Pauldrach2001}.} and employ the CMF radiative transfer for their line transitions. The radiative transfer for the background elements is mainly performed with the \citet{Sobolev1960} approximation, but the most important transitions can be done in the CMF as well \citep{Puls2005}. There is also a recent version of \textsc{Fastwind} that can treat all elements in the CMF \citep{Puls2020}, but this version has so far only been employed to perform mass-loss predictions \citep[e.g.,][]{Bjoerklund2021} and is not used in this work.

The input reference radius for all \textsc{Fastwind} models is the radius corresponding to an effective temperature for a Rosseland optical depth of $\tau_\text{Ross} = 2/3$. In the literature, the corresponding temperature is commonly termed $T_\text{eff}$, while the radius is denoted $R_\ast$. However, this designation is not unique among the different atmosphere codes. While we use the established label $T_\text{eff}$ to refer to the effective temperature at $\tau_\text{Ross} = 2/3$, the corresponding radius is denoted $R_{2/3}$ throughout this paper in order to avoid any confusion between the different codes that use the label $R_\ast$ for different radii.

To describe the strength of the wind, the mass-loss rate $\dot{M}$, terminal velocity $\varv_\infty,$ and clumping factor $f_\text{cl}$ (assuming optically thin clumping with no interclump medium) can be combined into the ``wind strength parameter'':
\begin{equation}
  \label{eq:qws}
   Q_\text{ws} = \frac{\dot{M} \sqrt{f_\text{cl}}}{M_\odot\,\mathrm{yr}^{-1}} \left( \frac{\mathrm{km}\,\mathrm{s}^{-1}}{\varv_\infty} \frac{R_\odot}{R_{2/3}} \right)^{3/2} 
,\end{equation}
which is a common input parameter for \textsc{Fastwind} models, in particular when calculating grids of models. The quantity $Q_\text{ws}$ was originally defined in \citet{Puls1996}, later adjusted for clumping \citep[e.g.,][]{Puls2008}, and is also known as ``optical depth invariant''. If the wind is optically thin, models with the same stellar parameters and the same $Q_\text{ws}$ yield very similar spectra, allowing a reduction of the calculation effort for model grids. For optically thick winds, the $\varv_\infty$ scaling changes and instead the ``transformed radius'' $R_\text{t}$ \citep{Schmutz1989} (or the ``transformed mass-loss rate'' $\dot{M}_\text{t}$ discussed in Sect.\,\ref{sec:discussion}) are better scaling quantities in this regime \citep[see, e.g.,][]{Bestenlehner2020}. In the subsequent sections, we briefly introduce the general concepts of all methods employing \textsc{Fastwind}. 

\subsubsection{F1 -- Optical/IACOB-GBAT}
  \label{sec:f1}

The F1 method makes use of a grid-based automatic tool (\textsc{gbat}) called \textsc{Iacob-Gbat} \citep{SimonDiaz2011}, which was developed as part of the \textsc{iacob} project \citep{SimonDiazHerrero2014} and is regularly applied there \citep[e.g.,][]{Holgado2018,Holgado2020}. \textsc{Iacob-Gbat} determines the goodness of a fit within a given grid of atmosphere models via a $\chi^2$ criterion applied on a list of selected, normalized lines. 

Only the optical H/He spectra are used in the F1 method, with H and He being the only explicit elements. A small grid of models was calculated for the SMC star, while a much larger grid is employed for the two LMC stars. The parameter range for the LMC model grid is listed in Table~\ref{tab:gbat_grid}. The $\chi^2$ calculation enables us to also estimate the uncertainties of the derived stellar (and wind) parameters. No wind clumping is included in any of the F1 models ($f_\text{cl} = 1$), meaning that any estimates of the mass-loss rate are only upper limits.

 \begin{table}[]
    \centering
    \caption{Parameter range in the \textsc{Fastwind} grid used in the F1 method}
    \begin{tabular}{lcc}\hline
    Parameter           &  range          & units \\ \hline
    \teff               &  25 -- 55       & kK    \\
    \logg               &  3.0 -- 4.2     & dex (cgs) \\
    $y_\text{He}$       &  0.06 -- 0.23   &       \\
    $\log Q_\text{ws}$  &  -15.0 -- -11.7 & dex (cgs) \\
    $\beta$             &  0.8-1.8        &     \\
    $\xi$               &  5 -- 20        &  km\,s$^{-1}$   \\
    \hline
    \end{tabular}
 \label{tab:gbat_grid}
 \tablefoot{    
    The grid marks the basis for the {\sc Iacob-Gbat} parameter determination (see Appendix\,\ref{sec:f1temp} for details). Grid models do not include clumping.
  }
\end{table}

\subsubsection{F2 and F3 -- Kiwi-GA: Optical and Optical+UV}
  \label{sec:Kiwi-GA}

The F2 and F3 methods use a similar approach  (see below) to derive the stellar and wind parameters plus the He and CNO abundances (by means of H, He, C, N, O, Si, P as explicit elements), but differ in the usage of the underlying data. In F2, only the optical data are taken into account, while F3 uses both optical and UV data. For the optical spectra, the normalization of \citet{Sana2024} is adopted, but the data are renormalized where the continuum clearly lies above unity. 
F2 and F3 make use of \textsc{Fastwind} \citep[v10.6,][]{Sundqvist2018} with optically thick wind clumping (macroclumping), combined with a genetic algorithm (GA) called Kiwi-GA\footnote{\url{https://github.com/sarahbrands/Kiwi-GA}} \citep{Brands2022}. Earlier forms of this method have been used in several analyses of massive star spectra \citep[e.g.,][]{Mokiem2005,Tramper2014,AbdulMasih2021,Brands2022}. 
Genetic algorithms are based on the concepts of natural selection and ``survival of the fittest''. First, an initial group of model input parameters is selected randomly from a given parameter space. The ``fitness'' of the resulting model spectra is tested against the data ---a stellar spectrum--- by computing a $\chi^2$ value for a selection of normalized lines,
thus deciding which parameters are selected for the next generation of models: parameters of the models with the lowest $\chi^2$ value have the greatest chance of being selected. With the new parameters, but also random ``mutations'' \footnote{Mutations are inserted to avoid a false convergence towards local but not global minima.}, models of the next generation are computed and their fitness is analyzed again. This process is repeated for $40-120$ generations, after which the algorithm converges to a set of best-fit parameters. The $\chi^2$ values further enable the calculation of uncertainties for the best-fitting model.
More details about Kiwi-GA are given in \citet{Brands2022}. For details regarding the uncertainty derivation, see  Brands et al. (in prep, part of the XShootU series). Requiring no model grid, but instead the calculation of new models on the fly, the GA concept for spectral fitting has so far only been combined with \textsc{Fastwind} atmosphere models due to their short computing times ($15-45$ minutes). 

Technically, the F3 analysis is not performed independently, but builds up on F2. 
In F2, $\beta$ is fixed to unity, and also the clumping parameters are fixed (for details, see Appendix~\ref{sec:mdetailwind}). 
F3 then allows us to vary $\beta$ and the full set of wind and clumping parameters in \textsc{Fastwind}, but fixes the projected rotational velocity and the $y_\text{He}$ ratio obtained in F2.

\subsubsection{F4 -- LMC optical model grid}
  \label{sec:f4}

The F4 method uses a grid-based approach, but is performed with a different set of models and a different pipeline than F1, though also minimizing the $\chi^2$ (see Appendix~\ref{sec:mdetailtgabu}). The underlying model grid has dedicated LMC abundances and is calculated with \textsc{Fastwind} v10.6, with H, He, C, N, O, and Si as explicit elements. The grid explores the $T_{\rm eff}$, $\log g$, $\dot{M,}$ and $y_\text{He}$ parameter space plus three CNO abundance combinations. The CNO abundance sets represent LMC baseline abundances \citep{Vink2023} plus semi and fully processed CNO composition due to the CNO-cycle according to the 60\,$M_{\odot}$ evolutionary track by \citet{Brott2011}. In the grid, a smooth wind is assumed (i.e., clumping factor $f_\textrm{cl} = 1$), the wind velocity field uses a fixed value of $\beta = 1.0$, and the microturbulence velocity is fixed to $10$\,km\,s$^{-1}$. In total, around 120\,000 stellar models were computed. Including convolutions for rotation results in a total number of about 1\,100\,000 synthetic spectra.

The spectroscopic analysis in F4 is solely based on the optical VLT/\xshooter\ data \citep[][eDR1]{Sana2024} and uses the spectral lines of H, \ion{He}{i-ii}, \ion{C}{ii-iv}, \ion{N}{ii-v,} and \ion{Si}{ii-iv} in the wavelength range of $\lambda\lambda$3800--7100~\AA. Further redward wavelengths have been ignored to avoid any impact of telluric lines on our results. Wavelengths of $< 3800\,$\AA\ were omitted to avoid spurious effects from the normalization around the Balmer jump. For reproducing the line spectra, the normalized spectra provided by \citet[eDR1]{Sana2024} were used without further renormalization. The uncertainties of the best-fitting model were derived with an empirical Bayesian approach and maximum a posterior approximations utilizing de-idealized models as described in detail in \citet{Bestenlehner2024}.

\subsection{CMFGEN (C methods)}\label{sec:cmfgen}

The stellar atmosphere code \textsc{CMFGEN} is a spherical, non-LTE code that was developed to model stars with strong winds \citep{Hillier1990,Hillier1998} that can also be optically thick in the continuum. Its radiative transfer is performed completely in the CMF with a detailed treatment for line-blanketing using a flexible superlevel approach \citep{Hillier1998}. While there is a time-dependent branch for the simulation of supernovae spectra \citep[e.g.,][]{Dessart2010}, the scheme we employ in this work assumes stationary outflows.

Unlike in \textsc{Fastwind}, \textsc{CMFGEN} uses the notation $R_\ast$ for the inner boundary in its input file. In general, $R_\ast$ will correspond to $\tau_\text{Ross} \gg 2/3$. The temperature $T_\text{eff} \equiv T_{2/3}$ and $\log g$ (also referring to $R_{2/3}$) are only a relevant input when iterating the density structure for the hydrostatic domain, which is necessary for objects such as the O stars studied in this work, where the spectrum is not completely formed in the wind.

\subsubsection{C1 - $\chi^2$ analysis}
\label{sect_cmfgen}

The C1 method is a grid-based approach with subsequent refining via additional model sets. The underlying grids of \cmfgen\ models are computed for stars in the Magellanic clouds, which will be presented in \citet{Marcolino2024}. The models of these grids are computed for scaled solar abundances (\mbox{CNO} and beyond). A microturbulence velocity of 10 \,km\,s$^{-1}$ is adopted for the line profiles in the CMF radiative transfer and in the spectrum calculation throughout the initial grids. No clumping is assumed in the initial grids ($f_\text{V} = 1$). Using a $\chi^2$ criterion on a series of optical-only lines (cf.~Appendix\,\ref{sec:mdetailtgabu}), the best-fitting grid model is determined. 

For the two LMC stars, the spectral fit is subsequently improved by calculating further models that can go beyond the grid assumptions for wind clumping, microturbulence, abundances, and so on. In this process, another $\chi^{2}$ analysis is performed on an extended line set including CNO lines. Finally, the wind parameters were adjusted to reproduce the wind (UV and H$\alpha$) spectral features. More details on these steps are given in Appendix\,\ref{sec:mdetail} and the whole methodology is presented in more depth in \citet[paper V in the series]{Martins2024}. Uncertainties are derived with the help of the obtained $\chi^{2}$ values from the different models.

\subsubsection{C2 - Individual fit}

The C2 method employs the traditional method of calculating a series of individual atmosphere models to obtain a reasonable manual (``by eye'') fit to the observed spectrum.
The initial assignment of $T_\text{eff}$, $L$, and $\log g$ is based on the spectral types of the sample stars and the O-star calibration by \citet{Martins2005}. No specific parameter restrictions
were made apart from testing only single-$\beta$ velocity laws  and assuming optically thin wind clumping. 

Only the two LMC stars are analyzed in the framework of C2. Initial abundance estimations for them employ Geneva evolutionary tracks from \citet{Eggenberger2021}, but further refinements were made when necessary. The uncertainties are estimated by comparing the best-fitting solution to models with varied parameters.

\subsection{PoWR (P methods)}\label{sec:powr}

The fundamental physical approach to the non-LTE stellar atmosphere modeling is similar between PoWR and \textsc{CMFGEN}, while the development and implementation of these codes are completely independent. 
While similarly designed for hot stars with significant winds, including those with optically thick continuum, the numerical approaches differ considerably, for example regarding the implementation of iron-line blanketing \citep{Graefener2002} or the determination of the temperature stratification \citep{Hamann2003}. \textsc{PoWR} provides the opportunity to calculate models completely from scratch, but the most common way to start the spectral analysis is to select a model from an older study or from a previously calculated grid \citep[e.g.,][]{Hainich2019}.

\subsubsection{Specific notations and hydrostatic domain treatment}

Due to its roots in the analysis of Wolf-Rayet stars, \textsc{PoWR} does not use $T_\text{eff} = T_\text{2/3}$ as an input parameter, but instead uses the inner boundary radius, termed $R_\ast$. This radius is associated to a maximum Rosseland continuum optical depth, which is typically set to $\tau_\text{Ross,cont} = 20$. As this is also the case for the present models, we denote the corresponding radius as $R_{20}$ to avoid any confusion from the different usages of the label $R_\ast$. The associated effective temperature ---typically termed $T_\ast$ in papers employing \textsc{PoWR} or \textsc{CMFGEN}--- is labeled $T_{20}$ here. The quantities $T_\text{eff}$ and $R_\text{2/3}$ are output parameters from converged models. The input surface gravity $\log g$ is also specified at $R_{20}$, but we instead provide the corresponding value for $g(R_{2/3})$ to ease the direct comparison. 

In the hydrostatic regime, \textsc{PoWR} integrates the hydrostatic equation using directly the radiative force $\Gamma_\text{rad} = a_\text{rad}/g$  calculated in the comoving frame to obtain the density and velocity stratification. When starting a new model, the initial integration is either performed with a mean $\Gamma_\text{rad}$ or a depth-dependent description is taken from an old model.

\subsubsection{P1/P2 - Individual fits with tailored models}

The P1 and P2 methods consist of a series of individual model calculations to obtain a reproduction of the observed spectrum that is deemed sufficient upon visual inspection. The initial models are selected from publicly available OB grids \citep{Hainich2019} with parameters as close as possible to the assumed stellar parameters. After constraining the rotational velocity (and macroturbulence), tailored models with adjusted stellar and wind parameters are calculated until the synthetic spectrum sufficiently reproduces the observed spectrum. The models further vary the depth-dependent optically thin clumping and the P2 models further include additional X-rays.

P1 and P2 differ slightly in their detailed assumptions and fixed inputs (e.g., in microturbulence entering the hydrostatic equation and the usage of the wind velocity law; see Appendix\,\ref{sec:mdetail} for details). P1 is only applied to the SMC star AzV\,377, while P2 is limited to the two LMC stars. The error margins quoted for P1 and P2 are determined by varying the individual stellar and wind parameters of the final synthetic model spectrum and include model parameters that still mimic the observed spectrum.

\section{Comparison of the spectral analyses}\label{sec:results}

\begin{table}
\centering
  \caption{Maximum spread among the main derived parameters obtained with the different methods.}\label{tab:spread}
  \begin{tabular}{lcccc}
 \hline\hline
  \rule{0cm}{2.2ex}Parameter &  AzV &  Sk &  Sk & Typical  \\ 
                             &  377 &  -69$^\circ$ 50 &  -66$^\circ$ 171 &  spread\tablefootmark{(a)}\!\!\!\! \\ \hline 
  \rule{0cm}{2.2ex}$\Delta T_\mathrm{eff}\,[\mathrm{K}]$                                  & $3250$  & $2850$  & $3150$   & $3000$ \\ 
  \rule{0cm}{2.2ex}$\Delta\!\log (g\,[\mathrm{cgs}])$                                       & $0.11$  & $0.19$  & $0.30$   & $0.20$ \\
  \rule{0cm}{2.2ex}$\Delta\!\log (L\,[L_\odot])$                                              & $0.09$  & $0.19$  & $0.16$   & $0.15$ \\
  \rule{0cm}{2.2ex}$\Delta\!~ R_{2/3}\,[R_\odot]$                                        & $1.21$  & $1.4$   & $5.1$    & $2.6$ \\
  \rule{0cm}{2.2ex}$\Delta \varv \sin i\,[\mathrm{km\,s^{-1}}]$\tablefootmark{(b)}\!\!\!          & $40$    & $20$    & $35$     & $30$   \\    
  \rule{0cm}{2.2ex}$\Delta \varv_\infty\,[\mathrm{km\,s^{-1}}]$\tablefootmark{(c)}\!\!\!          & $175$   & $300$   & $300$    & $250$   \\    
  \rule{0cm}{2.2ex}$\Delta\!\log (\dot{M}\,[M_\odot\,\mathrm{yr}^{-1}])$\tablefootmark{(c,d)}\!\!\!\!\! & $0.30$  & $0.45$  & $0.51$   & $0.42$ \\
  \rule{0cm}{2.2ex}$\Delta\!\log (Q_\ion{H}{i}\,[\mathrm{s}^{-1}])$                        & $0.17$  & $0.62$  & $0.61$   & $0.5$ \\
  \rule{0cm}{2.2ex}$\Delta\!\log (Q_\ion{He}{i}\,[\mathrm{s}^{-1}])$                       & $0.35$  & $0.82$  & $1.15$   & $0.8$ \\
\hline 
    \end{tabular}
    \tablefoot{
      \tablefoottext{a}{The last column lists a rounded average from all three objects, indicating a ``typical'' systematic uncertainty margin arising from the different codes and methods.}
      \tablefoottext{b}{Only accounting for methods where $\varv \sin i$ is not fixed.}
      \tablefoottext{c}{only incorporating values from methods that actually use the UV spectrum}
      \tablefoottext{d}{sensitive to clumping constraints/choices, see Sects.\,\ref{sec:clumping} and \ref{sec:clumpingdetails}.}
    }
\end{table}

For each of our three sample stars, the results from all methods applied to the specific object are listed in a separate table. For the O5-dwarf AzV 377, the results are listed in Table~\ref{table:paramsummary-azv377}. The values for the peculiar O7(n)(f)-star Sk\,-69$^\circ$ 50 are provided in Table~\ref{table:paramsummary-sk6950}, and the results for the O9-supergiant Sk\,-66$^\circ$ 171 are given in Table~\ref{table:paramsummary-sk66171}. A brief overview of the maximum spread in the derived fundamental parameters is shown in Table~\ref{tab:spread}. The last column further indicates a ``typical'' spread in each parameter arising from our sample. This value is calculated as a rounded average of the values for the different targets and approximately reflects the systematic uncertainty arising from the different analysis codes and methods.  

\subsection{Spectral line reproduction}\label{sec:speclines}

\begin{figure}
  \includegraphics[width=\columnwidth]{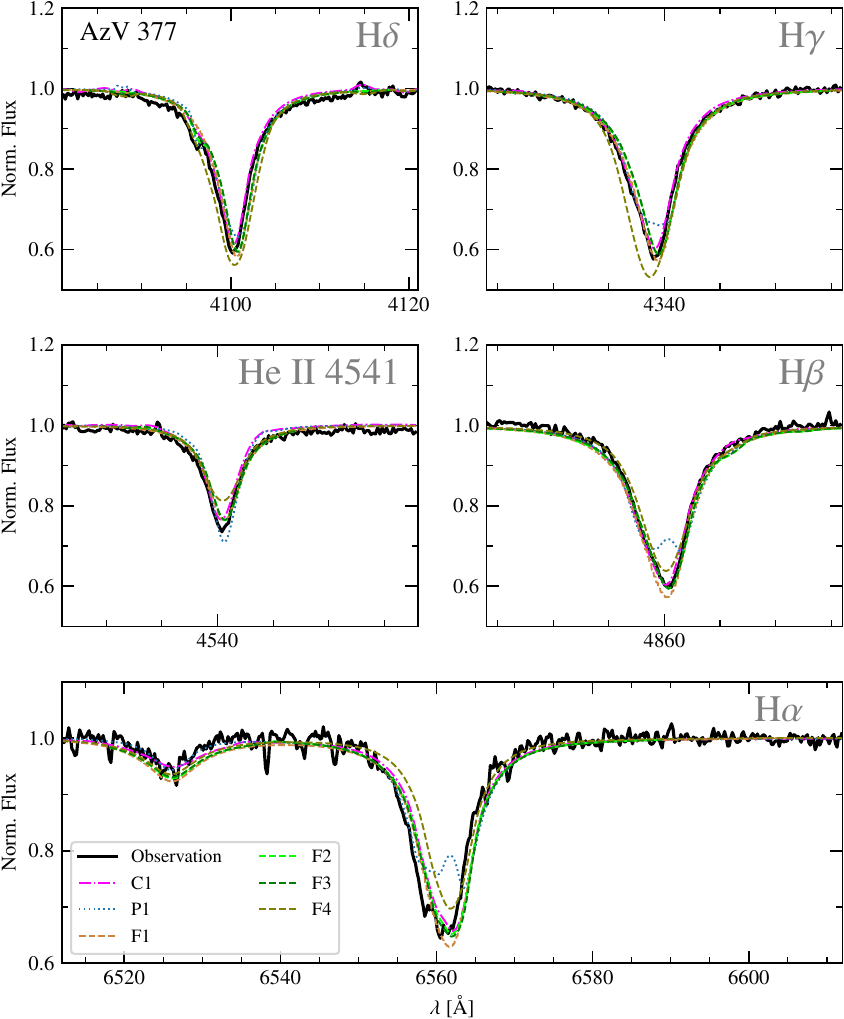}
  \caption{Comparison between the different methods in the Balmer lines and \ion{He}{ii}~$\lambda$4541 region for AzV 377.}
  \label{fig:balmer-AV377}
\end{figure}

\begin{figure}
  \includegraphics[width=\columnwidth]{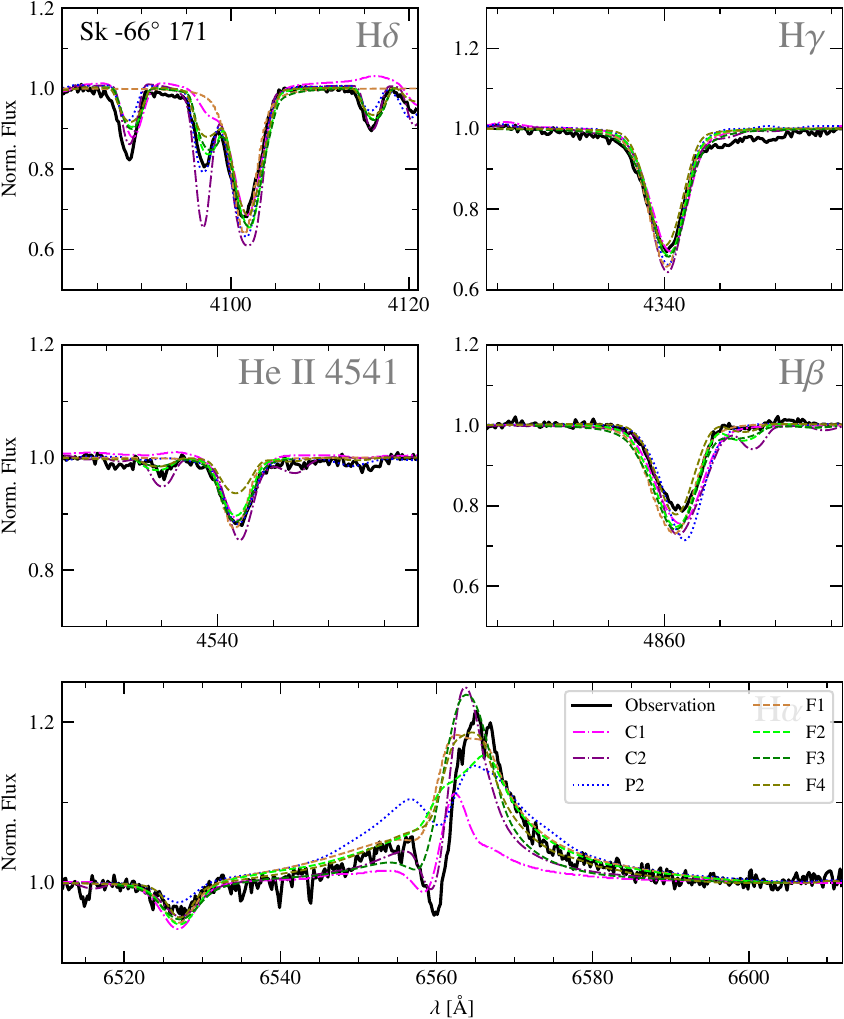}
  \caption{Comparison between the different methods in the Balmer lines and \ion{He}{ii}~$\lambda$4541 region for Sk\,-66$^{\circ}$ 171.}
  \label{fig:balmer-Sk66171}
\end{figure}

\begin{figure}
  \includegraphics[width=\columnwidth]{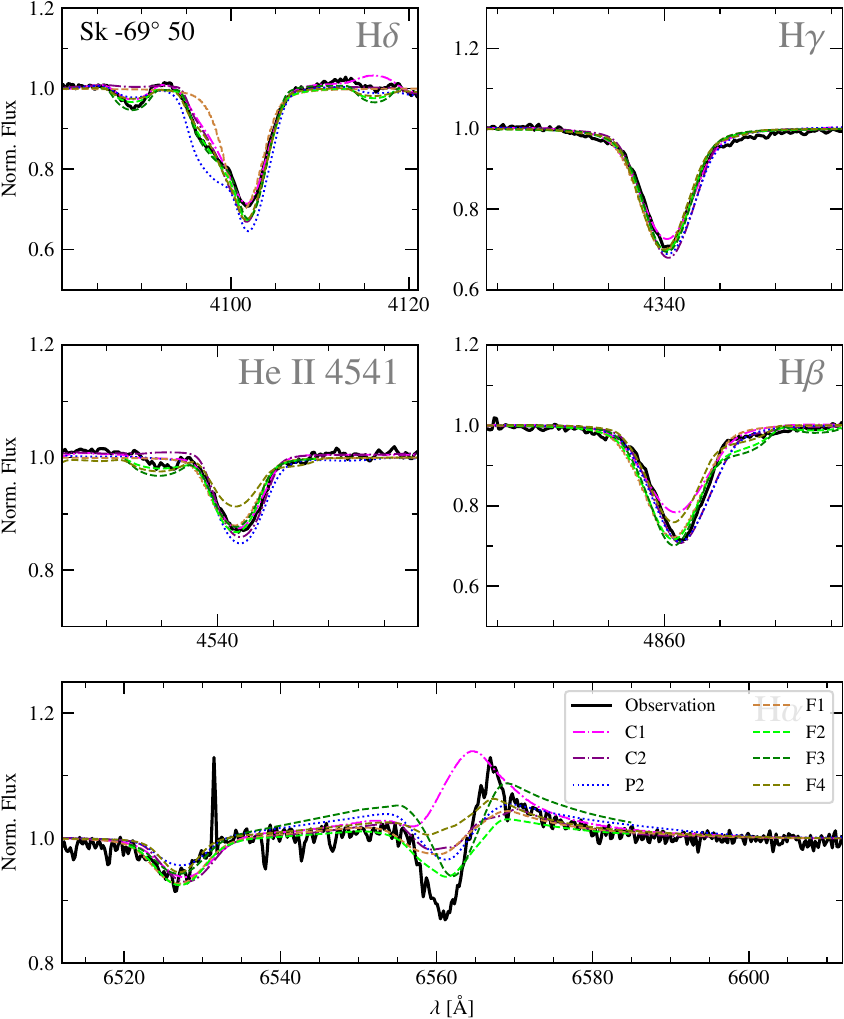}
  \caption{Comparison between the different methods in the Balmer lines and \ion{He}{ii}~$\lambda$4541 region for Sk\,-69$^{\circ}$ 50.}
  \label{fig:balmer-Sk6950}
\end{figure}

In Figs.\,\ref{fig:balmer-AV377} to \ref{fig:balmer-Sk6950}, we present panels of the first four Balmer lines and \ion{He}{ii}\,4541\,\AA\ for each target, showing the observation (in black) and the different synthetic spectra resulting from the different methods. Most of the lines are reproduced well by all of the different methods, with good agreement between them. 
In general, the grid-based fits provide a slightly poorer reproduction of the precise line shapes compared to the tailored and Kiwi-GA approaches, which is expected due to the finite spacing in the parameters.

For the more tailored approaches, the different reproduction of the line profiles does not reflect the ability of a certain code, as evident from the examples where the same underlying code yields different profile shapes. Instead, the panels illustrate the different choices made in the fitting process. This is especially evident when comparing the F2 and F3 results, which use the same code but take a different amount of data into account. Considering for example \object{AzV\,377}, the profile fits from the Kiwi-GA (F2), which only considers the optical spectrum, are quite similar to manually derived ones from the P1 method. When the UV spectra are also taken into account, the algorithm needs to make a compromise between both sets of data, and the overall fit of the shapes gets slightly worse.

\begin{figure}
  \includegraphics[width=\columnwidth]{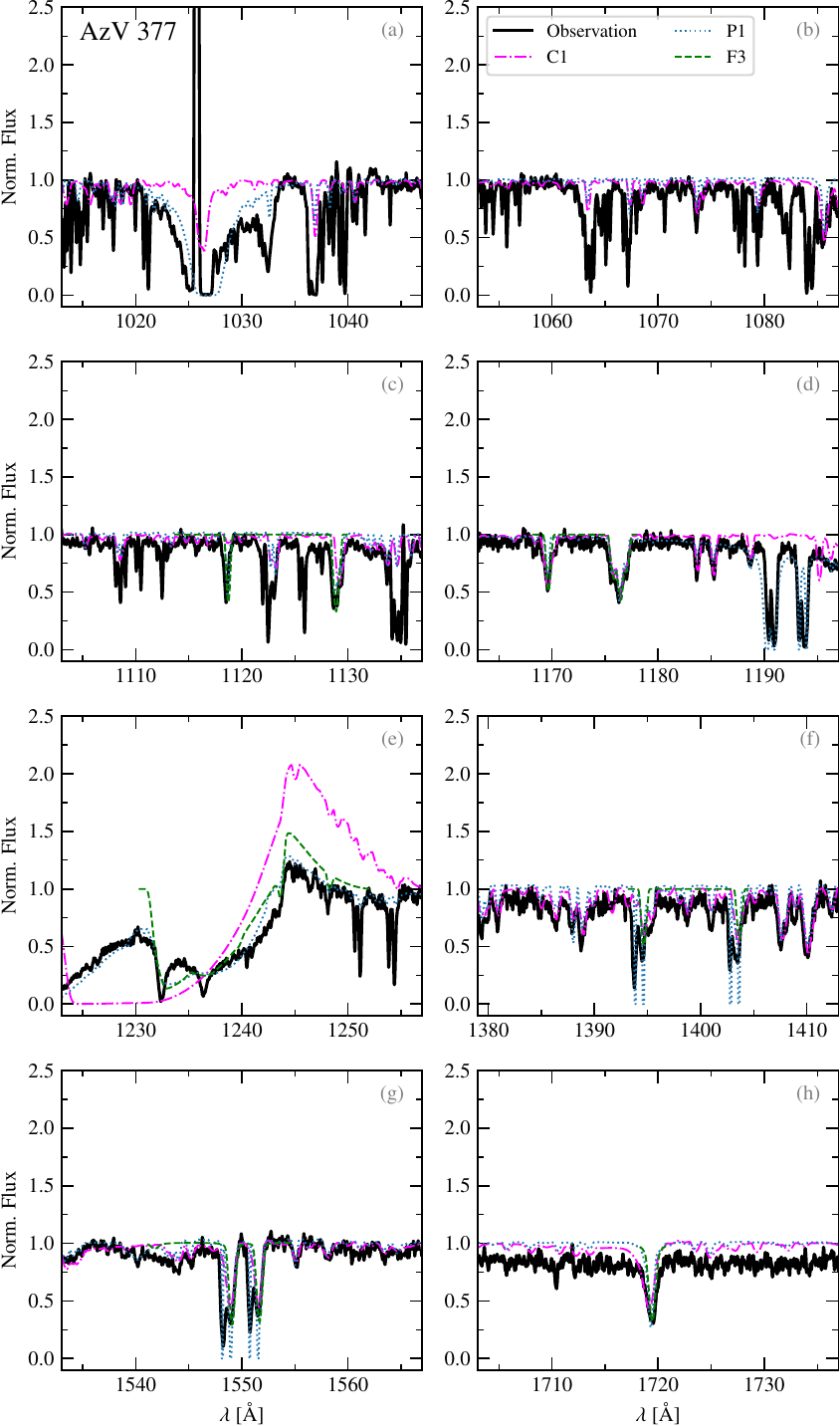}
  \caption{Comparison between the main UV profiles for AzV 377. The panels from (a) to (h) depict respectively the profiles of \ion{O}{vi} 1032/1038\,\AA, 
 \ion{S}{iv}\,1063/1073\AA\ (and \ion{He}{ii}\,1085\,\AA), \ion{P}{v}\,1118/1128\,\AA, \ion{C}{iii}\,1176\,\AA, \ion{N}{v}\,1238/1242\,\AA, \ion{Si}{iv}\,1394/1403\,\AA, \ion{C}{iv}\,1548/1551\,\AA, and \ion{N}{iv}\,1718\,\AA. Interstellar Ly$\alpha$ and Ly$\beta$ absorption affects some of the diagnostics; most notably the wing of \ion{N}{v}\,1238/1242\,\AA\ in case of larger terminal velocities (e.g., seen in AzV\,377).} 
  \label{fig:uv-AV377}
\end{figure}

\begin{figure}
  \includegraphics[width=\columnwidth]{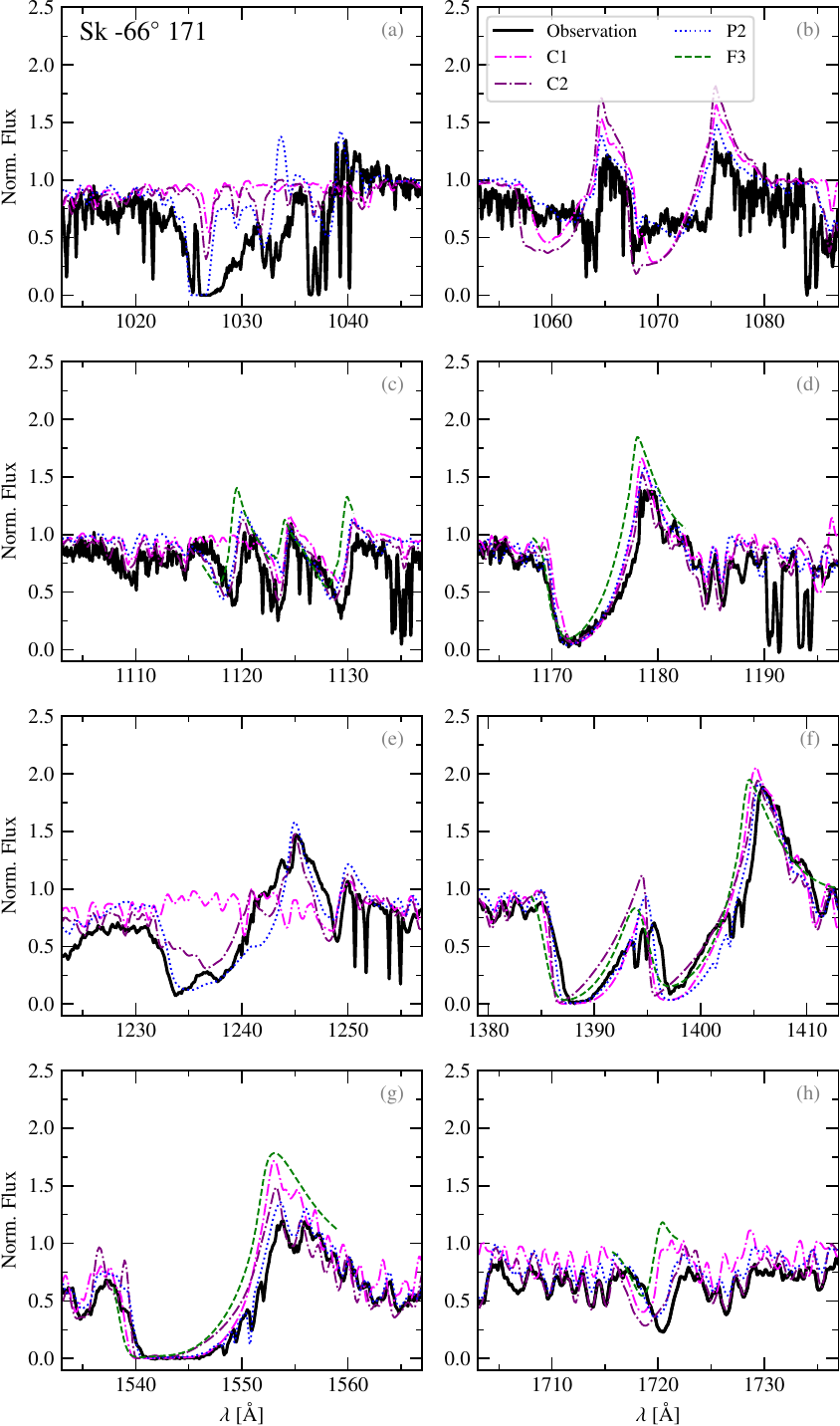}
  \caption{Comparison between the main UV profiles for Sk\,-66$^{\circ}$ 171. The spectral windows are the same as in Fig~\ref{fig:uv-AV377}, following the same order.}
  \label{fig:uv-Sk66171}
\end{figure}

\begin{figure}
  \includegraphics[width=\columnwidth]{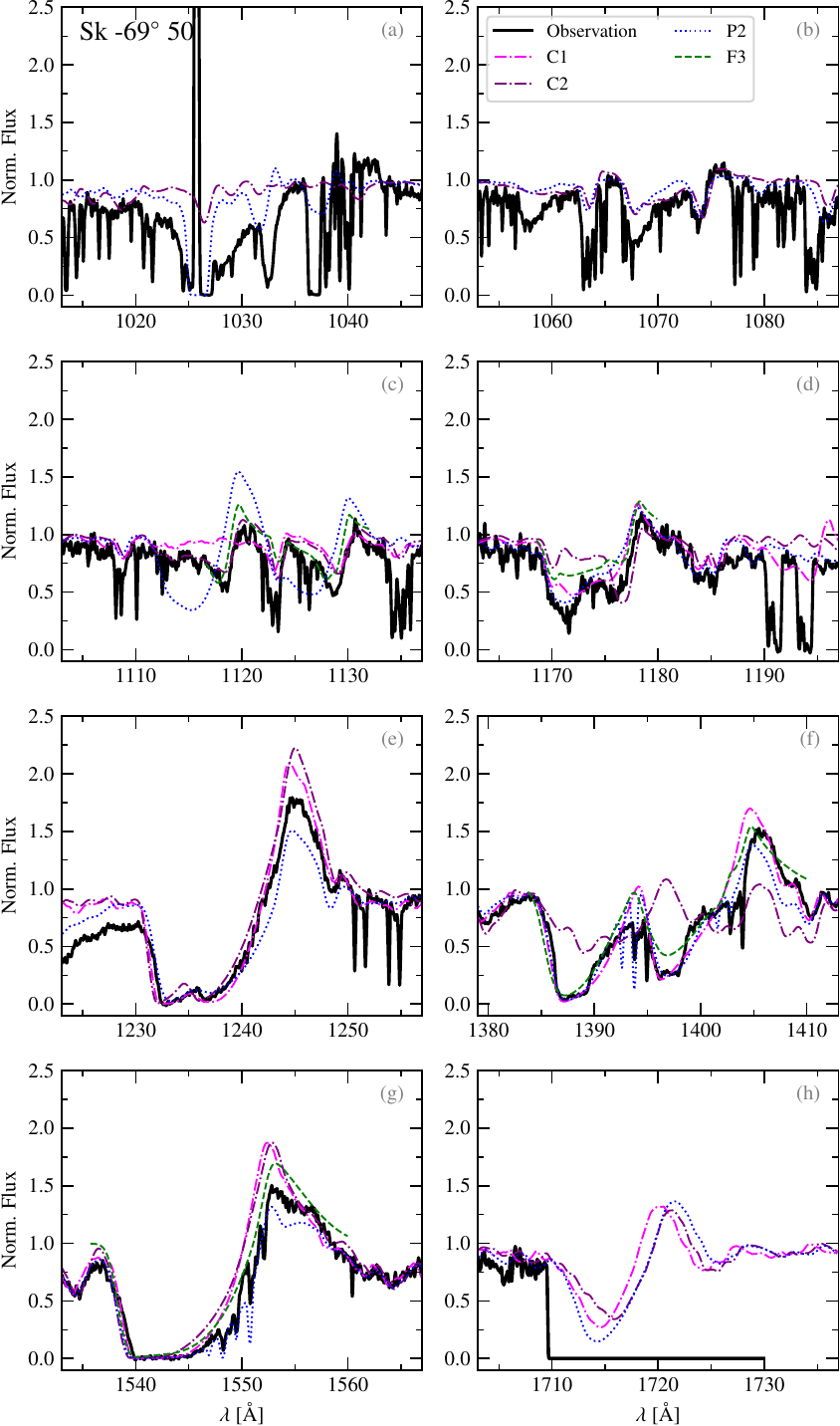}
  \caption{Comparison between the main UV profiles for Sk\,-69$^{\circ}$ 50. The spectral windows are the same as in Fig~\ref{fig:uv-AV377}, following the same order.}
  \label{fig:uv-Sk6950}
\end{figure}

In general, the reproduction of the H$\alpha$ profiles for the two LMC stars is not satisfactory. 
The variable nature of the H$\alpha$ profile in OB supergiants is well known \citep[e.g.,][]{Ebbets1982,Markova2005,Prinja2006} and the profiles tend to change even when the other diagnostics do not. Introducing different velocity laws or sophisticated clumping prescriptions with radial dependencies and/or optically thick clumps can improve the spectral fits \citep[e.g.,][]{Oskinova2007,Bouret2012,Surlan2013,Bernini-Peron2023,Ruebke2023}, but in particular pronounced P\,Cygni profiles in H$\alpha$ that differ in their velocity diagnostics from the UV profiles are challenging to reproduce. 
When examining and comparing the results for the different wavelength regimes, one further has to take into account that the available spectra for each object from FUSE, HST, and X-shooter were not taken simultaneously. Consequently, given the intrinsic wind variability, imperfections in the spectral reproductions between the wind-affected diagnostics from different regimes are to be expected.
The aim of this study is not to obtain a detailed reproduction of the H$\alpha$ profile as it does not significantly impact the total set of derived parameters, and would require an extensive additional modeling effort. The main UV diagnostic lines are featured in Figs.\,\ref{fig:uv-AV377}, \ref{fig:uv-Sk66171}, and \ref{fig:uv-Sk6950}, where the observations are compared to the synthetic spectral lines from the different methods. Unlike in the optical regime, the identification of the continuum level in the UV is cumbersome due to the forest of iron lines creating what is sometimes referred to as a ``pseudo continuum''. Several methods therefore work with flux-calibrated data in this regime, with the normalization  performed afterwards via the employed model. Therefore, discrepancies between observations and model near some of the diagnostic lines are not uncommon. Moreover, interstellar absorption affects the observation. While general reddening is accounted for in the models depicted in Figs.\,\ref{fig:uv-AV377}, \ref{fig:uv-Sk66171}, and \ref{fig:uv-Sk6950}, not all methods have applied line reddening for interstellar Ly$\alpha$ and Ly$\beta$ absorption on their synthetic spectra. In particular, Ly$\alpha$ absorption can become broad enough to affect the blue edge of \ion{N}{v}\,1238/1242\,\AA, limiting its $\varv_\infty$ diagnostic in some cases. Moreover, no correction for the considerable spectral imprint of the H$_2$ Lyman and Werner band lines below $1107\,$\AA\ has been performed, which needs to be taken into account when inspecting the panels for \ion{O}{vi} 1032/1038\,\AA\ and \ion{S}{iv}\,1063/1073\AA.

A complete overview of the spectral fitting results is illustrated in Figs.\,\ref{fig:master_AV377}, \ref{fig:master_SK6950}, and \ref{fig:master_SK66717} in the Appendix.

\subsection{Prominent spectral discrepancies}\label{sec:speclinediff}

\begin{figure}
  \includegraphics[width=\columnwidth]{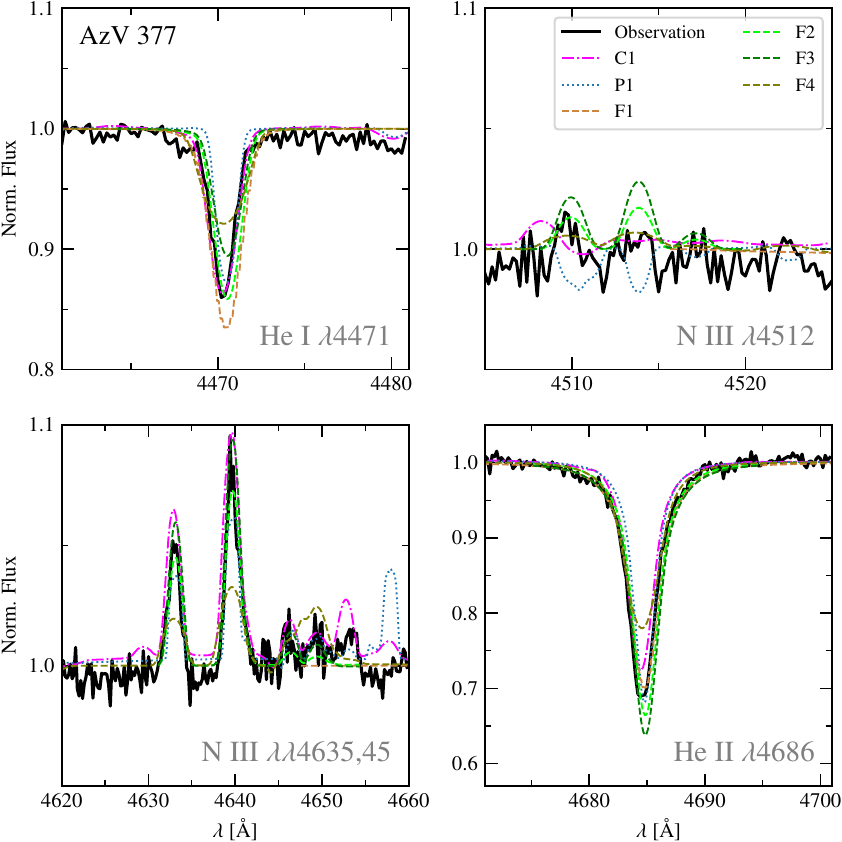}
  \caption{Comparison between the lines with more prominent disagreement on the optical for AzV 377.}
  \label{fig:dis-AV377}
\end{figure}

\begin{figure}
  \includegraphics[width=\columnwidth]{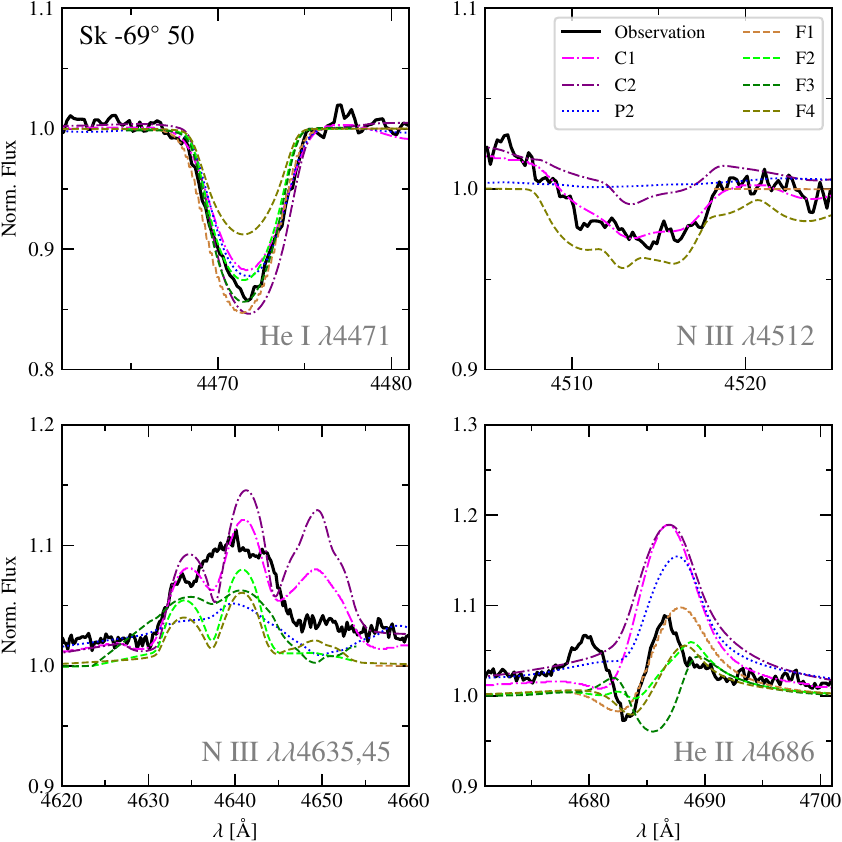}
  \caption{Comparison between the lines with more prominent disagreement on the optical for Sk\,-69$^{\circ}$ 50.}
  \label{fig:dis-Sk6950}
\end{figure}

\begin{figure}
  \includegraphics[width=\columnwidth]{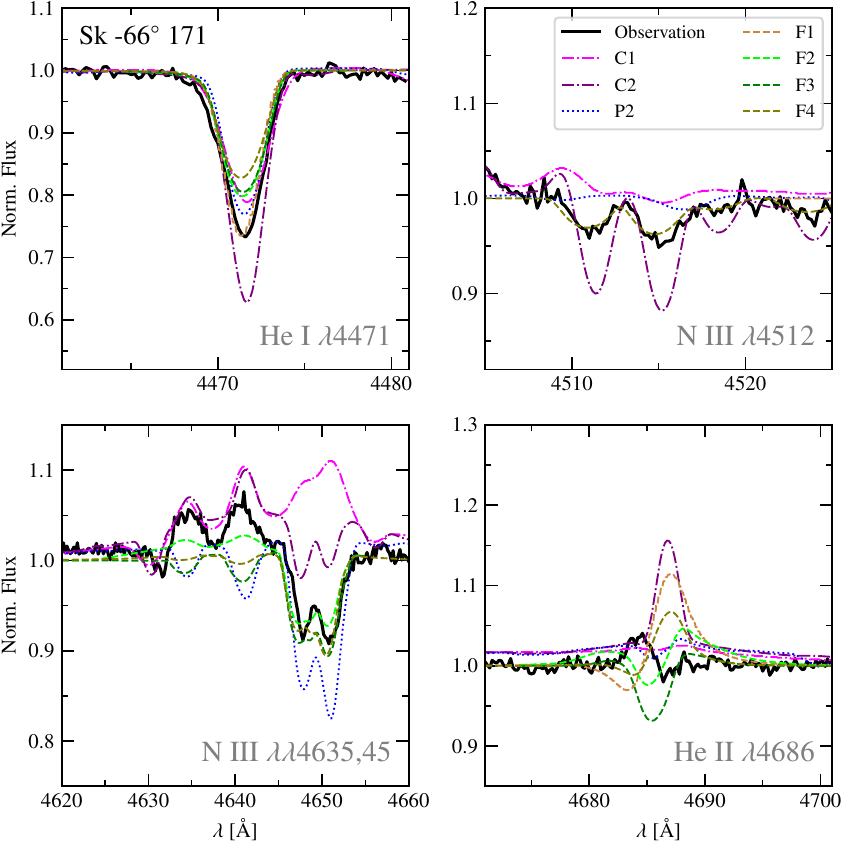}
  \caption{Comparison between the lines with more prominent disagreement on the optical for Sk\,-66$^{\circ}$ 171.}
  \label{fig:dis-Sk66171}
\end{figure}

Figures\,\ref{fig:dis-AV377}, \ref{fig:dis-Sk6950}, and \ref{fig:dis-Sk66171} show panels of those lines, which show considerable disagreement between the models in the blue optical region, which is usually the range with the best diagnostics for the stellar parameters of O stars. The depicted lines are \ion{He}{i}\,4471\,\AA, the \ion{N}{iii} multiplet between 4510 and 4524\,\AA, the \ion{N}{iii} complex around 4635\,\AA\ and 4645\,\AA, as well as \ion{He}{ii}\,4686\,\AA. All of these lines are sensitive to the wind onset region, which is one of the most uncertain regimes in stellar atmosphere calculations. In addition to its inherent physical uncertainties \citep[e.g.,][]{Cantiello2009,Sundqvist2011,Grassitelli2016,Schultz2023}, there can also be numerical artifacts stemming from connecting the (quasi-)hydrostatic domain to the wind domain described by the $\beta$-law. Consequently, lines that are formed in this onset region are subject to these inherent uncertainties and can for example be affected in their appearance when making minor changes to parameters related to the connection criteria, such as the assumed line broadening in the radiative transfer ($\varv_\text{Dop}$) or the $\beta$-value. Consequently, He lines should not be trusted blindly as a diagnostic for O stars between $30$ and $35\,$kK, which essentially covers both of our LMC targets in this work. The observed emission of some of the \ion{N}{iii} lines in Figs.\,\ref{fig:dis-AV377}, \ref{fig:dis-Sk6950}, and \ref{fig:dis-Sk66171} is also very sensitive to these connection settings. Moreover, there is an overlap of two resonance lines from \ion{N}{iii} and \ion{O}{iii} around $374\,$\AA\ in the extreme ultraviolet (EUV) that affects models at least in the temperature domain between $\sim$$33$ and $35\,$kK \citep[][]{RiveroGonzalez2011}. Minor details of the modeling approach affect the resulting optical lines, including the wavelengths in the atomic data, the broadening assumed in the radiative transfer, as well as the treatment of line overlaps \citep[see][for a more in-depth discussion]{RiveroGonzalez2011,Puls2020}. Consequently, there are notable issues in the reproduction of \ion{N}{iii} with different results for the different stars and no clear preference for any of the methods. The situation gets generally better for the hotter SMC O5 dwarf with the remaining differences in the He lines mainly arising from fixed-grid approaches.

For late O supergiants, \ion{He}{ii} recombination can either set in or be avoided even when parameters such as $T_\text{eff}$ or $\dot{M}$ undergo only minor changes. This effect is likely able to explain some the observed deficiencies for the \ion{He}{ii}\,4686\,\AA\ line, in particular for methods such as C2 or P2, which provide a good reproduction of the UV lines. UV and optical diagnostics can in practice favor slightly different temperatures, which cannot be resolved within a given atmosphere code version and thus demand a compromise in the fit, regardless of the specific method used. A similar problem exists for the \ion{N}{iii}\,4512\,\AA\ line. While such discrepancies between different wavelength regimes can arise due to the nonsimultaneous observations, they can also reflect limitations in the current treatment of 1D model atmospheres, for example with respect to the assumptions of a single wind velocity and ionization structure.

In the UV range, there is generally a good consensus between the methods taking this regime into account. Larger discrepancies mainly occur for high-ionization lines such as \ion{O}{vi}\,1032/1038\,\AA\ in the case of AzV 377 or \ion{N}{v}\,1238/1242 for the LMC stars. When calculating atmosphere models with the necessary $T_\text{eff}$ derived from the remaining diagnostics, the population of the levels corresponding to these lines is not sufficiently large to reproduce the observed strength. To remedy this shortcoming, the model codes have the ability to include additional X-rays\footnote{as observed in hot massive stars \citep[e.g.,][]{Chlebowski1991,Rauw2015,Crowther2022}, and suggested to result from shock emission induced by wind-instabilities \citep[e.g.,][and references therein]{Feldmeier1997}} in the wind, but not all methods make use of this (see Appendix\,\ref{sec:mdetail} for the handling of the individual methods). The \ion{P}{v}\,1118/1128\,\AA\ doublet is further known to be sensitive to optically thick clumping \citep[e.g.,][]{Sundqvist2011, Surlan2013}, which is not taken into account in most of the methods, except F3.

\subsection{Spectral energy distribution}\label{sec:sed}

\begin{figure}
  \includegraphics[width=\columnwidth]{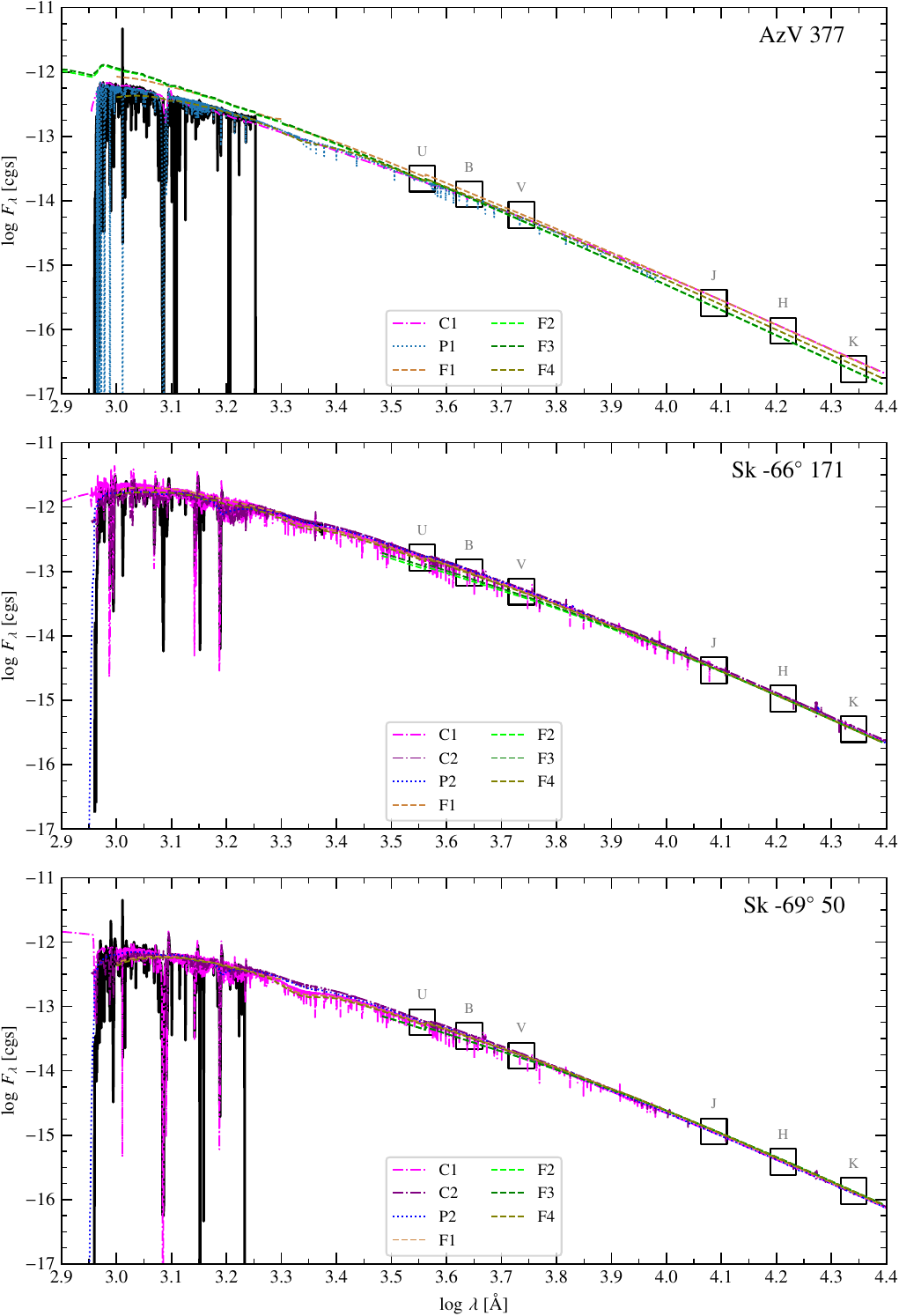}
  \caption{Modeled and observed SED of the targets. The flux points correspond to the magnitudes listed in Table~\ref{tab:sample}, following the same order (from bluer to redder). The observed flux-calibrated UV spectra correspond to those acquired by ULLYSES.}
  \label{fig:SED-stars}
\end{figure}

\begin{figure}
  \includegraphics[width=\columnwidth]{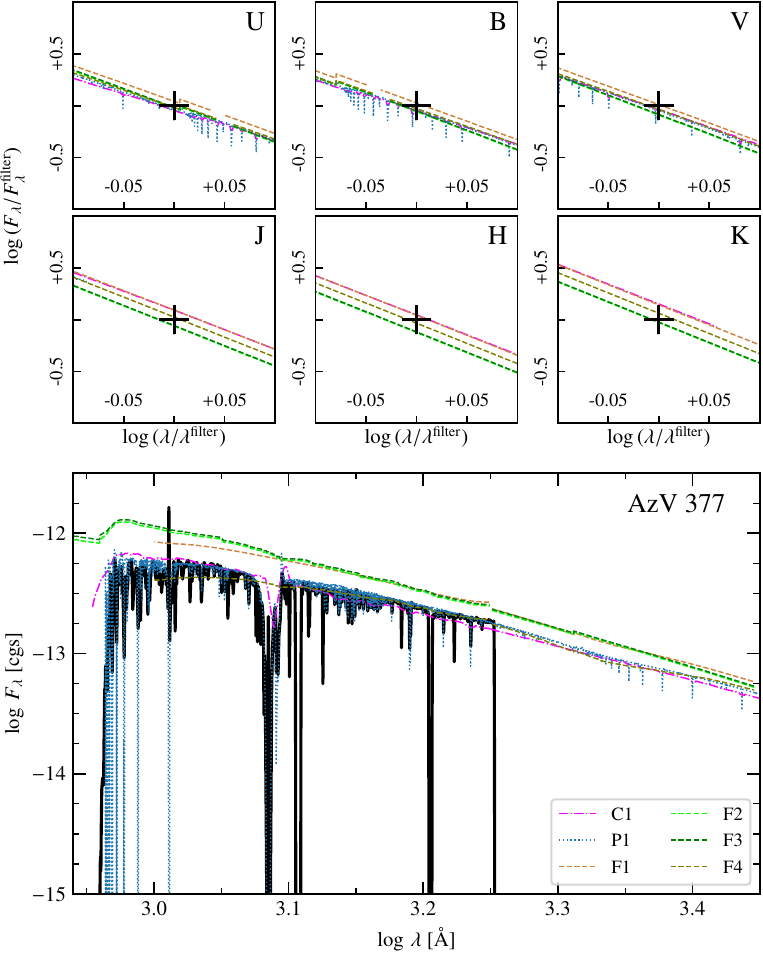}
  \caption{Zoom-in comparison of the SEDs from the different models for AzV\,377 around the applied photometry (small upper panels, crosses mark photometric measurements) and the flux-calibrated UV spectra (big lower panel).}
  \label{fig:SED-zoom-AV377}
\end{figure}

\begin{figure}
  \includegraphics[width=\columnwidth]{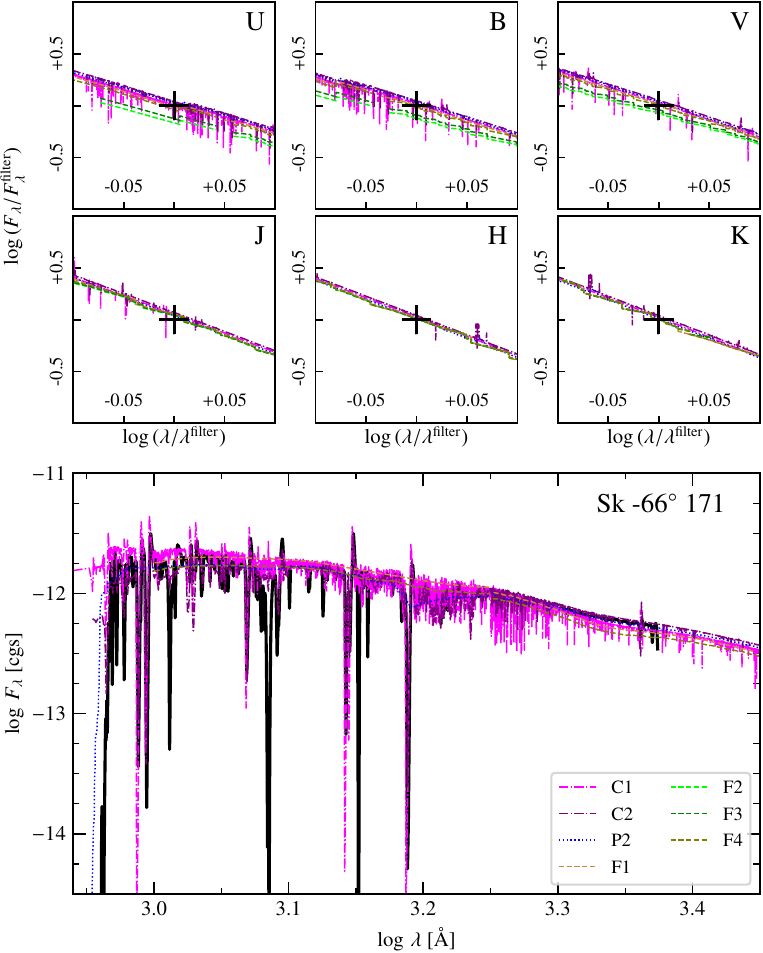}
  \caption{Same as Fig.\,\ref{fig:SED-zoom-AV377}, but for Sk\,-66$^{\circ}$ 171.}
  \label{fig:SED-zoom-Sk66171}
\end{figure}

\begin{figure}
  \includegraphics[width=\columnwidth]{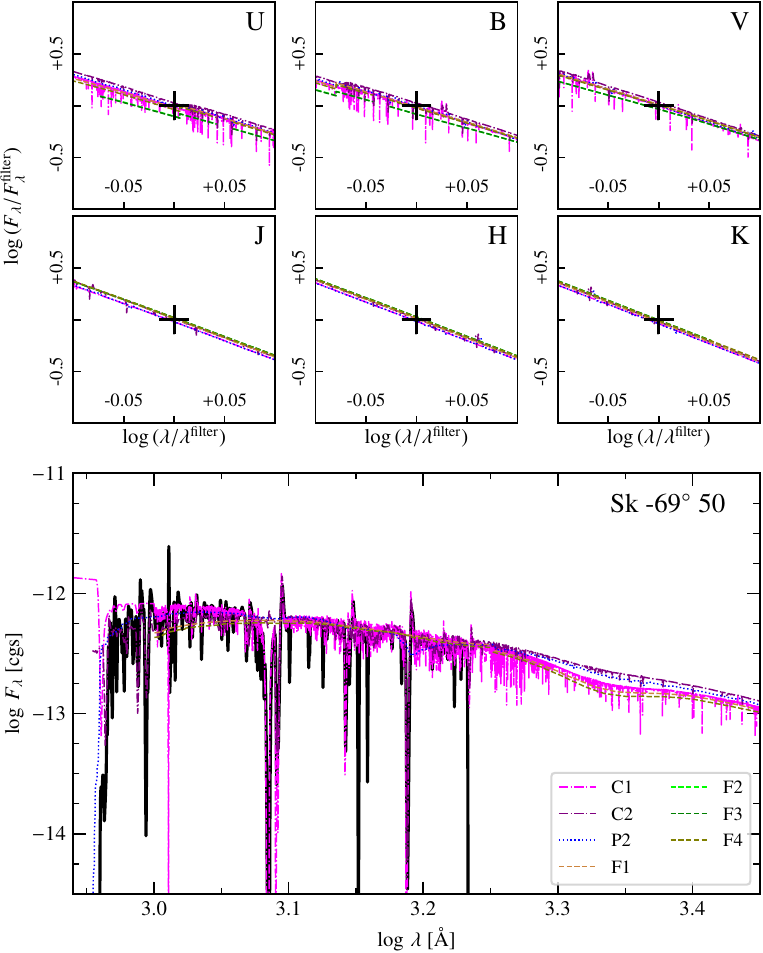}
  \caption{Same as Fig.\,\ref{fig:balmer-AV377}, but for Sk\,-69$^{\circ}$ 50.}
  \label{fig:SED-zoom-Sk6950}
\end{figure}

In Fig.\,\ref{fig:SED-stars}, we show the reproduction of the observed spectral energy distribution (SED) from flux-calibrated UV spectra plus optical and IR photometry with the different models for all three targets. Apart from the methods that do not take the UV spectra into account, all methods yield an acceptable reproduction of the SED, illustrating the phenomenon that all O stars essentially appear ``blue'', meaning that the flux we see in the optical and beyond simply maps what would be the Raleigh-Jeans of a blackbody (of $\approx$ $0.8 \dots 0.9\,T_\text{eff}$). Nevertheless, hot stars can also deviate significantly from a general black body shape. This is most obvious in the UV, where the large number of iron lines (``iron forest'') forms a ``false continuum'' and significantly alters the emerging shape of the flux distribution. Extending with their wavelengths far into the usually unobservable EUV, the large number of transitions from iron (and other elements with complex electron configurations) lead to a ``blanketing'' effect that alters not only the ionization and temperature structure of a hot star \citep[e.g.,][]{Dreizler1993,Hillier1998,Graefener2002,Lanz2003} but also the spectral shape, in that the continuum emission is enhanced at longer wavelengths \citep[e.g.,][]{Hummer1982,Abbott1985}.
For stars with stronger winds, the slope of the flux decline is further altered by additional free-free emission contributing to the continuum with the relative contributions getting larger for longer wavelengths. Consequently, the temperature determination for hot stars cannot be achieved with photometry \citep[see also][who in particular discuss the effect of blanketing]{Hummer1988}, but requires a detailed spectroscopic analysis with lines of different ionization stages acting as crucial temperature indicators (cf.\ the method descriptions in Sect.\,\ref{sec:fastwind}, \ref{sec:cmfgen}, \ref{sec:powr}). 

To investigate the reproduction of the SED in more detail, we show zoom-ins for the three targets around the photometric magnitudes and the flux-calibrated UV spectra in Figs.\,\ref{fig:SED-zoom-AV377}, \ref{fig:SED-zoom-Sk66171}, and \ref{fig:SED-zoom-Sk6950}. For most of the filters, there is excellent agreement between the model spectra and the photometry. Notable shifts occur in particular for the F1, F2, and F3 methods, which do not take the flux-calibrated UV spectra directly into account, but either use normalized parts of the UV spectrum (F3) or do not consider the UV part of the spectrum. To derive the luminosity, these methods use anchor magnitudes and a reddening law as described in Appendix~\ref{sec:mdetaillumi}. This seems to lead to a slight overestimation of the UV flux for the hottest target in the sample. Nevertheless, while the shift in Fig.\,\ref{fig:SED-zoom-AV377} appears quite dramatic, this is mostly a result of employing an extinction law that is not adjusted for the UV regime, and the difference in the derived luminosity is less than $0.1\,$dex compared to the other methods, which is a typical error margin (see also Table\,\ref{tab:spread}).

\subsection{Abundances}\label{sec:abundances}

Unless one purely relies on a fixed grid of models, the determination of the elemental abundances usually requires additional rounds of iteration among the necessary model calculations. In OB stars, the temperature in many cases cannot be sufficiently constrained without taking metal lines of different ionization stages into account. Consequently, abundance effects can overlap with temperature effects and in several (though not all) methods the finer tuning of the abundances is only performed once the main stellar parameters are robustly determined. Moreover, some of the involved grids only have a fixed set of abundances. Given that we do not aim to further ``tune'' the derived values after comparing our initially obtained values, we do not expect our abundances to be as robust as they usually would be in studies focusing on particular stars. Still, we can identify general trends and discrepancies between the analysis methods.

For both of the LMC O supergiants, almost all methods yield a He enrichment. Almost unanimously, all methods predict a He mass fraction of $\sim$$0.35$ for the O7(n)(f)p target, while the scatter is larger for the O9 Ia star with values reaching from almost zero enrichment ($X_\text{He} = 0.26$) up to $X_\text{He} = 0.39$. For the SMC O5 SMC dwarf, there is a similar scatter, interestingly now with different approaches yielding the higher enrichment of up to $X_\text{He} = 0.40$. Unless hydrogen is strongly depleted, the imprints of He enrichment are more subtle, making the determination more cumbersome than that of other abundances such as CNO.

All methods that determined CNO abundances found strong nitrogen enrichment for all of the studied targets.
Converted to mass fractions, the LMC baseline abundances \citep[cf.][]{Vink2023} are $X_\text{C} = 9.06 \cdot 10^{-4}$, $X_\text{N} = 1.11 \cdot 10^{-4}$, and $X_\text{O} = 2.96 \cdot 10^{-3}$, yielding a combined CNO abundance of $3.98 \cdot 10^{-3}$. For the SMC, values are $X_\text{C} = 2.34 \cdot 10^{-4}$, $X_\text{N} = 0.47 \cdot 10^{-4}$, and $X_\text{O} = 1.33 \cdot 10^{-3}$, yielding a total CNO mass fraction of $1.61 \cdot 10^{-3}$. With nitrogen mass fractions between $7.4 \cdot 10^{-4}$ (F2) and $1.9 \cdot 10^{-3}$ (C1), enrichment factors between 15 and 40 are found for the SMC dwarf AzV\,377. The situation is less clear for carbon, while the depletion of oxygen is clearly confirmed. The total CNO abundance found for AzV\,377 scatters between $0.9$ and $2.0$ times the baseline value, preventing any more robust conclusions as to whether our sample star is actually slightly more metal rich than is presumed to be typical for the SMC.

A similar scatter around the total CNO baseline abundances is found for the two LMC targets, with Sk\,-69$^{\circ}$ 50 yielding slightly higher factors ($0.8\dots2.4$) than Sk\,-66$^{\circ}$ 171 ($0.78\dots1.96$). Nevertheless, the two targets are quite different in their nitrogen enrichment, which is found to be much higher for the O7(n)(f)p target, where most methods yield enrichment factors of $\sim$$40$ (except C2) compared to more moderate factors of $\sim$$10$ for the O9 supergiants. Clearly, Sk\,-69$^{\circ}$ 50 seems to be the most evolved target in our sample, as all methods find carbon to be depleted, while the situation is less clear for the other two sample stars.

\subsection{Ionizing fluxes}\label{sec:ionflux}

All of our sample stars show a considerable flux beyond the hydrogen ionization edge. As also evident from the tabulated results (Tables \ref{table:paramsummary-azv377}, \ref{table:paramsummary-sk6950}, and \ref{table:paramsummary-sk66171}), the ionizing fluxes depend strongly on the temperature and luminosity of the stars with higher temperatures and luminosity yielding higher fluxes. The fluxes beyond the \ion{He}{ii} ionization edge are more complicated, as they also strongly depend on the wind density. For denser winds, even very hot stars can yield essentially no \ion{He}{ii} ionizing flux. This is the case for the two supergiants in the sample, for which the models formally yield photon fluxes of up to $\sim$$10^{42}\,\mathrm{s}^{-1}$. These values are orders of magnitude below considerable contributors such as early O dwarfs \citep[e.g.,][]{Smith2002,Martins2021}, hot, thin-wind Wolf-Rayet stars \citep[e.g.,][]{Crowther2006,Sander2023}, or luminous envelope-stripped stars below the WR regime \citep[e.g.,][]{Goetberg2023,Ramachandran2023}. We note that the absolute numbers of the reported magnitude must be taken with care as these low values can be subject to numerical uncertainties, for example if models are optically thick up to the outer boundary at some of the corresponding wavelengths. The studied O5 dwarf has a \ion{He}{ii}-ionizing photon flux of $\sim$$10^{43}\dots10^{44}\,\mathrm{s}^{-1}$, which is actually in line with expectations for the derived HRD position \citep{Martins2021}.

\section{Discussion}\label{sec:discussion}

\subsection{HRD position and evolutionary status}

\begin{figure}[t]
  \includegraphics[width=\columnwidth]{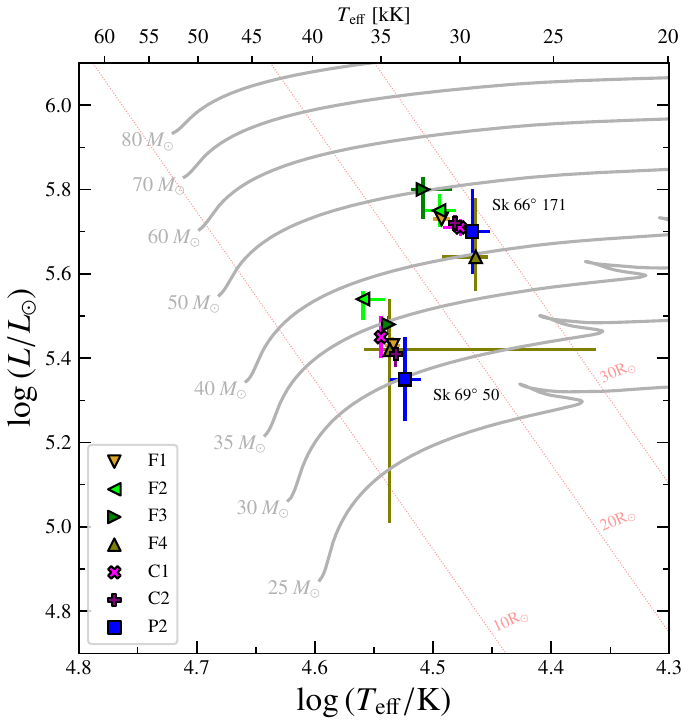}
  \caption{HRD with the obtained positions for the two LMC stars Sk\,-69$^{\circ}$ 50 and Sk\,-66$^{\circ}$ 171 -- indicated by the respective labels. For comparison, tracks from \citet[up to $M_\text{init} = 50\,M_\odot$]{Brott2011} and \citet[from $60\,M_\odot$]{Koehler2015} are shown.}
  \label{fig:hrd-lmc}
\end{figure}

\begin{figure}[t]
  \includegraphics[width=\columnwidth]{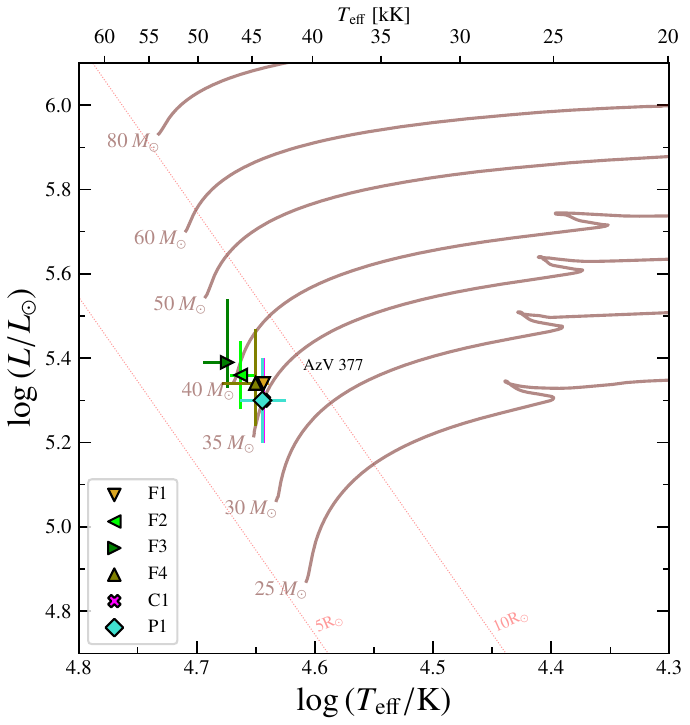}
  \caption{HRD with the obtained positions for the SMC star AzV 377. For comparison, tracks from \citet[up to $M_\text{init} = 60\,M_\odot$]{Brott2011} and \citet[from $80\,M_\odot$]{Koehler2015} are shown.}
  \label{fig:hrd-smc}
\end{figure}

In Figs.\,\ref{fig:hrd-lmc} and \ref{fig:hrd-smc}, we provide an overview of all the obtained positions of our three sample stars with the different methods in the Hertzsprung–Russell diagram (HRD). For all targets, there is a noticeable trend that hotter solutions tend to come with higher luminosities. This can be understood when considering the SED fit (cf.\ Sect.\,\ref{sec:sed}). If one aims to fit the same photometric SED with a hotter model atmosphere, this enforces a slightly higher reddening and thus a higher luminosity\footnote{The precise determination order of the values ($T_\text{eff}$, $L$, $E_{B-V}$) is dependent on the method, with some calculating the reddening upfront and others only after $T_\text{eff}$ is found. See Appendix~\ref{sec:mdetaillumi} for details.}. 
In particular, we see the F2 and F3 methods yielding the highest luminosities (and P1/P2 the lowest), in line with our findings for the SED fits (cf.\ Sect.\,\ref{sec:sed}).
The obtained stellar parameters are therefore not independent and changes in one parameter can propagate into other quantities. Given the high number of input parameters into stellar atmosphere models and the nontrivial effects of their variation, only the most obvious ones can be calculated in the form of a rigorous error propagation. In all other cases, the only feasible option is to assume a larger error than obtained from statistical considerations. 
As for cool-star atmospheres, systematic errors are usually not considered at all. These can arise for example because of uncertain atomic data, method-inherent approximations, or code-specific numerical treatments.

\subsection{Mass discrepancy}

\begin{figure}
  \includegraphics[width=\columnwidth]{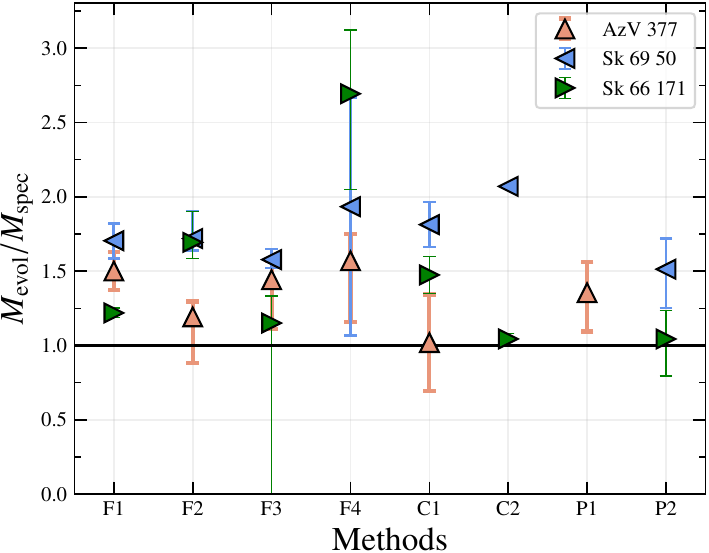}
  \caption{Ratio of the determined spectroscopic masses to evolutionary masses (based on \citealp{Brott2011} or \citealp{Koehler2015} respectively) for the different methods and sample stars.}
  \label{fig:mspec-mevol}
\end{figure}

From comparing the derived positions in the HRD, one can derive ``evolutionary masses'' $M_\text{evol}$, assuming that a given set of tracks (and the interpolation between them) sufficiently describe the history of the stars. Using the tracks from \citet{Brott2011} and \citet{Koehler2015}, also shown in the HRDs (Figs.\,\ref{fig:hrd-lmc}, \ref{fig:hrd-smc}), we derived $M_\text{evol}$ for each of our three sample stars and compare them to the spectroscopically derived mass $M_\text{spec}$ in Fig.\,\ref{fig:mspec-mevol}, employing the same $\chi^2$ approach as in \citet{Bernini-Peron2023}. The age-dependent $M_\text{evol}$ is generally lower than the zero-age main sequence mass $M_\text{init}$ because of the decreasing mass along each evolutionary track. In all cases, the ratio between $M_\text{evol}$ and $M_\text{spec}$ is $\geq 1.0$, meaning that the masses inferred from the evolutionary tracks are similar to or higher than the ones from spectroscopy. This so-called ``mass discrepancy'' is a long-standing issue in the analysis of hot stars \citep[e.g.,][]{Herrero1992,Markova2018}. Interestingly, detailed spectral analyses for detached pre-interaction binaries \citep[e.g.,][]{Mahy2017,Mahy2020} obtained spectroscopic masses that match evolutionary estimates, raising questions about whether or not the evolutionary status presumed in the tracks applies to all of our analyzed sample stars.

With the determination of the radius $R_{2/3}$ from $L$ and $T_\text{eff}$, the different luminosities for example also affect the derived spectroscopic masses. 
The highest ratio occurs for the F4 analysis of Sk\,-66$^{\circ}$ 171 and is a consequence of the HRD ``outlier'' position, where F4 yielded a much lower luminosity than the other methods. Disregarding this point, the ratios for AzV\,377 and Sk\,-66$^{\circ}$ 171 are relatively moderate, with values ranging between 
only $1.0$ and $\sim$$1.6$. In particular, we note that for the late O supergiant Sk\,-66$^{\circ}$ 171 the more tailored methods (F3, C2, P2) yield the best matches. In contrast, the other LMC star, Sk\,-69$^{\circ}$ 50, has a systematic shift and never shows a ratio below $1.5$. It is therefore likely that this star is not sufficiently described by the evolutionary tracks. Moreover, the origin of the class of Onfp stars of which Sk\,-69$^{\circ}$ 50 is a member has been subject to speculation, including the suggestion that these objects are products of stellar mergers \citep{Walborn2010}. 
For the SMC O5 dwarf AzV\,377, the evolutionary situation is less clear, with the methods scattering between good agreement and notable discrepancy. 

To remove the mass discrepancies, the derived $\log g$ values would have to be larger by $0.1$ to $0.3$\,dex. Even when neglecting Sk\,-69$^{\circ}$ 50 due to its probably more evolved status, the necessary increase would have to be up to $0.18$\,dex to account for a mass discrepancy factor of $1.5$. One ingredient that could increase $\log g$ is the inclusion of a turbulence term in the hydrostatic equation (cf.\, Eq.\,\ref{eq:loggturb}), but so far only one of the codes applied in this work (PoWR) does that. \citet{Markova2018} studied the mass discrepancy of Galactic O-type stars using both \textsc{CMFGEN} and \textsc{Fastwind}, finding comparable discrepancy values with the two codes. Hence, while it is too early to derive a clear tendency and there are prominent exceptions such as the C2 method for AzV\,377 yielding only a small discrepancy (cf.\ Fig.\,\ref{fig:mspec-mevol}), the inclusion of a turbulence term in the hydrostatic equation and its resulting increase in $\log g$ could mark an important step to minimize the occurrence of mass discrepancies. A more focused study on the inclusion of different microturbulent velocity values in the hydrostatic equation ---further motivated by recent 2D simulation results from \citet{Debnath2024}--- would be necessary to better judge this effect.

\subsection{Wind parameters}

\subsubsection{Terminal velocities}

\begin{figure}
  \includegraphics[width=\columnwidth]{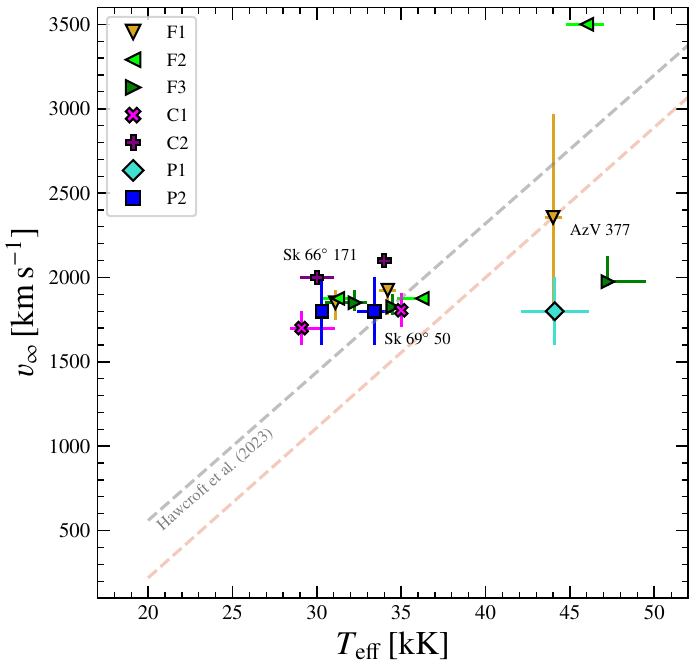}
  \caption{Comparison of the derived (F1, F2: assumed) terminal wind velocity $\varv_\infty$ versus the derived effective temperature. The dashed lines denote the LMC (gray) and SMC (salmon) relations from \citet{Hawcroft2023}.}
  \label{fig:vinf-teff}
\end{figure}

With the direct availability of the terminal wind velocity from the UV spectra, all of the methods making use of these data obtain very similar values for $\varv_\infty$. Notably, all of our targets show terminal velocities between $\sim$$1800$ and $2000\,\mathrm{km}\,\mathrm{s}^{-1}$. This is not expected from their spectral types, as evident also from Fig.\,\ref{fig:vinf-teff}, where we plot the derived values for $\varv_\infty$ as a function of $T_\text{eff}$ and compare them to the trends derived in \citet[paper III of the XShootU series]{Hawcroft2023}. The methods assuming a terminal velocity (F1, F2, see Appendix~\ref{sec:mdetailwind}) consequently overestimate $\varv_\infty$ for the SMC O5 dwarf AzV\,377, while the two LMC stars are closer to the expected relation. The low terminal velocity for the SMC star is surprising given its high temperature and ---as we discuss below--- does not coincide with a higher mass-loss rate.

\begin{figure}
  \includegraphics[width=\columnwidth]{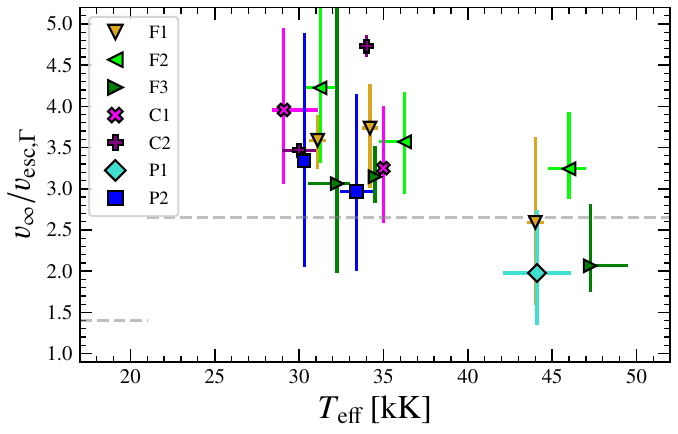}
  \caption{Ratio between the terminal wind velocity $\varv_\infty$ and the effective escape velocity $\varv_{\text{esc},\Gamma}$ as a function of $T_\text{eff}$. The dashed gray line shows the empirical results from \citet{Kudritzki2000}.}
  \label{fig:vinf-vesc}
\end{figure}

The values for the two LMC stars align well with the $\varv_\infty(T_\text{eff})$ trend reported by \citet{Hawcroft2023}. In particular, the value for Sk\,-69$^{\circ}$ 50 matches perfectly, while the value for the late O supergiant Sk\,-66$^{\circ}$ 171 is a slightly higher than the value from the trend formula. To compare our findings with the predictions from the mCAK theory for radiation-driven winds of OB stars \citep{Castor1975,Pauldrach1986}, we also plot the ratio of $\varv_\infty$ to the effective escape velocity,
\begin{equation}
  \label{eq:vescg}
  \varv_{\text{esc},\Gamma} := \sqrt{\frac{2 G M}{R} \left( 1 - \Gamma_\text{e}\right)}
,\end{equation}
in Fig.\,\ref{fig:vinf-vesc}. Usually, Eq.\,\eqref{eq:vescg} is evaluated at $R = R_{2/3}$, with $M$ being the stellar mass, $\Gamma_\text{e} \propto L/M$ denoting the classic Eddington parameter taking only electron scattering opacity into account, and $G$ the gravitational constant. The determination of $\varv_{\text{esc},\Gamma}$ is subject to a variety of error propagations, resulting from the uncertainties in for example the determination of $\log g$ and the luminosity $L$. Consequently, the derived $\varv_\infty/\varv_{\text{esc},\Gamma}$ ratios in Fig.\,\ref{fig:vinf-vesc} show considerable error bars, with the smallest bars actually resulting from incomplete error estimations (e.g., for C2). Considering the large error bars, one could argue that the obtained values are not in conflict with the presumed ratio of 2.65 times the escape ratio found by \citet{Kudritzki2000}, who slightly updated the value from the factor 2.6 found by \citet{Lamers1995}. However, there is a systematic trend for the two LMC stars towards higher ratios, which is qualitatively in line with the mass discrepancy trend found, namely the tendency to determine lower spectroscopic masses for these stars than what would be inferred from evolutionary tracks. For the SMC dwarf, the opposite trend is seen, with the two methods determining $\varv_\infty$ yielding ratios of $\sim$$2.1$. This actually aligns nicely with the expected metallicity dependence of $\varv_\infty \propto Z^{0.2}$ for $\varv_\infty$ from \citet{Vink2021} and \citet{Hawcroft2021}. Assuming $0.5\,Z_\odot$ for the LMC and $0.14\,Z_\odot$ for the SMC, the ratio of $2.6$ is expected to reduce to $2.02$, which is well in line with the findings for AzV\,377.

\subsubsection{Mass-loss rates}

\begin{figure}
  \includegraphics[width=\columnwidth]{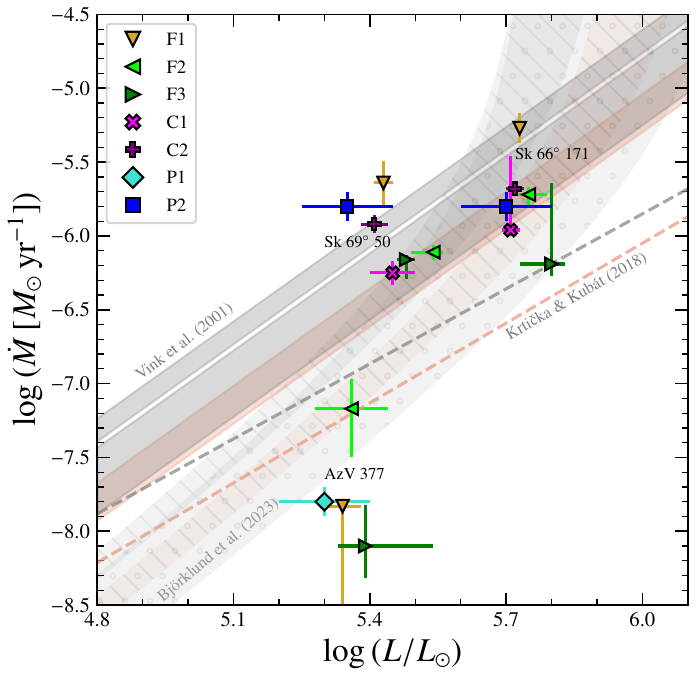}
  \caption{Mass-loss rate $\dot{M}$ versus stellar luminosity $L$ for our sample stars analyzed with the different methods. For comparison, we also plot the mass-loss recipes from \citet[solid shading]{vink2001}, \citet[dashed lines]{Krticka2018}, and \citet[hatched shading]{Bjoerklund2021}. The SMC comparison data are drawn in light red.}
  \label{fig:mdot-l}
\end{figure}

The mass-loss rates determined by the different methods are shown in Fig.\,\ref{fig:mdot-l}, where we also plot three predictions from the literature, namely \citet{vink2001}, \citet{Krticka2018}, and \citet{Bjoerklund2023}. For F1, the reported value should be considered as an upper limit due to the use of unclumped wind models\footnote{Model F4 "only" provides values for $Q_\text{ws}$, also using unclumped models, whilst actual mass-loss rates have not been calculated.}. To account for the fact that the formulae by \citet{vink2001} and \citet{Bjoerklund2023} have additional dependencies in addition to those on luminosity and metallicity, we are shading areas that account for the maximum and minimum values of the remaining parameters. 
With the exception of the model F2, which does not account for the UV spectrum, all methods determine mass-loss rates for the SMC dwarf AzV\,377 that fall even below the \citet{Bjoerklund2023} predictions. This seems to be in line with earlier findings of dwarfs in the SMC by \citet{Bouret2003}, \citet{Ramachandran2019}, and \citet{Rickard2022}.

For the LMC targets, the situation is different. The mass-loss rate derived for the late-O supergiant Sk\,-66$^{\circ}$ 171 agrees with the \citet{vink2001} predictions, but also with the \citet{Bjoerklund2023} recipe as this formula turns upwards for higher luminositites. For Sk\,-69$^{\circ}$ 50, there are two groups of solutions resulting from the assumption of either low and moderate clumping ($f_\text{cl} \leq 5$) coinciding with larger mass-loss rates, or strongly clumped solutions ($f_\text{cl} \geq 10$) and correspondingly lower values for $\dot{M}$. Depending on assumptions or results for clumping parameters and stratification, the derived mass-loss rates are either about $0.3\,$dex lower than predicted by \citet{vink2001} or even slightly higher than the prediction. The high-clumping solutions for Sk\,-69$^{\circ}$ 50 further align with the \citet{Bjoerklund2023} predictions. In its simple $\dot{M}(L)$ form, that is, without any temperature dependency, the \citet{Krticka2018} formula is not able to reproduce the derived values and stays between the LMC and SMC solutions. Within this very limited sample, it is hard to draw any robust conclusions about the mass-loss rates, but the results underline the complexity of the situation with our different sample stars spanning from an SMC star with a relatively weak wind to an O7 target that probably exceeds the wind expectations for the LMC.

\begin{figure}
  \includegraphics[width=\columnwidth]{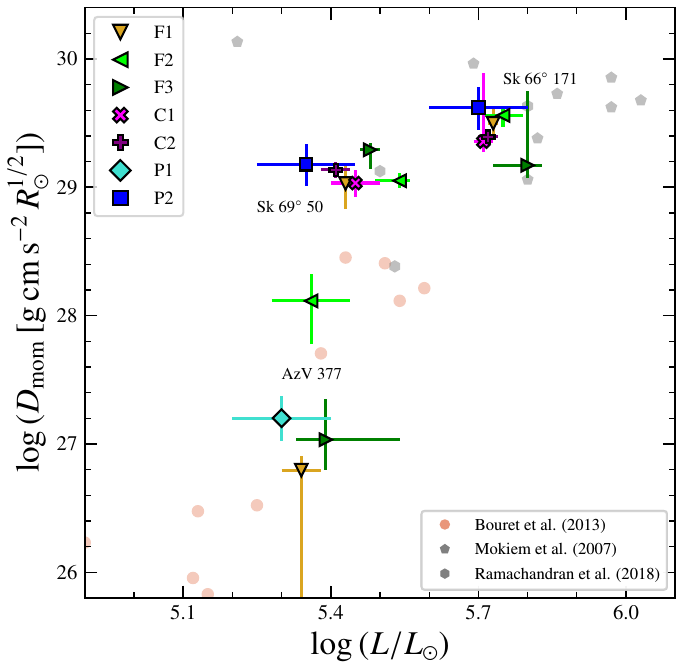}
  \caption{Modified wind momentum rate versus stellar luminosity $L$ for our sample stars analyzed with the different methods.}
  \label{fig:dmom-l}
\end{figure}

\begin{figure}
  \includegraphics[width=\columnwidth]{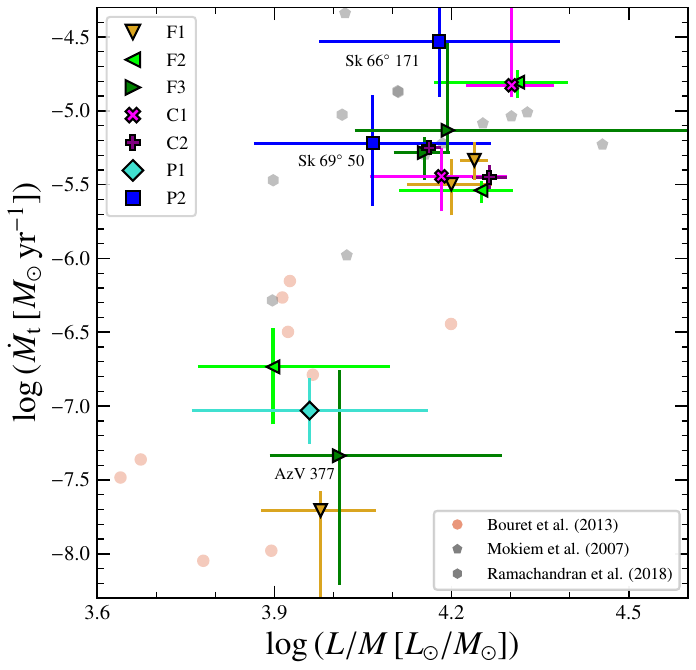}
  \caption{Transformed mass-loss rates versus the ratio between luminosity $L$ and spectroscopic mass $M$ for our sample stars analyzed with the different methods.}
  \label{fig:mdtr-ldm}
\end{figure}

In addition to the raw mass-loss rate, it is worth also considering the modified wind momentum rate,
\begin{equation}
  D_\text{mom} = \dot{M} \varv_\infty \sqrt{R/R_\odot}
,\end{equation}
which is expected to be proportional to some (positive) power of the stellar luminosity $L/L_\odot$ \citep[e.g.,][]{Kudritzki2000}. In Fig.\,\ref{fig:dmom-l}, we show the modified wind momentum rates $D_\text{mom}$, taking into account the clumping-adjusted mass-loss rates $\dot{M} \sqrt{f_\text{cl}}$ , rather than their raw values. By doing so, we lift the split between the two groups seen for Sk\,-69$^{\circ}$ 50 in Fig.\,\ref{fig:mdot-l} and both LMC targets now get very similar values in $D_\text{mom}$ from most of the methods, with the remaining spread being mainly due to the differences in determining $\log L$. The results for the SMC dwarf, in contrast, remain slightly more spread, but we have to consider that only two methods (F3 and P1) performed a full UV+optical study for this target and the two results agree within their error bars. For comparison, we also show results from \citet{Bouret2013} for the SMC as well as \citet{Mokiem2007} and \citet{Ramachandran2018} for the LMC. The data compiled in \citet{Mokiem2007} also contain analysis results from \citet{Crowther2002} and \citet{Massey2005}. In general, our derived quantities fall within the obtained literature values. 

Finally, we also plot the transformed mass-loss rate, 
\begin{equation}
  \dot{M}_\text{t} = \dot{M} \sqrt{f_\text{cl}} \frac{1000\,\mathrm{km\,s}^{-1}}{\varv_\infty} \left(\frac{10^6\,L_\odot}{L}\right)^{3/4}
,\end{equation}
defined by \citet{Graefener2013} as a function of $L/M$ in Fig.\,\ref{fig:mdtr-ldm}. The quantity describes the mass-loss rate the star might have if it had $10^6\,L_\odot$ and an unclumped wind with $\varv_\infty = 1000\,\mathrm{km}\,\mathrm{s}^{-1}$. Similar to the $D_\text{mom}$ plot (Fig.\,\ref{fig:dmom-l}), the SMC dwarf ends up in a very different regime than the two LMC targets, which cluster more than in the $D_\text{mom}$ plane. While the absolute values for $\dot{M}_\text{t}$ are relatively close to the regime obtained in Wolf-Rayet studies \citep{Sander2020,Sander2023}, neither of our two LMC targets would qualify as an Of/WN star as this would require H$\beta$ to show a P\,Cygni profile \citep{Crowther2011}, which is not observed. The two LMC stars are also too cool to yield notable \ion{He}{ii} ionizing flux (cf.\ Sect.\,\ref{sec:ionflux}), reflecting that the characteristic values for classical WR stars ($-4.5$) are not transferable to this parameter regime.

\section{Conclusions and perspectives for forthcoming papers}\label{sec:perspectives}

In this work, we present an analysis of three O stars from the ULLYSES and XShootU sample, namely the SMC O5V((f)) dwarf AzV\,377 as well as the LMC O stars Sk\,-69$^{\circ}$ 50 (O7(n)(f)p) and Sk\,-66$^{\circ}$ 171 (O9Ia). We analyze these targets using a variety of different methods, applying different model atmosphere codes (FASTWIND, CMFGEN, PoWR), ranging from grid-based approaches to tailored spectral fits. Some methods are only applied to some of the targets and some only take the optical spectra into account, thereby skipping a detailed determination of the wind parameters only accessible from the UV. This study was performed as a ``blind test'', meaning that each method was performed without prior knowledge of any outcomes from the other methods. The study is not intended to be a benchmarking of any particular atmosphere code or method, but our aim is to provide an overview of the ``natural'' spread of results obtained with the different existing approaches in the field. (No fine tuning of the results was performed after comparing the resulting parameters.) Moreover, our detailed descriptions of the individual methods serve as an introduction of the different techniques applied within the ``XShooting ULLYSES'' collaboration.

Overall, the different applied methods show a reasonable amount of agreement for the obtained parameters. Nevertheless, a spread of up to $3\,$kK is obtained for the effective temperatures across all three targets, with the GA-based method tending to yield slightly higher values. The differences in $\log g$ are on the order of $0.1\,$dex for the SMC dwarf and up to $0.2\,$dex for the LMC O stars. One ingredient to minimize the $\log g$ discrepancies could be the inclusion of a microturbulent velocity $\xi$ in the hydrostatic treatment, which is so far only possible in one of the three atmosphere codes employed in this work (PoWR). The inclusion of $\xi > 0$ demands higher $\log g$ values and thereby also has the potential to reduce current discrepancies between spectroscopic and evolutionary masses (hinted, e.g., by the good agreement for Sk\,-66$^{\circ}$ 171). However, we also obtain some spread in this ``mass discrepancy'' between methods that do not include turbulence in the hydrostatic solution, and therefore more in-depth studies on this topic are required. Differences in the adopted $E(B-V)$ values tend to be up to $0.1\,$dex and can affect the derived luminosities. The spread in reddening is mainly caused by the genetic algorithm methods, which tend to yield higher reddening values and consequently also higher luminosities. The remaining methods, regardless of the applied atmosphere code, differ by less than $0.05\,$dex in $E(B-V)$.

The inspection of our three sample targets illustrates the wide range of regimes found among the O stars in the ULLYSES sample. The SMC dwarf AzV\,377 exhibits a very low mass-loss rate, below current theoretical predictions, while the two LMC targets show comparably strong winds aligning with or even exceeding current recipes. The wind and the abundance analysis as well as the derived mass discrepancy suggest that the O7(n)(f)p star Sk\,-69$^{\circ}$ 50 is a particularly evolved object that might no longer be in the stage of central hydrogen burning. The two other targets are also nitrogen-enriched, but the methods differ in their conclusion regarding a potential mass discrepancy. 

Overall, the different methods scatter slightly more than expected, which we attribute in particular to the ``blind test'' scenario. 
Several of the heterogeneous choices among the various methods in the current study (e.g., the selection of the clumping law or differences in the reddening approach) could be avoided in a more coordinated effort. 
Hence, the scatter obtained in this work should reflect the amount of scatter to be expected when ``blindly'' combining data from different literature sources.
For coordinated efforts within the collaboration, we recommend  harmonize assumptions and selections as much as possible in order to minimize uncertainties in future studies. Our derived value spreads also provide a good indicator for future comparisons of data from heterogeneous sources.

\begin{acknowledgements}
  This manuscript has been inspired by discussions during the workshop ``ULLYSES - new horizons in massive star spectroscopy'', hosted and supported by the Lorentz Center, Leiden, Netherlands.
  Based on observations obtained with the NASA/ESA \textit{Hubble} Space Telescope, retrieved from the Mikulski Archive for Space Telescopes (MAST) at the Space Telescope Science Institute (STScI). STScI is operated by the Association of Universities for Research in Astronomy, Inc. under NASA contract NAS 5-26555. 
  AACS, MBP, VR, RRL, and CJKL are supported by the Deutsche Forschungsgemeinschaft (DFG - German Research Foundation) in the form of an Emmy Noether Research Group -- Project-ID 445674056 (SA4064/1-1, PI Sander).
  AACS and RRL acknowledge funding by the Deutsche Forschungsgemeinschaft (DFG, German Research Foundation) -- Project-ID 138713538 -- SFB 881 (``The Milky Way System'', subproject P04). AACS, MBP, VR, and ECS are further supported by funding from the Federal Ministry of Education and Research (BMBF) and the Baden-Württemberg Ministry of Science as part of the Excellence Strategy of the German Federal and State Governments.
  ECS acknowledges financial support by the Federal Ministry for Economic Affairs and Climate Action (BMWK) via the German Aerospace Center (Deutsches Zentrum f\"ur Luft- und Raumfahrt, DLR) grant 50 OR 2306 (PI: Ramachandran/Sander). CJKL gratefully acknowledges support from the International Max Planck Research School for Astronomy and Cosmic Physics at the University of Heidelberg in the form of an IMPRS PhD fellowship.
  PAC and JMB are supported by the Science and Technology Facilities Council research grant ST/V000853/1 (PI. V. Dhillon). DP acknowledges financial support by the Deutsches Zentrum f\"ur Luft und Raumfahrt (DLR) grant FKZ 50OR2005.
  RK acknowledges financial support via the Heisenberg Research Grant funded by the German Research Foundation (DFG) under grant no.~KU 2849/9.
  JSV gratefully acknowledges support from STFC via grant ST/V000233/1. OM acknowledges support from the project RVO:67985815 of the Astronomical Institute of the Czech Academy of Sciences. A.u.-D. acknowledges NASA ATP grant number 80NSSC22K0628 and support by NASA through Chandra Award number TM4-25001A issued by the Chandra X-ray Observatory 27 Center, which is operated by the Smithsonian Astrophysical Observatory for and on behalf of NASA under contract NAS8-03060. 
  ACGM acknowledges the support from the Polish National Science Centre grant Maestro (2018/30/A/ST9/00050) and from the Max Planck Society through a "Partner Group" grant.
  F.N. acknowledges support by grants PID2019-105552RB-C41 and PID2022-137779OB-C41 funded by MCIN/AEI/10.13039/501100011033 by "ERDF A way of making Europe".
  SRB acknowledges support by NextGeneration EU/PRTR and MIU (UNI/551/2021) through grant Margarita Salas-ULL. 
  This work used the Dutch national e-infrastructure with the support of the SURF Cooperative using grant no. EINF-3257. 
\end{acknowledgements}

\bibliographystyle{aa}
\bibliography{bib}

\begin{thebibliography}{178}
\expandafter\ifx\csname natexlab\endcsname\relax\def\natexlab#1{#1}\fi

\bibitem[{{Abbott} \& {Hummer}(1985)}]{Abbott1985}
{Abbott}, D.~C. \& {Hummer}, D.~G. 1985, \apj, 294, 286

\bibitem[{{Abbott} {et~al.}(2021){Abbott}, {Abbott}, {Abraham}, {Acernese}, {Ackley}, {Adams}, {Adams}, {Adhikari}, {Adya}, {Affeldt}, {Agathos}, {Agatsuma}, {Aggarwal}, {Aguiar}, {Aiello}, {Ain}, {Ajith}, {Allen}, {Allocca}, {Altin}, {Amato}, {Anand}, {Ananyeva}, {Anderson}, {Anderson}, {Angelova}, {Ansoldi}, {Antelis}, {Antier}, {Appert}, {Arai}, {Araya}, {Areeda}, {Ar{\`e}ne}, {Arnaud}, {Aronson}, {Arun}, {Asali}, {Ascenzi}, {Ashton}, {Aston}, {Astone}, {Aubin}, {Aufmuth}, {AultONeal}, {Austin}, {Avendano}, {Babak}, {Badaracco}, {Bader}, {Bae}, {Baer}, {Bagnasco}, {Baird}, {Ball}, {Ballardin}, {Ballmer}, {Bals}, {Balsamo}, {Baltus}, {Banagiri}, {Bankar}, {Bankar}, {Barayoga}, {Barbieri}, {Barish}, {Barker}, {Barneo}, {Barnum}, {Barone}, {Barr}, {Barsotti}, {Barsuglia}, {Barta}, {Bartlett}, {Bartos}, {Bassiri}, {Basti}, {Bawaj}, {Bayley}, {Bazzan}, {Becher}, {B{\'e}csy}, {Bedakihale}, {Bejger}, {Belahcene}, {Beniwal}, {Benjamin}, {Bennett}, {Bentley}, {Bergamin}, {Berger}, {Bergmann}, {Bernuzzi}, {Berry},
  {Bersanetti}, {Bertolini}, {Betzwieser}, {Bhandare}, {Bhandari}, {Bhattacharjee}, {Bidler}, {Bilenko}, {Billingsley}, {Birney}, {Birnholtz}, {Biscans}, {Bischi}, {Biscoveanu}, {Bisht}, {Bitossi}, {Bizouard}, {Blackburn}, {Blackman}, {Blair}, {Blair}, {Blair}, {Blanch}, {Bobba}, {Bode}, {Boer}, {Boetzel}, {Bogaert}, {Boldrini}, {Bondu}, {Bonilla}, {Bonnand}, {Booker}, {Boom}, {Bork}, {Boschi}, {Bose}, {Bossilkov}, {Boudart}, {Bouffanais}, {Bozzi}, {Bradaschia}, {Brady}, {Bramley}, {Branchesi}, {Brau}, {Breschi}, {Briant}, {Briggs}, {Brighenti}, {Brillet}, {Brinkmann}, {Brockill}, {Brooks}, {Brooks}, {Brown}, {Brunett}, {Bruno}, {Bruntz}, {Buikema}, {Bulik}, {Bulten}, {Buonanno}, {Buscicchio}, {Buskulic}, {Byer}, {Cabero}, {Cadonati}, {Caesar}, {Cagnoli}, {Cahillane}, {Calder{\'o}n Bustillo}, {Callaghan}, {Callister}, {Calloni}, {Camp}, {Canepa}, {Cannon}, {Cao}, {Cao}, {Carapella}, {Carbognani}, {Carney}, {Carpinelli}, {Carullo}, {Carver}, {Casanueva Diaz}, {Casentini}, {Caudill}, {Cavagli{\`a}}, {Cavalier},
  {Cavalieri}, {Cella}, {Cerd{\'a}-Dur{\'a}n}, {Cesarini}, {Chaibi}, {Chakravarti}, {Chan}, {Chan}, {Chandra}, {Chanial}, {Chao}, {Charlton}, {Chase}, {Chassande-Mottin}, {Chatterjee}, {Chattopadhyay}, {Chaturvedi}, {Chatziioannou}, {Chen}, {Chen}, {Chen}, {Chen}, {Cheng}, {Cheong}, {Chia}, {Chiadini}, {Chierici}, {Chincarini}, {Chiummo}, {Cho}, {Cho}, {Cho}, {Choate}, {Christensen}, {Chu}, {Chua}, {Chung}, {Chung}, {Ciani}, {Ciecielag}, {Cie{\'s}lar}, {Cifaldi}, {Ciobanu}, {Ciolfi}, {Cipriano}, {Cirone}, {Clara}, {Clark}, {Clark}, {Clarke}, {Clearwater}, {Clesse}, {Cleva}, {Coccia}, {Cohadon}, {Cohen}, {Colleoni}, {Collette}, {Collins}, {Colpi}, {Constancio}, {Conti}, {Cooper}, {Corban}, {Corbitt}, {Cordero-Carri{\'o}n}, {Corezzi}, {Corley}, {Cornish}, {Corre}, {Corsi}, {Cortese}, {Costa}, {Cotesta}, {Coughlin}, {Coughlin}, {Coulon}, {Countryman}, {Couvares}, {Covas}, {Coward}, {Cowart}, {Coyne}, {Coyne}, {Creighton}, {Creighton}, {Croquette}, {Crowder}, {Cudell}, {Cullen}, {Cumming}, {Cummings},
  {Cunningham}, {Cuoco}, {Curylo}, {Dal Canton}, {D{\'a}lya}, {Dana}, {DaneshgaranBajastani}, {D'Angelo}, {Danilishin}, {D'Antonio}, {Danzmann}, {Darsow-Fromm}, {Dasgupta}, {Datrier}, {Dattilo}, {Dave}, {Davier}, {Davies}, {Davis}, {Daw}, {Dean}, {DeBra}, {Deenadayalan}, {Degallaix}, {De Laurentis}, {Del{\'e}glise}, {Del Favero}, {De Lillo}, {De Lillo}, {Del Pozzo}, {DeMarchi}, {De Matteis}, {D'Emilio}, {Demos}, {Denker}, {Dent}, {Depasse}, {De Pietri}, {De Rosa}, {De Rossi}, {DeSalvo}, {de Varona}, {Dhurandhar}, {D{\'\i}az}, {Diaz-Ortiz}, {Didio}, {Dietrich}, {Di Fiore}, {DiFronzo}, {Di Giorgio}, {Di Giovanni}, {Di Giovanni}, {Di Girolamo}, {Di Lieto}, {Ding}, {Di Pace}, {Di Palma}, {Di Renzo}, {Divakarla}, {Dmitriev}, {Doctor}, {D'Onofrio}, {Donovan}, {Dooley}, {Doravari}, {Dorrington}, {Downes}, {Drago}, {Driggers}, {Du}, {Ducoin}, {Dupej}, {Durante}, {D'Urso}, {Duverne}, {Dwyer}, {Easter}, {Eddolls}, {Edelman}, {Edo}, {Edy}, {Effler}, {Eichholz}, {Eikenberry}, {Eisenmann}, {Eisenstein}, {Ejlli}, {Errico},
  {Essick}, {Estell{\'e}s}, {Estevez}, {Etienne}, {Etzel}, {Evans}, {Evans}, {Ewing}, {Fafone}, {Fair}, {Fairhurst}, {Fan}, {Farah}, {Farinon}, {Farr}, {Farr}, {Fauchon-Jones}, {Favata}, {Fays}, {Fazio}, {Feicht}, {Fejer}, {Feng}, {Fenyvesi}, {Ferguson}, {Fernandez-Galiana}, {Ferrante}, {Ferreira}, {Fidecaro}, {Figura}, {Fiori}, {Fiorucci}, {Fishbach}, {Fisher}, {Fishner}, {Fittipaldi}, {Fitz-Axen}, {Fiumara}, {Flaminio}, {Floden}, {Flynn}, {Fong}, {Font}, {Forsyth}, {Fournier}, {Frasca}, {Frasconi}, {Frei}, {Freise}, {Frey}, {Frey}, {Fritschel}, {Frolov}, {Fronz{\'e}}, {Fulda}, {Fyffe}, {Gabbard}, {Gadre}, {Gaebel}, {Gair}, {Gais}, {Galaudage}, {Gamba}, {Ganapathy}, {Ganguly}, {Gaonkar}, {Garaventa}, {Garc{\'\i}a-Quir{\'o}s}, {Garufi}, {Gateley}, {Gaudio}, {Gayathri}, {Gemme}, {Gennai}, {George}, {George}, {Gergely}, {Ghonge}, {Ghosh}, {Ghosh}, {Ghosh}, {Giacomazzo}, {Giacoppo}, {Giaime}, {Giardina}, {Gibson}, {Gier}, {Gill}, {Giri}, {Glanzer}, {Gleckl}, {Godwin}, {Goetz}, {Goetz}, {Gohlke}, {Goncharov},
  {Gonz{\'a}lez}, {Gopakumar}, {Gossan}, {Gosselin}, {Gouaty}, {Grace}, {Grado}, {Granata}, {Granata}, {Grant}, {Gras}, {Grassia}, {Gray}, {Gray}, {Greco}, {Green}, {Green}, {Gretarsson}, {Griggs}, {Grignani}, {Grimaldi}, {Grimes}, {Grimm}, {Grote}, {Grunewald}, {Gruning}, {Guerrero}, {Guidi}, {Guimaraes}, {Guix{\'e}}, {Gulati}, {Guo}, {Gupta}, {Gupta}, {Gupta}, {Gustafson}, {Gustafson}, {Guzman}, {Haegel}, {Halim}, {Hall}, {Hamilton}, {Hammond}, {Haney}, {Hanke}, {Hanks}, {Hanna}, {Hannuksela}, {Hannuksela}, {Hansen}, {Hansen}, {Hanson}, {Harder}, {Hardwick}, {Haris}, {Harms}, {Harry}, {Harry}, {Hartwig}, {Hasskew}, {Haster}, {Haughian}, {Hayes}, {Healy}, {Heidmann}, {Heintze}, {Heinze}, {Heinzel}, {Heitmann}, {Hellman}, {Hello}, {Helmling-Cornell}, {Hemming}, {Hendry}, {Heng}, {Hennes}, {Hennig}, {Hennig}, {Hernandez Vivanco}, {Heurs}, {Hild}, {Hill}, {Hines}, {Hochheim}, {Hofgard}, {Hofman}, {Hohmann}, {Holgado}, {Holland}, {Hollows}, {Holmes}, {Holt}, {Holz}, {Hopkins}, {Horst}, {Hough}, {Howell}, {Hoy},
  {Hoyland}, {Huang}, {H{\"u}bner}, {Huddart}, {Huerta}, {Hughey}, {Hui}, {Husa}, {Huttner}, {Hutzler}, {Huxford}, {Huynh-Dinh}, {Idzkowski}, {Iess}, {Imperato}, {Inchauspe}, {Ingram}, {Intini}, {Isi}, {Iyer}, {JaberianHamedan}, {Jacqmin}, {Jadhav}, {Jadhav}, {James}, {Jani}, {Janssens}, {Janthalur}, {Jaranowski}, {Jariwala}, {Jaume}, {Jenkins}, {Jeunon}, {Jiang}, {Johns}, {Jones}, {Jones}, {Jones}, {Jones}, {Jones}, {Jonker}, {Ju}, {Junker}, {Kalaghatgi}, {Kalogera}, {Kamai}, {Kandhasamy}, {Kang}, {Kanner}, {Kapadia}, {Kapasi}, {Karathanasis}, {Karki}, {Kashyap}, {Kasprzack}, {Kastaun}, {Katsanevas}, {Katsavounidis}, {Katzman}, {Kawabe}, {K{\'e}f{\'e}lian}, {Keitel}, {Key}, {Khadka}, {Khalili}, {Khan}, {Khan}, {Khazanov}, {Khetan}, {Khursheed}, {Kijbunchoo}, {Kim}, {Kim}, {Kim}, {Kim}, {Kim}, {Kim}, {Kimball}, {King}, {Kinley-Hanlon}, {Kirchhoff}, {Kissel}, {Kleybolte}, {Klimenko}, {Knowles}, {Knyazev}, {Koch}, {Koehlenbeck}, {Koekoek}, {Koley}, {Kolstein}, {Komori}, {Kondrashov}, {Kontos}, {Koper},
  {Korobko}, {Korth}, {Kovalam}, {Kozak}, {Kr{\"a}mer}, {Kringel}, {Krishnendu}, {Kr{\'o}lak}, {Kuehn}, {Kumar}, {Kumar}, {Kumar}, {Kumar}, {Kuns}, {Kwang}, {Lackey}, {Laghi}, {Lalande}, {Lam}, {Lamberts}, {Landry}, {Lane}, {Lang}, {Lange}, {Lantz}, {Lanza}, {La Rosa}, {Lartaux-Vollard}, {Lasky}, {Laxen}, {Lazzarini}, {Lazzaro}, {Leaci}, {Leavey}, {Lecoeuche}, {Lee}, {Lee}, {Lee}, {Lee}, {Lehmann}, {Leon}, {Leroy}, {Letendre}, {Levin}, {Li}, {Li}, {Li}, {Li}, {Li}, {Linde}, {Linker}, {Linley}, {Littenberg}, {Liu}, {Liu}, {Llorens-Monteagudo}, {Lo}, {Lockwood}, {London}, {Longo}, {Lorenzini}, {Loriette}, {Lormand}, {Losurdo}, {Lough}, {Lousto}, {Lovelace}, {L{\"u}ck}, {Lumaca}, {Lundgren}, {Ma}, {Macas}, {MacInnis}, {Macleod}, {MacMillan}, {Macquet}, {Maga{\~n}a Hernandez}, {Maga{\~n}a-Sandoval}, {Magazz{\`u}}, {Magee}, {Majorana}, {Maksimovic}, {Maliakal}, {Malik}, {Man}, {Mandic}, {Mangano}, {Mansell}, {Manske}, {Mantovani}, {Mapelli}, {Marchesoni}, {Marion}, {M{\'a}rka}, {M{\'a}rka}, {Markakis},
  {Markosyan}, {Markowitz}, {Maros}, {Marquina}, {Marsat}, {Martelli}, {Martin}, {Martin}, {Martinez}, {Martinez}, {Martynov}, {Masalehdan}, {Mason}, {Massera}, {Masserot}, {Massinger}, {Masso-Reid}, {Mastrogiovanni}, {Matas}, {Mateu-Lucena}, {Matichard}, {Matiushechkina}, {Mavalvala}, {Maynard}, {McCann}, {McCarthy}, {McClelland}, {McCormick}, {McCuller}, {McGuire}, {McIsaac}, {McIver}, {McManus}, {McRae}, {McWilliams}, {Meacher}, {Meadors}, {Mehmet}, {Mehta}, {Melatos}, {Melchor}, {Mendell}, {Menendez-Vazquez}, {Mercer}, {Mereni}, {Merfeld}, {Merilh}, {Merritt}, {Merzougui}, {Meshkov}, {Messenger}, {Messick}, {Metzdorff}, {Meyers}, {Meylahn}, {Mhaske}, {Miani}, {Miao}, {Michaloliakos}, {Michel}, {Middleton}, {Milano}, {Miller}, {Miller}, {Millhouse}, {Mills}, {Milotti}, {Milovich-Goff}, {Minazzoli}, {Minenkov}, {Mir}, {Mishkin}, {Mishra}, {Mistry}, {Mitra}, {Mitrofanov}, {Mitselmakher}, {Mittleman}, {Mo}, {Mogushi}, {Mohapatra}, {Mohite}, {Molina}, {Molina-Ruiz}, {Mondin}, {Montani}, {Moore}, {Moraru},
  {Morawski}, {Moreno}, {Morisaki}, {Mours}, {Mow-Lowry}, {Mozzon}, {Muciaccia}, {Mukherjee}, {Mukherjee}, {Mukherjee}, {Mukherjee}, {Mukund}, {Mullavey}, {Munch}, {Mu{\~n}iz}, {Murray}, {Nadji}, {Nagar}, {Nardecchia}, {Naticchioni}, {Nayak}, {Neil}, {Neilson}, {Nelemans}, {Nelson}, {Nery}, {Neunzert}, {Ng}, {Ng}, {Nguyen}, {Nguyen}, {Nguyen}, {Nichols}, {Nissanke}, {Nocera}, {Noh}, {North}, {Nothard}, {Nuttall}, {Oberling}, {O'Brien}, {O'Dell}, {Oganesyan}, {Ogin}, {Oh}, {Oh}, {Ohme}, {Ohta}, {Okada}, {Olivetto}, {Oppermann}, {Oram}, {O'Reilly}, {Ormiston}, {Ormsby}, {Ortega}, {O'Shaughnessy}, {Ossokine}, {Osthelder}, {Ottaway}, {Overmier}, {Owen}, {Pace}, {Pagano}, {Page}, {Pagliaroli}, {Pai}, {Pai}, {Palamos}, {Palashov}, {Palomba}, {Pan}, {Panda}, {Pang}, {Pankow}, {Pannarale}, {Pant}, {Paoletti}, {Paoli}, {Paolone}, {Parker}, {Pascucci}, {Pasqualetti}, {Passaquieti}, {Passuello}, {Patel}, {Patricelli}, {Payne}, {Pechsiri}, {Pedraza}, {Pegoraro}, {Pele}, {Penn}, {Perego}, {Perez}, {P{\'e}rigois},
  {Perreca}, {Perri{\`e}s}, {Petermann}, {Petterson}, {Pfeiffer}, {Pham}, {Phukon}, {Piccinni}, {Pichot}, {Piendibene}, {Piergiovanni}, {Pierini}, {Pierro}, {Pillant}, {Pilo}, {Pinard}, {Pinto}, {Piotrzkowski}, {Pirello}, {Pitkin}, {Placidi}, {Plastino}, {Pluchar}, {Poggiani}, {Polini}, {Pong}, {Ponrathnam}, {Popolizio}, {Porter}, {Poverman}, {Powell}, {Pracchia}, {Prajapati}, {Prasai}, {Prasanna}, {Pratten}, {Prestegard}, {Principe}, {Prodi}, {Prokhorov}, {Prosposito}, {Puecher}, {Punturo}, {Puosi}, {Puppo}, {P{\"u}rrer}, {Qi}, {Quetschke}, {Quinonez}, {Quitzow-James}, {Raab}, {Raaijmakers}, {Radkins}, {Radulesco}, {Raffai}, {Rafferty}, {Rail}, {Raja}, {Rajan}, {Rajbhandari}, {Rakhmanov}, {Ramirez}, {Ramirez}, {Ramos-Buades}, {Rana}, {Rao}, {Rapagnani}, {Rapol}, {Ratto}, {Raymond}, {Razzano}, {Read}, {Regimbau}, {Rei}, {Reid}, {Reitze}, {Rettegno}, {Ricci}, {Richardson}, {Richardson}, {Richardson}, {Ricker}, {Riemenschneider}, {Riles}, {Rizzo}, {Robertson}, {Robinet}, {Rocchi}, {Rocha}, {Rodriguez},
  {Rodriguez-Soto}, {Rolland}, {Rollins}, {Roma}, {Romanelli}, {Romano}, {Romel}, {Romero}, {Romero-Shaw}, {Romie}, {Ronchini}, {Rose}, {Rose}, {Rose}, {Rosell}, {Rosi{\'n}ska}, {Rosofsky}, {Ross}, {Rowan}, {Rowlinson}, {Roy}, {Roy}, {Ruggi}, {Ryan}, {Sachdev}, {Sadecki}, {Sakellariadou}, {Salafia}, {Salconi}, {Saleem}, {Samajdar}, {Sanchez}, {Sanchez}, {Sanchez}, {Sanchis-Gual}, {Sanders}, {Santiago}, {Santos}, {Saravanan}, {Sarin}, {Sassolas}, {Sathyaprakash}, {Sauter}, {Savage}, {Savant}, {Sawant}, {Sayah}, {Schaetzl}, {Schale}, {Scheel}, {Scheuer}, {Schindler-Tyka}, {Schmidt}, {Schnabel}, {Schofield}, {Sch{\"o}nbeck}, {Schreiber}, {Schulte}, {Schutz}, {Schwarm}, {Schwartz}, {Scott}, {Scott}, {Seglar-Arroyo}, {Seidel}, {Sellers}, {Sengupta}, {Sennett}, {Sentenac}, {Sequino}, {Sergeev}, {Setyawati}, {Shaffer}, {Shahriar}, {Sharifi}, {Sharma}, {Sharma}, {Shawhan}, {Shen}, {Shikauchi}, {Shink}, {Shoemaker}, {Shoemaker}, {Shukla}, {ShyamSundar}, {Sieniawska}, {Sigg}, {Singer}, {Singh}, {Singh}, {Singha},
  {Singhal}, {Sintes}, {Sipala}, {Skliris}, {Slagmolen}, {Slaven-Blair}, {Smetana}, {Smith}, {Smith}, {Somala}, {Son}, {Soni}, {Sorazu}, {Sordini}, {Sorrentino}, {Sorrentino}, {Soulard}, {Souradeep}, {Sowell}, {Spencer}, {Spera}, {Srivastava}, {Srivastava}, {Staats}, {Stachie}, {Steer}, {Steinke}, {Steinlechner}, {Steinlechner}, {Steinmeyer}, {Stevenson}, {Stolle-McAllister}, {Stops}, {Stover}, {Strain}, {Stratta}, {Strunk}, {Sturani}, {Stuver}, {S{\"u}dbeck}, {Sudhagar}, {Sudhir}, {Suh}, {Summerscales}, {Sun}, {Sun}, {Sunil}, {Sur}, {Suresh}, {Sutton}, {Swinkels}, {Szczepa{\'n}czyk}, {Tacca}, {Tait}, {Talbot}, {Tanasijczuk}, {Tanner}, {Tao}, {Tapia}, {Tapia San Martin}, {Tasson}, {Taylor}, {Tenorio}, {Terkowski}, {Thirugnanasambandam}, {Thomas}, {Thomas}, {Thomas}, {Thompson}, {Thondapu}, {Thorne}, {Thrane}, {Tiwari}, {Tiwari}, {Tiwari}, {Toland}, {Tolley}, {Tonelli}, {Tornasi}, {Torres-Forn{\'e}}, {Torrie}, {Tosta e Melo}, {T{\"o}yr{\"a}}, {Tran}, {Trapananti}, {Travasso}, {Traylor}, {Tringali},
  {Tripathee}, {Trovato}, {Trudeau}, {Tsai}, {Tsang}, {Tse}, {Tso}, {Tsukada}, {Tsuna}, {Tsutsui}, {Turconi}, {Ubhi}, {Udall}, {Ueno}, {Ugolini}, {Unnikrishnan}, {Urban}, {Usman}, {Utina}, {Vahlbruch}, {Vajente}, {Vajpeyi}, {Valdes}, {Valentini}, {Valsan}, {van Bakel}, {Beuzekom}, {van den Brand}, {Van Den Broeck}, {Vander-Hyde}, {van der Schaaf}, {van Heijningen}, {Vardaro}, {Vargas}, {Varma}, {Vass}, {Vas{\'u}th}, {Vecchio}, {Vedovato}, {Veitch}, {Veitch}, {Venkateswara}, {Venneberg}, {Venugopalan}, {Verkindt}, {Verma}, {Veske}, {Vetrano}, {Vicer{\'e}}, {Viets}, {Villa-Ortega}, {Vinet}, {Vitale}, {Vo}, {Vocca}, {Vorvick}, {Vyatchanin}, {Wade}, {Wade}, {Wade}, {Walet}, {Walker}, {Wallace}, {Wallace}, {Walsh}, {Wang}, {Wang}, {Wang}, {Wang}, {Ward}, {Warner}, {Was}, {Washington}, {Watchi}, {Weaver}, {Wei}, {Weinert}, {Weinstein}, {Weiss}, {Wellmann}, {Wen}, {We{\ss}els}, {Westhouse}, {Wette}, {Whelan}, {White}, {White}, {Whiting}, {Whittle}, {Wilken}, {Williams}, {Williams}, {Williamson}, {Willis}, {Willke},
  {Wilson}, {Wimmer}, {Winkler}, {Wipf}, {Woan}, {Woehler}, {Wofford}, {Wong}, {Wrangel}, {Wright}, {Wu}, {Wysocki}, {Xiao}, {Yamamoto}, {Yang}, {Yang}, {Yang}, {Yap}, {Yeeles}, {Yoon}, {Yu}, {Yu}, {Yuen}, {Zadro{\.z}ny}, {Zanolin}, {Zelenova}, {Zendri}, {Zevin}, {Zhang}, {Zhang}, {Zhang}, {Zhang}, {Zhao}, {Zhao}, {Zhou}, {Zhou}, {Zhu}, {Zimmerman}, {Zucker}, {Zweizig}, {LIGO Scientific Collaboration}, \& {Virgo Collaboration}}]{Abbott2021}
{Abbott}, R., {Abbott}, T.~D., {Abraham}, S., {et~al.} 2021, \apjl, 913, L7

\bibitem[{{Abdul-Masih} {et~al.}(2021){Abdul-Masih}, {Sana}, {Hawcroft}, {Almeida}, {Brands}, {de Mink}, {Justham}, {Langer}, {Mahy}, {Marchant}, {Menon}, {Puls}, \& {Sundqvist}}]{AbdulMasih2021}
{Abdul-Masih}, M., {Sana}, H., {Hawcroft}, C., {et~al.} 2021, \aap, 651, A96

\bibitem[{{Aerts} {et~al.}(2009){Aerts}, {Puls}, {Godart}, \& {Dupret}}]{Aerts2009}
{Aerts}, C., {Puls}, J., {Godart}, M., \& {Dupret}, M.~A. 2009, \aap, 508, 409

\bibitem[{{Arellano-C{\'o}rdova} {et~al.}(2022){Arellano-C{\'o}rdova}, {Berg}, {Chisholm}, {Arrabal Haro}, {Dickinson}, {Finkelstein}, {Leclercq}, {Rogers}, {Simons}, {Skillman}, {Trump}, \& {Kartaltepe}}]{Arellano-Cordova2022}
{Arellano-C{\'o}rdova}, K.~Z., {Berg}, D.~A., {Chisholm}, J., {et~al.} 2022, \apjl, 940, L23

\bibitem[{{Arrabal Haro} {et~al.}(2023){Arrabal Haro}, {Dickinson}, {Finkelstein}, {Fujimoto}, {Fern{\'a}ndez}, {Kartaltepe}, {Jung}, {Cole}, {Burgarella}, {Chworowsky}, {Hutchison}, {Morales}, {Papovich}, {Simons}, {Amor{\'\i}n}, {Backhaus}, {Bagley}, {Bisigello}, {Calabr{\`o}}, {Castellano}, {Cleri}, {Dav{\'e}}, {Dekel}, {Ferguson}, {Fontana}, {Gawiser}, {Giavalisco}, {Harish}, {Hathi}, {Hirschmann}, {Holwerda}, {Huertas-Company}, {Koekemoer}, {Larson}, {Lucas}, {Mobasher}, {P{\'e}rez-Gonz{\'a}lez}, {Pirzkal}, {Rose}, {Santini}, {Trump}, {de la Vega}, {Wang}, {Weiner}, {Wilkins}, {Yang}, {Yung}, \& {Zavala}}]{ArrabalHaro2023}
{Arrabal Haro}, P., {Dickinson}, M., {Finkelstein}, S.~L., {et~al.} 2023, \apjl, 951, L22

\bibitem[{{Baum} {et~al.}(1992){Baum}, {Hamann}, {Koesterke}, \& {Wessolowski}}]{Baum1992}
{Baum}, E., {Hamann}, W.~R., {Koesterke}, L., \& {Wessolowski}, U. 1992, \aap, 266, 402

\bibitem[{{Bernini-Peron} {et~al.}(2023){Bernini-Peron}, {Marcolino}, {Sander}, {Bouret}, {Ramachandran}, {Saling}, {Schneider}, {Oskinova}, \& {Najarro}}]{Bernini-Peron2023}
{Bernini-Peron}, M., {Marcolino}, W.~L.~F., {Sander}, A.~A.~C., {et~al.} 2023, \aap, 677, A50

\bibitem[{{Bestenlehner}(2022)}]{bestenlehner2022}
{Bestenlehner}, J.~M. 2022, in Massive Stars Near and Far, ed. J.~{Mackey}, J.~S. {Vink}, \& N.~{St-Louis}, Vol. Proc.\ IAU Symp.\ 361, arXiv:2209.00998

\bibitem[{{Bestenlehner} {et~al.}(2022){Bestenlehner}, {Crowther}, {Broos}, {Pollock}, \& {Townsley}}]{Bestenlehner2022a}
{Bestenlehner}, J.~M., {Crowther}, P.~A., {Broos}, P.~S., {Pollock}, A. M.~T., \& {Townsley}, L.~K. 2022, \mnras, 510, 6133

\bibitem[{{Bestenlehner} {et~al.}(2020){Bestenlehner}, {Crowther}, {Caballero-Nieves}, {Schneider}, {Sim{\'o}n-D{\'\i}az}, {Brands}, {de Koter}, {Gr{\"a}fener}, {Herrero}, {Langer}, {Lennon}, {Ma{\'\i}z Apell{\'a}niz}, {Puls}, \& {Vink}}]{Bestenlehner2020}
{Bestenlehner}, J.~M., {Crowther}, P.~A., {Caballero-Nieves}, S.~M., {et~al.} 2020, \mnras, 499, 1918

\bibitem[{{Bestenlehner} {et~al.}(2024){Bestenlehner}, {En{\ss}lin}, {Bergemann}, {Crowther}, {Greiner}, \& {Selig}}]{Bestenlehner2024}
{Bestenlehner}, J.~M., {En{\ss}lin}, T., {Bergemann}, M., {et~al.} 2024, \mnras, 528, 6735

\bibitem[{{Bestenlehner} {et~al.}(2014){Bestenlehner}, {Gr{\"a}fener}, {Vink}, {Najarro}, {de Koter}, {Sana}, {Evans}, {Crowther}, {H{\'e}nault-Brunet}, {Herrero}, {Langer}, {Schneider}, {Sim{\'o}n-D{\'\i}az}, {Taylor}, \& {Walborn}}]{Bestenlehner2014}
{Bestenlehner}, J.~M., {Gr{\"a}fener}, G., {Vink}, J.~S., {et~al.} 2014, \aap, 570, A38

\bibitem[{{Bj{\"o}rklund} {et~al.}(2021){Bj{\"o}rklund}, {Sundqvist}, {Puls}, \& {Najarro}}]{Bjoerklund2021}
{Bj{\"o}rklund}, R., {Sundqvist}, J.~O., {Puls}, J., \& {Najarro}, F. 2021, \aap, 648, A36

\bibitem[{{Bj{\"o}rklund} {et~al.}(2023){Bj{\"o}rklund}, {Sundqvist}, {Singh}, {Puls}, \& {Najarro}}]{Bjoerklund2023}
{Bj{\"o}rklund}, R., {Sundqvist}, J.~O., {Singh}, S.~M., {Puls}, J., \& {Najarro}, F. 2023, \aap, 676, A109

\bibitem[{{Boco} {et~al.}(2021){Boco}, {Lapi}, {Chruslinska}, {Donevski}, {Sicilia}, \& {Danese}}]{Boco2021}
{Boco}, L., {Lapi}, A., {Chruslinska}, M., {et~al.} 2021, \apj, 907, 110

\bibitem[{{Bonanos} {et~al.}(2009){Bonanos}, {Massa}, {Sewilo}, {Lennon}, {Panagia}, {Smith}, {Meixner}, {Babler}, {Bracker}, {Meade}, {Gordon}, {Hora}, {Indebetouw}, \& {Whitney}}]{Bonanos+2009}
{Bonanos}, A.~Z., {Massa}, D.~L., {Sewilo}, M., {et~al.} 2009, \aj, 138, 1003

\bibitem[{{Bouchet} {et~al.}(1985){Bouchet}, {Lequeux}, {Maurice}, {Prevot}, \& {Prevot-Burnichon}}]{Bouchet+1985}
{Bouchet}, P., {Lequeux}, J., {Maurice}, E., {Prevot}, L., \& {Prevot-Burnichon}, M.~L. 1985, \aap, 149, 330

\bibitem[{{Bouret} {et~al.}(2012){Bouret}, {Hillier}, {Lanz}, \& {Fullerton}}]{Bouret2012}
{Bouret}, J.~C., {Hillier}, D.~J., {Lanz}, T., \& {Fullerton}, A.~W. 2012, \aap, 544, A67

\bibitem[{{Bouret} {et~al.}(2005){Bouret}, {Lanz}, \& {Hillier}}]{Bouret2005}
{Bouret}, J.~C., {Lanz}, T., \& {Hillier}, D.~J. 2005, \aap, 438, 301

\bibitem[{{Bouret} {et~al.}(2003){Bouret}, {Lanz}, {Hillier}, {Heap}, {Hubeny}, {Lennon}, {Smith}, \& {Evans}}]{Bouret2003}
{Bouret}, J.~C., {Lanz}, T., {Hillier}, D.~J., {et~al.} 2003, \apj, 595, 1182

\bibitem[{{Bouret} {et~al.}(2013){Bouret}, {Lanz}, {Martins}, {Marcolino}, {Hillier}, {Depagne}, \& {Hubeny}}]{Bouret2013}
{Bouret}, J.~C., {Lanz}, T., {Martins}, F., {et~al.} 2013, \aap, 555, A1

\bibitem[{{Bouret} {et~al.}(2021){Bouret}, {Martins}, {Hillier}, {Marcolino}, {Rocha-Pinto}, {Georgy}, {Lanz}, \& {Hubeny}}]{Bouret2021}
{Bouret}, J.~C., {Martins}, F., {Hillier}, D.~J., {et~al.} 2021, \aap, 647, A134

\bibitem[{{Brands} {et~al.}(2022){Brands}, {de Koter}, {Bestenlehner}, {Crowther}, {Sundqvist}, {Puls}, {Caballero-Nieves}, {Abdul-Masih}, {Driessen}, {Garc{\'\i}a}, {Geen}, {Gr{\"a}fener}, {Hawcroft}, {Kaper}, {Keszthelyi}, {Langer}, {Sana}, {Schneider}, {Shenar}, \& {Vink}}]{Brands2022}
{Brands}, S.~A., {de Koter}, A., {Bestenlehner}, J.~M., {et~al.} 2022, \aap, 663, A36

\bibitem[{{Brott} {et~al.}(2011){Brott}, {de Mink}, {Cantiello}, {Langer}, {de Koter}, {Evans}, {Hunter}, {Trundle}, \& {Vink}}]{Brott2011}
{Brott}, I., {de Mink}, S.~E., {Cantiello}, M., {et~al.} 2011, \aap, 530, A115

\bibitem[{{Cantiello} {et~al.}(2009){Cantiello}, {Langer}, {Brott}, {de Koter}, {Shore}, {Vink}, {Voegler}, {Lennon}, \& {Yoon}}]{Cantiello2009}
{Cantiello}, M., {Langer}, N., {Brott}, I., {et~al.} 2009, \aap, 499, 279

\bibitem[{{Carneiro} {et~al.}(2016){Carneiro}, {Puls}, {Sundqvist}, \& {Hoffmann}}]{Carneiro2016}
{Carneiro}, L.~P., {Puls}, J., {Sundqvist}, J.~O., \& {Hoffmann}, T.~L. 2016, \aap, 590, A88

\bibitem[{{Casagrande} \& {VandenBerg}(2014)}]{Casagrande2014}
{Casagrande}, L. \& {VandenBerg}, D.~A. 2014, \mnras, 444, 392

\bibitem[{{Castor} {et~al.}(1975){Castor}, {Abbott}, \& {Klein}}]{Castor1975}
{Castor}, J.~I., {Abbott}, D.~C., \& {Klein}, R.~I. 1975, \apj, 195, 157

\bibitem[{{Castro} {et~al.}(2018){Castro}, {Crowther}, {Evans}, {Mackey}, {Castro-Rodriguez}, {Vink}, {Melnick}, \& {Selman}}]{Castro2018}
{Castro}, N., {Crowther}, P.~A., {Evans}, C.~J., {et~al.} 2018, \aap, 614, A147

\bibitem[{{Chlebowski} \& {Garmany}(1991)}]{Chlebowski1991}
{Chlebowski}, T. \& {Garmany}, C.~D. 1991, \apj, 368, 241

\bibitem[{{Cioni} {et~al.}(2019){Cioni}, {Storm}, {Bell}, {Lemasle}, {Niederhofer}, {Bestenlehner}, {El Youssoufi}, {Feltzing}, {Gonz{\'a}lez-Fern{\'a}ndez}, {Grebel}, {Hobbs}, {Irwin}, {Jablonka}, {Koch}, {Schnurr}, {Schmidt}, \& {Steinmetz}}]{Cioni2019}
{Cioni}, M. . R.~L., {Storm}, J., {Bell}, C.~P.~M., {et~al.} 2019, The Messenger, 175, 54

\bibitem[{{Cioni} {et~al.}(2011){Cioni}, {Clementini}, {Girardi}, {Guandalini}, {Gullieuszik}, {Miszalski}, {Moretti}, {Ripepi}, {Rubele}, {Bagheri}, {Bekki}, {Cross}, {de Blok}, {de Grijs}, {Emerson}, {Evans}, {Gibson}, {Gonzales-Solares}, {Groenewegen}, {Irwin}, {Ivanov}, {Lewis}, {Marconi}, {Marquette}, {Mastropietro}, {Moore}, {Napiwotzki}, {Naylor}, {Oliveira}, {Read}, {Sutorius}, {van Loon}, {Wilkinson}, \& {Wood}}]{Cioni2011}
{Cioni}, M. R.~L., {Clementini}, G., {Girardi}, L., {et~al.} 2011, \aap, 527, A116

\bibitem[{{Crowther} {et~al.}(2022){Crowther}, {Broos}, {Townsley}, {Pollock}, {Tehrani}, \& {Gagn{\'e}}}]{Crowther2022}
{Crowther}, P.~A., {Broos}, P.~S., {Townsley}, L.~K., {et~al.} 2022, \mnras, 515, 4130

\bibitem[{{Crowther} {et~al.}(2016){Crowther}, {Caballero-Nieves}, {Bostroem}, {Ma{\'\i}z Apell{\'a}niz}, {Schneider}, {Walborn}, {Angus}, {Brott}, {Bonanos}, {de Koter}, {de Mink}, {Evans}, {Gr{\"a}fener}, {Herrero}, {Howarth}, {Langer}, {Lennon}, {Puls}, {Sana}, \& {Vink}}]{Crowther2016}
{Crowther}, P.~A., {Caballero-Nieves}, S.~M., {Bostroem}, K.~A., {et~al.} 2016, \mnras, 458, 624

\bibitem[{{Crowther} \& {Hadfield}(2006)}]{Crowther2006}
{Crowther}, P.~A. \& {Hadfield}, L.~J. 2006, \aap, 449, 711

\bibitem[{{Crowther} {et~al.}(2002){Crowther}, {Hillier}, {Evans}, {Fullerton}, {De Marco}, \& {Willis}}]{Crowther2002}
{Crowther}, P.~A., {Hillier}, D.~J., {Evans}, C.~J., {et~al.} 2002, \apj, 579, 774

\bibitem[{{Crowther} \& {Walborn}(2011)}]{Crowther2011}
{Crowther}, P.~A. \& {Walborn}, N.~R. 2011, \mnras, 416, 1311

\bibitem[{{Curti} {et~al.}(2023){Curti}, {D'Eugenio}, {Carniani}, {Maiolino}, {Sandles}, {Witstok}, {Baker}, {Bennett}, {Piotrowska}, {Tacchella}, {Charlot}, {Nakajima}, {Maheson}, {Mannucci}, {Amiri}, {Arribas}, {Belfiore}, {Bonaventura}, {Bunker}, {Chevallard}, {Cresci}, {Curtis-Lake}, {Hayden-Pawson}, {Jones}, {Kumari}, {Laseter}, {Looser}, {Marconi}, {Maseda}, {Scholtz}, {Smit}, {{\"U}bler}, \& {Wallace}}]{Curti2023}
{Curti}, M., {D'Eugenio}, F., {Carniani}, S., {et~al.} 2023, \mnras, 518, 425

\bibitem[{{Cutri} {et~al.}(2003){Cutri}, {Skrutskie}, {van Dyk}, {Beichman}, {Carpenter}, {Chester}, {Cambresy}, {Evans}, {Fowler}, {Gizis}, {Howard}, {Huchra}, {Jarrett}, {Kopan}, {Kirkpatrick}, {Light}, {Marsh}, {McCallon}, {Schneider}, {Stiening}, {Sykes}, {Weinberg}, {Wheaton}, {Wheelock}, \& {Zacarias}}]{Cutri+2003}
{Cutri}, R.~M., {Skrutskie}, M.~F., {van Dyk}, S., {et~al.} 2003, VizieR Online Data Catalog, II/246

\bibitem[{{Cutri} {et~al.}(2012){Cutri}, {Skrutskie}, {van Dyk}, {Beichman}, {Carpenter}, {Chester}, {Cambresy}, {Evans}, {Fowler}, {Gizis}, {Howard}, {Huchra}, {Jarrett}, {Kopan}, {Kirkpatrick}, {Light}, {Marsh}, {McCallon}, {Schneider}, {Stiening}, {Sykes}, {Weinberg}, {Wheaton}, {Wheelock}, \& {Zacharias}}]{Cutri2012}
{Cutri}, R.~M., {Skrutskie}, M.~F., {van Dyk}, S., {et~al.} 2012, VizieR Online Data Catalog, II/281

\bibitem[{{Cutri} {et~al.}(2021){Cutri}, {Wright}, {Conrow}, {Fowler}, {Eisenhardt}, {Grillmair}, {Kirkpatrick}, {Masci}, {McCallon}, {Wheelock}, {Fajardo-Acosta}, {Yan}, {Benford}, {Harbut}, {Jarrett}, {Lake}, {Leisawitz}, {Ressler}, {Stanford}, {Tsai}, {Liu}, {Helou}, {Mainzer}, {Gettngs}, {Gonzalez}, {Hoffman}, {Marsh}, {Padgett}, {Skrutskie}, {Beck}, {Papin}, \& {Wittman}}]{Cutri2013}
{Cutri}, R.~M., {Wright}, E.~L., {Conrow}, T., {et~al.} 2021, VizieR Online Data Catalog, II/328

\bibitem[{{Debnath} {et~al.}(2024){Debnath}, {Sundqvist}, {Moens}, {Van der Sijpt}, {Verhamme}, \& {Poniatowski}}]{Debnath2024}
{Debnath}, D., {Sundqvist}, J.~O., {Moens}, N., {et~al.} 2024, \aap, 684, A177

\bibitem[{{Dessart} \& {Hillier}(2010)}]{Dessart2010}
{Dessart}, L. \& {Hillier}, D.~J. 2010, \mnras, 405, 2141

\bibitem[{{Dreizler} \& {Werner}(1993)}]{Dreizler1993}
{Dreizler}, S. \& {Werner}, K. 1993, \aap, 278, 199

\bibitem[{{Ebbets}(1982)}]{Ebbets1982}
{Ebbets}, D. 1982, \apjs, 48, 399

\bibitem[{{Eggenberger} {et~al.}(2021){Eggenberger}, {Ekstr{\"o}m}, {Georgy}, {Martinet}, {Pezzotti}, {Nandal}, {Meynet}, {Buldgen}, {Salmon}, {Haemmerl{\'e}}, {Maeder}, {Hirschi}, {Yusof}, {Groh}, {Farrell}, {Murphy}, \& {Choplin}}]{Eggenberger2021}
{Eggenberger}, P., {Ekstr{\"o}m}, S., {Georgy}, C., {et~al.} 2021, \aap, 652, A137

\bibitem[{{Evans} {et~al.}(2004{\natexlab{a}}){Evans}, {Howarth}, {Irwin}, {Burnley}, \& {Harries}}]{Evans2004a}
{Evans}, C.~J., {Howarth}, I.~D., {Irwin}, M.~J., {Burnley}, A.~W., \& {Harries}, T.~J. 2004{\natexlab{a}}, \mnras, 353, 601

\bibitem[{{Evans} {et~al.}(2006){Evans}, {Lennon}, {Smartt}, \& {Trundle}}]{Evans2006}
{Evans}, C.~J., {Lennon}, D.~J., {Smartt}, S.~J., \& {Trundle}, C. 2006, \aap, 456, 623

\bibitem[{{Evans} {et~al.}(2004{\natexlab{b}}){Evans}, {Lennon}, {Walborn}, {Trundle}, \& {Rix}}]{Evans2004b}
{Evans}, C.~J., {Lennon}, D.~J., {Walborn}, N.~R., {Trundle}, C., \& {Rix}, S.~A. 2004{\natexlab{b}}, \pasp, 116, 909

\bibitem[{{Evans} {et~al.}(2011){Evans}, {Taylor}, {H{\'e}nault-Brunet}, {Sana}, {de Koter}, {Sim{\'o}n-D{\'\i}az}, {Carraro}, {Bagnoli}, {Bastian}, {Bestenlehner}, {Bonanos}, {Bressert}, {Brott}, {Campbell}, {Cantiello}, {Clark}, {Costa}, {Crowther}, {de Mink}, {Doran}, {Dufton}, {Dunstall}, {Friedrich}, {Garcia}, {Gieles}, {Gr{\"a}fener}, {Herrero}, {Howarth}, {Izzard}, {Langer}, {Lennon}, {Ma{\'\i}z Apell{\'a}niz}, {Markova}, {Najarro}, {Puls}, {Ramirez}, {Sab{\'\i}n-Sanjuli{\'a}n}, {Smartt}, {Stroud}, {van Loon}, {Vink}, \& {Walborn}}]{Evans2011}
{Evans}, C.~J., {Taylor}, W.~D., {H{\'e}nault-Brunet}, V., {et~al.} 2011, \aap, 530, A108

\bibitem[{{Feldmeier} {et~al.}(1997){Feldmeier}, {Puls}, \& {Pauldrach}}]{Feldmeier1997}
{Feldmeier}, A., {Puls}, J., \& {Pauldrach}, A.~W.~A. 1997, \aap, 322, 878

\bibitem[{{Fitzpatrick}(1988)}]{Fitzpatrick1988}
{Fitzpatrick}, E.~L. 1988, \apj, 335, 703

\bibitem[{{Fitzpatrick}(1999)}]{Fitzpatrick1999}
{Fitzpatrick}, E.~L. 1999, \pasp, 111, 63

\bibitem[{{Fitzpatrick} {et~al.}(2019){Fitzpatrick}, {Massa}, {Gordon}, {Bohlin}, \& {Clayton}}]{Fitzpatrick2019}
{Fitzpatrick}, E.~L., {Massa}, D., {Gordon}, K.~D., {Bohlin}, R., \& {Clayton}, G.~C. 2019, \apj, 886, 108

\bibitem[{{Gaia Collaboration} {et~al.}(2023){Gaia Collaboration}, {Vallenari}, {Brown}, {Prusti}, {de Bruijne}, {Arenou}, {Babusiaux}, {Biermann}, {Creevey}, {Ducourant}, \& et~al.}]{GAIA2023}
{Gaia Collaboration}, {Vallenari}, A., {Brown}, A.~G.~A., {et~al.} 2023, \aap, 674, A1

\bibitem[{{Garcia} \& {Bianchi}(2004)}]{Garcia2004}
{Garcia}, M. \& {Bianchi}, L. 2004, \apj, 606, 497

\bibitem[{{Gordon} {et~al.}(2003){Gordon}, {Clayton}, {Misselt}, {Landolt}, \& {Wolff}}]{Gordon2003}
{Gordon}, K.~D., {Clayton}, G.~C., {Misselt}, K.~A., {Landolt}, A.~U., \& {Wolff}, M.~J. 2003, \apj, 594, 279

\bibitem[{{Gormaz-Matamala} {et~al.}(2021){Gormaz-Matamala}, {Cur{\'e}}, {Hillier}, {Najarro}, {Kub{\'a}tov{\'a}}, \& {Kub{\'a}t}}]{GormazMatamala2021}
{Gormaz-Matamala}, A.~C., {Cur{\'e}}, M., {Hillier}, D.~J., {et~al.} 2021, \apj, 920, 64

\bibitem[{{G{\"o}tberg} {et~al.}(2023){G{\"o}tberg}, {Drout}, {Ji}, {Groh}, {Ludwig}, {Crowther}, {Smith}, {de Koter}, \& {de Mink}}]{Goetberg2023}
{G{\"o}tberg}, Y., {Drout}, M.~R., {Ji}, A.~P., {et~al.} 2023, \apj, 959, 125

\bibitem[{{Graczyk} {et~al.}(2020){Graczyk}, {Pietrzy{\'n}ski}, {Thompson}, {Gieren}, {Zgirski}, {Villanova}, {G{\'o}rski}, {Wielg{\'o}rski}, {Karczmarek}, {Narloch}, {Pilecki}, {Taormina}, {Smolec}, {Suchomska}, {Gallenne}, {Nardetto}, {Storm}, {Kudritzki}, {Ka{\l}uszy{\'n}ski}, \& {Pych}}]{Graczyk2020}
{Graczyk}, D., {Pietrzy{\'n}ski}, G., {Thompson}, I.~B., {et~al.} 2020, \apj, 904, 13

\bibitem[{{Gr{\"a}fener} \& {Hamann}(2005)}]{Graefener2005}
{Gr{\"a}fener}, G. \& {Hamann}, W.~R. 2005, \aap, 432, 633

\bibitem[{{Gr{\"a}fener} {et~al.}(2002){Gr{\"a}fener}, {Koesterke}, \& {Hamann}}]{Graefener2002}
{Gr{\"a}fener}, G., {Koesterke}, L., \& {Hamann}, W.~R. 2002, \aap, 387, 244

\bibitem[{{Gr{\"a}fener} \& {Vink}(2013)}]{Graefener2013}
{Gr{\"a}fener}, G. \& {Vink}, J.~S. 2013, \aap, 560, A6

\bibitem[{{Grassitelli} {et~al.}(2016){Grassitelli}, {Fossati}, {Langer}, {Sim{\'o}n-D{\'\i}az}, {Castro}, \& {Sanyal}}]{Grassitelli2016}
{Grassitelli}, L., {Fossati}, L., {Langer}, N., {et~al.} 2016, \aap, 593, A14

\bibitem[{{Green} {et~al.}(2012){Green}, {Froning}, {Osterman}, {Ebbets}, {Heap}, {Leitherer}, {Linsky}, {Savage}, {Sembach}, {Shull}, {Siegmund}, {Snow}, {Spencer}, {Stern}, {Stocke}, {Welsh}, {B{\'e}land}, {Burgh}, {Danforth}, {France}, {Keeney}, {McPhate}, {Penton}, {Andrews}, {Brownsberger}, {Morse}, \& {Wilkinson}}]{Green2012}
{Green}, J.~C., {Froning}, C.~S., {Osterman}, S., {et~al.} 2012, \apj, 744, 60

\bibitem[{{Gvaramadze} {et~al.}(2018){Gvaramadze}, {Kniazev}, {Maryeva}, \& {Berdnikov}}]{Gvaramadze2018}
{Gvaramadze}, V.~V., {Kniazev}, A.~Y., {Maryeva}, O.~V., \& {Berdnikov}, L.~N. 2018, \mnras, 474, 1412

\bibitem[{{Gvaramadze} {et~al.}(2019){Gvaramadze}, {Maryeva}, {Kniazev}, {Alexashov}, {Castro}, {Langer}, \& {Katkov}}]{Gvaramadze2019}
{Gvaramadze}, V.~V., {Maryeva}, O.~V., {Kniazev}, A.~Y., {et~al.} 2019, \mnras, 482, 4408

\bibitem[{{Hainich} {et~al.}(2019){Hainich}, {Ramachandran}, {Shenar}, {Sander}, {Todt}, {Gruner}, {Oskinova}, \& {Hamann}}]{Hainich2019}
{Hainich}, R., {Ramachandran}, V., {Shenar}, T., {et~al.} 2019, \aap, 621, A85

\bibitem[{{Hamann} \& {Gr{\"a}fener}(2003)}]{Hamann2003}
{Hamann}, W.~R. \& {Gr{\"a}fener}, G. 2003, \aap, 410, 993

\bibitem[{{Hamann} \& {Koesterke}(1998)}]{Hamann1998}
{Hamann}, W.~R. \& {Koesterke}, L. 1998, \aap, 335, 1003

\bibitem[{{Hawcroft} {et~al.}(2021){Hawcroft}, {Sana}, {Mahy}, {Sundqvist}, {Abdul-Masih}, {Bouret}, {Brands}, {de Koter}, {Driessen}, \& {Puls}}]{Hawcroft2021}
{Hawcroft}, C., {Sana}, H., {Mahy}, L., {et~al.} 2021, \aap, 655, A67

\bibitem[{{Hawcroft} {et~al.}(2023){Hawcroft}, {Sana}, {Mahy}, {Sundqvist}, {de Koter}, {Crowther}, {Bestenlehner}, {Brands}, {David-Uraz}, {Decin}, {Erba}, {Garcia}, {Hamann}, {Herrero}, {Ignace}, {Kee}, {Kub{\'a}tov{\'a}}, {Lefever}, {Moffat}, {Najarro}, {Oskinova}, {Pauli}, {Prinja}, {Puls}, {Sander}, {Shenar}, {St-Louis}, {ud-Doula}, \& {Vink}}]{Hawcroft2023}
{Hawcroft}, C., {Sana}, H., {Mahy}, L., {et~al.} 2023, \aap, in press, arXiv:2303.12165

\bibitem[{{Heap} {et~al.}(2006){Heap}, {Lanz}, \& {Hubeny}}]{Heap2006}
{Heap}, S.~R., {Lanz}, T., \& {Hubeny}, I. 2006, \apj, 638, 409

\bibitem[{{Herrero} {et~al.}(1992){Herrero}, {Kudritzki}, {Vilchez}, {Kunze}, {Butler}, \& {Haser}}]{Herrero1992}
{Herrero}, A., {Kudritzki}, R.~P., {Vilchez}, J.~M., {et~al.} 1992, \aap, 261, 209

\bibitem[{{Hillier}(1990)}]{Hillier1990}
{Hillier}, D.~J. 1990, \aap, 231, 116

\bibitem[{{Hillier} {et~al.}(2003){Hillier}, {Lanz}, {Heap}, {Hubeny}, {Smith}, {Evans}, {Lennon}, \& {Bouret}}]{Hillier2003}
{Hillier}, D.~J., {Lanz}, T., {Heap}, S.~R., {et~al.} 2003, \apj, 588, 1039

\bibitem[{{Hillier} \& {Miller}(1998)}]{Hillier1998}
{Hillier}, D.~J. \& {Miller}, D.~L. 1998, \apj, 496, 407

\bibitem[{{Holgado} {et~al.}(2018){Holgado}, {Sim{\'o}n-D{\'\i}az}, {Barb{\'a}}, {Puls}, {Herrero}, {Castro}, {Garcia}, {Ma{\'\i}z Apell{\'a}niz}, {Negueruela}, \& {Sab{\'\i}n-Sanjuli{\'a}n}}]{Holgado2018}
{Holgado}, G., {Sim{\'o}n-D{\'\i}az}, S., {Barb{\'a}}, R.~H., {et~al.} 2018, \aap, 613, A65

\bibitem[{{Holgado} {et~al.}(2020){Holgado}, {Sim{\'o}n-D{\'\i}az}, {Haemmerl{\'e}}, {Lennon}, {Barb{\'a}}, {Cervi{\~n}o}, {Castro}, {Herrero}, {Meynet}, \& {Arias}}]{Holgado2020}
{Holgado}, G., {Sim{\'o}n-D{\'\i}az}, S., {Haemmerl{\'e}}, L., {et~al.} 2020, \aap, 638, A157

\bibitem[{{Howarth}(1983)}]{Howarth1983}
{Howarth}, I.~D. 1983, \mnras, 203, 301

\bibitem[{{Hummer}(1982)}]{Hummer1982}
{Hummer}, D.~G. 1982, \apj, 257, 724

\bibitem[{{Hummer} {et~al.}(1988){Hummer}, {Abbott}, {Voels}, \& {Bohannan}}]{Hummer1988}
{Hummer}, D.~G., {Abbott}, D.~C., {Voels}, S.~A., \& {Bohannan}, B. 1988, \apj, 328, 704

\bibitem[{{Isserstedt}(1975)}]{Isserstedt1975}
{Isserstedt}, J. 1975, \aaps, 19, 259

\bibitem[{{Kimble} {et~al.}(1998){Kimble}, {Woodgate}, {Bowers}, {Kraemer}, {Kaiser}, {Gull}, {Heap}, {Danks}, {Boggess}, {Green}, {Hutchings}, {Jenkins}, {Joseph}, {Linsky}, {Maran}, {Moos}, {Roesler}, {Timothy}, {Weistrop}, {Grady}, {Loiacono}, {Brown}, {Brumfield}, {Content}, {Feinberg}, {Isaacs}, {Krebs}, {Krueger}, {Melcher}, {Rebar}, {Vitagliano}, {Yagelowich}, {Meyer}, {Hood}, {Argabright}, {Becker}, {Bottema}, {Breyer}, {Bybee}, {Christon}, {Delamere}, {Dorn}, {Downey}, {Driggers}, {Ebbets}, {Gallegos}, {Garner}, {Hetlinger}, {Lettieri}, {Ludtke}, {Michika}, {Nyquist}, {Rose}, {Stocker}, {Sullivan}, {Van Houten}, {Woodruff}, {Baum}, {Hartig}, {Balzano}, {Biagetti}, {Blades}, {Bohlin}, {Clampin}, {Doxsey}, {Ferguson}, {Goudfrooij}, {Hulbert}, {Kutina}, {McGrath}, {Lindler}, {Beck}, {Feggans}, {Plait}, {Sandoval}, {Hill}, {Collins}, {Cornett}, {Fowler}, {Hill}, {Landsman}, {Malumuth}, {Standley}, {Blouke}, {Grusczak}, {Reed}, {Robinson}, {Valenti}, \& {Wolfe}}]{Kimble1998}
{Kimble}, R.~A., {Woodgate}, B.~E., {Bowers}, C.~W., {et~al.} 1998, \apjl, 492, L83

\bibitem[{{K{\"o}hler} {et~al.}(2015){K{\"o}hler}, {Langer}, {de Koter}, {de Mink}, {Crowther}, {Evans}, {Gr{\"a}fener}, {Sana}, {Sanyal}, {Schneider}, \& {Vink}}]{Koehler2015}
{K{\"o}hler}, K., {Langer}, N., {de Koter}, A., {et~al.} 2015, \aap, 573, A71

\bibitem[{{Krti{\v{c}}ka} \& {Kub{\'a}t}(2018)}]{Krticka2018}
{Krti{\v{c}}ka}, J. \& {Kub{\'a}t}, J. 2018, \aap, 612, A20

\bibitem[{{Kudritzki} \& {Puls}(2000)}]{Kudritzki2000}
{Kudritzki}, R.-P. \& {Puls}, J. 2000, \araa, 38, 613

\bibitem[{{Lamers} {et~al.}(1995){Lamers}, {Snow}, \& {Lindholm}}]{Lamers1995}
{Lamers}, H. J.~G.~L.~M., {Snow}, T.~P., \& {Lindholm}, D.~M. 1995, \apj, 455, 269

\bibitem[{{Lanz} {et~al.}(1996){Lanz}, {de Koter}, {Hubeny}, \& {Heap}}]{Lanz1996}
{Lanz}, T., {de Koter}, A., {Hubeny}, I., \& {Heap}, S.~R. 1996, \apj, 465, 359

\bibitem[{{Lanz} \& {Hubeny}(2003)}]{Lanz2003}
{Lanz}, T. \& {Hubeny}, I. 2003, \apjs, 146, 417

\bibitem[{{Lanz} \& {Hubeny}(2007)}]{Lanz2007}
{Lanz}, T. \& {Hubeny}, I. 2007, \apjs, 169, 83

\bibitem[{{Lecroq} {et~al.}(2024){Lecroq}, {Charlot}, {Bressan}, {Bruzual}, {Costa}, {Iorio}, {Spera}, {Mapelli}, {Chen}, {Chevallard}, \& {Dall'Amico}}]{Lecroq2024}
{Lecroq}, M., {Charlot}, S., {Bressan}, A., {et~al.} 2024, \mnras, 527, 9480

\bibitem[{{Mahy} {et~al.}(2017){Mahy}, {Damerdji}, {Gosset}, {Nitschelm}, {Eenens}, {Sana}, \& {Klotz}}]{Mahy2017}
{Mahy}, L., {Damerdji}, Y., {Gosset}, E., {et~al.} 2017, \aap, 607, A96

\bibitem[{{Mahy} {et~al.}(2020){Mahy}, {Sana}, {Abdul-Masih}, {Almeida}, {Langer}, {Shenar}, {de Koter}, {de Mink}, {de Wit}, {Grin}, {Evans}, {Moffat}, {Schneider}, {Barb{\'a}}, {Clark}, {Crowther}, {Gr{\"a}fener}, {Lennon}, {Tramper}, \& {Vink}}]{Mahy2020}
{Mahy}, L., {Sana}, H., {Abdul-Masih}, M., {et~al.} 2020, \aap, 634, A118

\bibitem[{{Ma{\'\i}z Apell{\'a}niz} {et~al.}(2014){Ma{\'\i}z Apell{\'a}niz}, {Evans}, {Barb{\'a}}, {Gr{\"a}fener}, {Bestenlehner}, {Crowther}, {Garc{\'\i}a}, {Herrero}, {Sana}, {Sim{\'o}n-D{\'\i}az}, {Taylor}, {van Loon}, {Vink}, \& {Walborn}}]{JMaiz2014}
{Ma{\'\i}z Apell{\'a}niz}, J., {Evans}, C.~J., {Barb{\'a}}, R.~H., {et~al.} 2014, \aap, 564, A63

\bibitem[{{Marcolino} {et~al.}(2024){Marcolino}, {Bouret}, {Martins}, \& {Hillier}}]{Marcolino2024}
{Marcolino}, W., {Bouret}, J.~C., {Martins}, F., \& {Hillier}, D.~J. 2024, \aap, arXiv:2408.10364

\bibitem[{{Markova} {et~al.}(2018){Markova}, {Puls}, \& {Langer}}]{Markova2018}
{Markova}, N., {Puls}, J., \& {Langer}, N. 2018, \aap, 613, A12

\bibitem[{{Markova} {et~al.}(2005){Markova}, {Puls}, {Scuderi}, \& {Markov}}]{Markova2005}
{Markova}, N., {Puls}, J., {Scuderi}, S., \& {Markov}, H. 2005, \aap, 440, 1133

\bibitem[{{Martins} {et~al.}(2024){Martins}, {Bouret}, {Hillier}, {Brands}, {Crowther}, {Herrero}, {Najarro}, {Pauli}, {Puls}, {Ramachandran}, {Sander}, {Vink}, \& {the XshootU collaboration}}]{Martins2024}
{Martins}, F., {Bouret}, J.~C., {Hillier}, D.~J., {et~al.} 2024, \aap, in press, arXiv:2405.01267

\bibitem[{{Martins} {et~al.}(2012{\natexlab{a}}){Martins}, {Escolano}, {Wade}, {Donati}, {Bouret}, \& {MIMES Collaboration}}]{martins12b}
{Martins}, F., {Escolano}, C., {Wade}, G.~A., {et~al.} 2012{\natexlab{a}}, \aap, 538, A29

\bibitem[{{Martins} {et~al.}(2015){Martins}, {Herv{\'e}}, {Bouret}, {Marcolino}, {Wade}, {Neiner}, {Alecian}, {Grunhut}, \& {Petit}}]{martins15}
{Martins}, F., {Herv{\'e}}, A., {Bouret}, J.~C., {et~al.} 2015, \aap, 575, A34

\bibitem[{{Martins} {et~al.}(2012{\natexlab{b}}){Martins}, {Mahy}, {Hillier}, \& {Rauw}}]{Martins2012}
{Martins}, F., {Mahy}, L., {Hillier}, D.~J., \& {Rauw}, G. 2012{\natexlab{b}}, \aap, 538, A39

\bibitem[{{Martins} \& {Palacios}(2021)}]{Martins2021}
{Martins}, F. \& {Palacios}, A. 2021, \aap, 645, A67

\bibitem[{{Martins} \& {Plez}(2006)}]{mp06}
{Martins}, F. \& {Plez}, B. 2006, \aap, 457, 637

\bibitem[{{Martins} {et~al.}(2005){Martins}, {Schaerer}, \& {Hillier}}]{Martins2005}
{Martins}, F., {Schaerer}, D., \& {Hillier}, D.~J. 2005, \aap, 436, 1049

\bibitem[{{Maryeva} {et~al.}(2014){Maryeva}, {Zhuchkov}, \& {Malogolovets}}]{Maryeva2014}
{Maryeva}, O., {Zhuchkov}, R., \& {Malogolovets}, E. 2014, \pasa, 31, e020

\bibitem[{{Massey}(2002)}]{Massey2002}
{Massey}, P. 2002, \apjs, 141, 81

\bibitem[{{Massey} {et~al.}(2004){Massey}, {Bresolin}, {Kudritzki}, {Puls}, \& {Pauldrach}}]{Massey2004}
{Massey}, P., {Bresolin}, F., {Kudritzki}, R.~P., {Puls}, J., \& {Pauldrach}, A.~W.~A. 2004, \apj, 608, 1001

\bibitem[{{Massey} {et~al.}(2013){Massey}, {Neugent}, {Hillier}, \& {Puls}}]{Massey2013}
{Massey}, P., {Neugent}, K.~F., {Hillier}, D.~J., \& {Puls}, J. 2013, \apj, 768, 6

\bibitem[{{Massey} {et~al.}(2005){Massey}, {Puls}, {Pauldrach}, {Bresolin}, {Kudritzki}, \& {Simon}}]{Massey2005}
{Massey}, P., {Puls}, J., {Pauldrach}, A.~W.~A., {et~al.} 2005, \apj, 627, 477

\bibitem[{{Massey} {et~al.}(2009){Massey}, {Zangari}, {Morrell}, {Puls}, {DeGioia-Eastwood}, {Bresolin}, \& {Kudritzki}}]{Massey2009}
{Massey}, P., {Zangari}, A.~M., {Morrell}, N.~I., {et~al.} 2009, \apj, 692, 618

\bibitem[{{Misselt} {et~al.}(1999){Misselt}, {Clayton}, \& {Gordon}}]{Misselt1999}
{Misselt}, K.~A., {Clayton}, G.~C., \& {Gordon}, K.~D. 1999, \apj, 515, 128

\bibitem[{{Mokiem} {et~al.}(2005){Mokiem}, {de Koter}, {Puls}, {Herrero}, {Najarro}, \& {Villamariz}}]{Mokiem2005}
{Mokiem}, M.~R., {de Koter}, A., {Puls}, J., {et~al.} 2005, \aap, 441, 711

\bibitem[{{Mokiem} {et~al.}(2007){Mokiem}, {de Koter}, {Vink}, {Puls}, {Evans}, {Smartt}, {Crowther}, {Herrero}, {Langer}, {Lennon}, {Najarro}, \& {Villamariz}}]{Mokiem2007}
{Mokiem}, M.~R., {de Koter}, A., {Vink}, J.~S., {et~al.} 2007, \aap, 473, 603

\bibitem[{{Moos} {et~al.}(2000){Moos}, {Cash}, {Cowie}, {Davidsen}, {Dupree}, {Feldman}, {Friedman}, {Green}, {Green}, {Gry}, {Hutchings}, {Jenkins}, {Linsky}, {Malina}, {Michalitsianos}, {Savage}, {Shull}, {Siegmund}, {Snow}, {Sonneborn}, {Vidal-Madjar}, {Willis}, {Woodgate}, {York}, {Ake}, {Andersson}, {Andrews}, {Barkhouser}, {Bianchi}, {Blair}, {Brownsberger}, {Cha}, {Chayer}, {Conard}, {Fullerton}, {Gaines}, {Grange}, {Gummin}, {Hebrard}, {Kriss}, {Kruk}, {Mark}, {McCarthy}, {Morbey}, {Murowinski}, {Murphy}, {Oegerle}, {Ohl}, {Oliveira}, {Osterman}, {Sahnow}, {Saisse}, {Sembach}, {Weaver}, {Welsh}, {Wilkinson}, \& {Zheng}}]{Moos2000}
{Moos}, H.~W., {Cash}, W.~C., {Cowie}, L.~L., {et~al.} 2000, \apjl, 538, L1

\bibitem[{{M{\"u}ller} \& {Vink}(2008)}]{MuellerVink2008}
{M{\"u}ller}, P.~E. \& {Vink}, J.~S. 2008, \aap, 492, 493

\bibitem[{{Najarro} {et~al.}(2006){Najarro}, {Hillier}, {Puls}, {Lanz}, \& {Martins}}]{Najarro+2006}
{Najarro}, F., {Hillier}, D.~J., {Puls}, J., {Lanz}, T., \& {Martins}, F. 2006, \aap, 456, 659

\bibitem[{{Neijssel} {et~al.}(2019){Neijssel}, {Vigna-G{\'o}mez}, {Stevenson}, {Barrett}, {Gaebel}, {Broekgaarden}, {de Mink}, {Sz{\'e}csi}, {Vinciguerra}, \& {Mandel}}]{Neijssel2019}
{Neijssel}, C.~J., {Vigna-G{\'o}mez}, A., {Stevenson}, S., {et~al.} 2019, \mnras, 490, 3740

\bibitem[{{Oskinova} {et~al.}(2007){Oskinova}, {Hamann}, \& {Feldmeier}}]{Oskinova2007}
{Oskinova}, L.~M., {Hamann}, W.~R., \& {Feldmeier}, A. 2007, \aap, 476, 1331

\bibitem[{{Owocki}(2008)}]{Owocki2008}
{Owocki}, S.~P. 2008, in Clumping in Hot-Star Winds, ed. W.-R. {Hamann}, A.~{Feldmeier}, \& L.~M. {Oskinova}, 121

\bibitem[{{Page} {et~al.}(2012){Page}, {Brindle}, {Talavera}, {Still}, {Rosen}, {Yershov}, {Ziaeepour}, {Mason}, {Cropper}, {Breeveld}, {Loiseau}, {Mignani}, {Smith}, \& {Murdin}}]{Page2012}
{Page}, M.~J., {Brindle}, C., {Talavera}, A., {et~al.} 2012, \mnras, 426, 903

\bibitem[{{Pauldrach} {et~al.}(1986){Pauldrach}, {Puls}, \& {Kudritzki}}]{Pauldrach1986}
{Pauldrach}, A., {Puls}, J., \& {Kudritzki}, R.~P. 1986, \aap, 164, 86

\bibitem[{{Pauldrach} {et~al.}(2001){Pauldrach}, {Hoffmann}, \& {Lennon}}]{Pauldrach2001}
{Pauldrach}, A.~W.~A., {Hoffmann}, T.~L., \& {Lennon}, M. 2001, \aap, 375, 161

\bibitem[{{Pietrzy{\'n}ski} {et~al.}(2019){Pietrzy{\'n}ski}, {Graczyk}, {Gallenne}, {Gieren}, {Thompson}, {Pilecki}, {Karczmarek}, {G{\'o}rski}, {Suchomska}, {Taormina}, {Zgirski}, {Wielg{\'o}rski}, {Ko{\l}aczkowski}, {Konorski}, {Villanova}, {Nardetto}, {Kervella}, {Bresolin}, {Kudritzki}, {Storm}, {Smolec}, \& {Narloch}}]{Pietrzynski2019}
{Pietrzy{\'n}ski}, G., {Graczyk}, D., {Gallenne}, A., {et~al.} 2019, \nat, 567, 200

\bibitem[{{Pontoppidan} {et~al.}(2022){Pontoppidan}, {Barrientes}, {Blome}, {Braun}, {Brown}, {Carruthers}, {Coe}, {DePasquale}, {Espinoza}, {Marin}, {Gordon}, {Henry}, {Hustak}, {James}, {Jenkins}, {Koekemoer}, {LaMassa}, {Law}, {Lockwood}, {Moro-Martin}, {Mullally}, {Pagan}, {Player}, {Proffitt}, {Pulliam}, {Ramsay}, {Ravindranath}, {Reid}, {Robberto}, {Sabbi}, {Ubeda}, {Balogh}, {Flanagan}, {Gardner}, {Hasan}, {Meinke}, \& {Nota}}]{Pontoppidan2022}
{Pontoppidan}, K.~M., {Barrientes}, J., {Blome}, C., {et~al.} 2022, \apjl, 936, L14

\bibitem[{{Prinja} {et~al.}(2006){Prinja}, {Markova}, {Scuderi}, \& {Markov}}]{Prinja2006}
{Prinja}, R.~K., {Markova}, N., {Scuderi}, S., \& {Markov}, H. 2006, \aap, 457, 987

\bibitem[{{Puls} {et~al.}(1996){Puls}, {Kudritzki}, {Herrero}, {Pauldrach}, {Haser}, {Lennon}, {Gabler}, {Voels}, {Vilchez}, {Wachter}, \& {Feldmeier}}]{Puls1996}
{Puls}, J., {Kudritzki}, R.~P., {Herrero}, A., {et~al.} 1996, \aap, 305, 171

\bibitem[{{Puls} {et~al.}(2020){Puls}, {Najarro}, {Sundqvist}, \& {Sen}}]{Puls2020}
{Puls}, J., {Najarro}, F., {Sundqvist}, J.~O., \& {Sen}, K. 2020, \aap, 642, A172

\bibitem[{{Puls} {et~al.}(2005){Puls}, {Urbaneja}, {Venero}, {Repolust}, {Springmann}, {Jokuthy}, \& {Mokiem}}]{Puls2005}
{Puls}, J., {Urbaneja}, M.~A., {Venero}, R., {et~al.} 2005, \aap, 435, 669

\bibitem[{{Puls} {et~al.}(2008){Puls}, {Vink}, \& {Najarro}}]{Puls2008}
{Puls}, J., {Vink}, J.~S., \& {Najarro}, F. 2008, \aapr, 16, 209

\bibitem[{{Ramachandran} {et~al.}(2018{\natexlab{a}}){Ramachandran}, {Hainich}, {Hamann}, {Oskinova}, {Shenar}, {Sander}, {Todt}, \& {Gallagher}}]{Ramachandran2018Of}
{Ramachandran}, V., {Hainich}, R., {Hamann}, W.~R., {et~al.} 2018{\natexlab{a}}, \aap, 609, A7

\bibitem[{{Ramachandran} {et~al.}(2018{\natexlab{b}}){Ramachandran}, {Hamann}, {Hainich}, {Oskinova}, {Shenar}, {Sander}, {Todt}, \& {Gallagher}}]{Ramachandran2018}
{Ramachandran}, V., {Hamann}, W.~R., {Hainich}, R., {et~al.} 2018{\natexlab{b}}, \aap, 615, A40

\bibitem[{{Ramachandran} {et~al.}(2019){Ramachandran}, {Hamann}, {Oskinova}, {Gallagher}, {Hainich}, {Shenar}, {Sander}, {Todt}, \& {Fulmer}}]{Ramachandran2019}
{Ramachandran}, V., {Hamann}, W.~R., {Oskinova}, L.~M., {et~al.} 2019, \aap, 625, A104

\bibitem[{{Ramachandran} {et~al.}(2023){Ramachandran}, {Klencki}, {Sander}, {Pauli}, {Shenar}, {Oskinova}, \& {Hamann}}]{Ramachandran2023}
{Ramachandran}, V., {Klencki}, J., {Sander}, A.~A.~C., {et~al.} 2023, \aap, 674, L12

\bibitem[{{Rauw} {et~al.}(2015){Rauw}, {Naz{\'e}}, {Wright}, {Drake}, {Guarcello}, {Prinja}, {Peck}, {Albacete Colombo}, {Herrero}, {Kobulnicky}, {Sciortino}, \& {Vink}}]{Rauw2015}
{Rauw}, G., {Naz{\'e}}, Y., {Wright}, N.~J., {et~al.} 2015, \apjs, 221, 1

\bibitem[{{Rickard} {et~al.}(2022){Rickard}, {Hainich}, {Hamann}, {Oskinova}, {Prinja}, {Ramachandran}, {Pauli}, {Todt}, {Sander}, {Shenar}, {Chu}, \& {Gallagher}}]{Rickard2022}
{Rickard}, M.~J., {Hainich}, R., {Hamann}, W.~R., {et~al.} 2022, \aap, 666, A189

\bibitem[{{Rivero Gonz{\'a}lez} {et~al.}(2011){Rivero Gonz{\'a}lez}, {Puls}, \& {Najarro}}]{RiveroGonzalez2011}
{Rivero Gonz{\'a}lez}, J.~G., {Puls}, J., \& {Najarro}, F. 2011, \aap, 536, A58

\bibitem[{{Rivero Gonz{\'a}lez} {et~al.}(2012){Rivero Gonz{\'a}lez}, {Puls}, {Najarro}, \& {Brott}}]{RiveroGonzalez2012}
{Rivero Gonz{\'a}lez}, J.~G., {Puls}, J., {Najarro}, F., \& {Brott}, I. 2012, \aap, 537, A79

\bibitem[{{Rix} {et~al.}(2004){Rix}, {Pettini}, {Leitherer}, {Bresolin}, {Kudritzki}, \& {Steidel}}]{Rix2004}
{Rix}, S.~A., {Pettini}, M., {Leitherer}, C., {et~al.} 2004, \apj, 615, 98

\bibitem[{{Robertson} {et~al.}(2023){Robertson}, {Tacchella}, {Johnson}, {Hainline}, {Whitler}, {Eisenstein}, {Endsley}, {Rieke}, {Stark}, {Alberts}, {Dressler}, {Egami}, {Hausen}, {Rieke}, {Shivaei}, {Williams}, {Willmer}, {Arribas}, {Bonaventura}, {Bunker}, {Cameron}, {Carniani}, {Charlot}, {Chevallard}, {Curti}, {Curtis-Lake}, {D'Eugenio}, {Jakobsen}, {Looser}, {L{\"u}tzgendorf}, {Maiolino}, {Maseda}, {Rawle}, {Rix}, {Smit}, {{\"U}bler}, {Willott}, {Witstok}, {Baum}, {Bhatawdekar}, {Boyett}, {Chen}, {de Graaff}, {Florian}, {Helton}, {Hviding}, {Ji}, {Kumari}, {Lyu}, {Nelson}, {Sandles}, {Saxena}, {Suess}, {Sun}, {Topping}, \& {Wallace}}]{Robertson2023}
{Robertson}, B.~E., {Tacchella}, S., {Johnson}, B.~D., {et~al.} 2023, Nature Astronomy, 7, 611

\bibitem[{{Roman-Duval} {et~al.}(2020){Roman-Duval}, {Proffitt}, {Taylor}, {Monroe}, {Fischer}, {Fischer}, {Fullerton}, {Aloisi}, {Britt}, {Busko}, {Carlberg}, {De Rosa}, {Jedrzejewski}, {Lockwood}, {Frazer}, {Hernandez}, {James}, {Oliveira}, {Plesha}, {Riedel}, {Riley}, {Sahnow}, {Sankrit}, {Shaw}, {Smith}, {Sohn}, {Som}, {Ubeda}, \& {Welty}}]{Roman-Duval2020}
{Roman-Duval}, J., {Proffitt}, C.~R., {Taylor}, J.~M., {et~al.} 2020, Research Notes of the American Astronomical Society, 4, 205

\bibitem[{{R{\"u}bke} {et~al.}(2023){R{\"u}bke}, {Herrero}, \& {Puls}}]{Ruebke2023}
{R{\"u}bke}, K., {Herrero}, A., \& {Puls}, J. 2023, \aap, 679, A19

\bibitem[{{Sana} {et~al.}(2024){Sana}, {Tramper}, {Abdul-Masih}, {Blomme}, {Dsilva}, {Maravelias}, {Martins}, {Mehner}, {Wofford}, {Banyard}, {Barbosa}, {Bestenlehner}, {Hawcroft}, {Hillier}, {Todt}, {Larkin}, {Mahy}, {Najarro}, {Ramachandran}, {Ramirez-Tannus}, {Rubio-Diez}, {Sander}, {Shenar}, {Vink}, {Backs}, {Brands}, {Crowther}, {Decin}, {de Koter}, {Hamann}, {Kehrig}, {Kuiper}, {Oskinova}, {Pauli}, {Sundqvist}, {Verhamme}, \& {the XSHOOT-U collaboration}}]{Sana2024}
{Sana}, H., {Tramper}, F., {Abdul-Masih}, M., {et~al.} 2024, \aap, in press, arXiv:2402.16987

\bibitem[{{Sander} {et~al.}(2015){Sander}, {Shenar}, {Hainich}, {G{\'\i}menez-Garc{\'\i}a}, {Todt}, \& {Hamann}}]{Sander2015}
{Sander}, A., {Shenar}, T., {Hainich}, R., {et~al.} 2015, \aap, 577, A13

\bibitem[{{Sander} {et~al.}(2017){Sander}, {Hamann}, {Todt}, {Hainich}, \& {Shenar}}]{Sander2017}
{Sander}, A.~A.~C., {Hamann}, W.~R., {Todt}, H., {Hainich}, R., \& {Shenar}, T. 2017, \aap, 603, A86

\bibitem[{{Sander} {et~al.}(2023){Sander}, {Lefever}, {Poniatowski}, {Ramachandran}, {Sabhahit}, \& {Vink}}]{Sander2023}
{Sander}, A.~A.~C., {Lefever}, R.~R., {Poniatowski}, L.~G., {et~al.} 2023, \aap, 670, A83

\bibitem[{{Sander} \& {Vink}(2020)}]{Sander2020}
{Sander}, A. A.~C. \& {Vink}, J.~S. 2020, \mnras, 499, 873

\bibitem[{{Santolaya-Rey} {et~al.}(1997){Santolaya-Rey}, {Puls}, \& {Herrero}}]{SantolayaRey1997}
{Santolaya-Rey}, A.~E., {Puls}, J., \& {Herrero}, A. 1997, \aap, 323, 488

\bibitem[{{Schaerer} {et~al.}(2022){Schaerer}, {Marques-Chaves}, {Barrufet}, {Oesch}, {Izotov}, {Naidu}, {Guseva}, \& {Brammer}}]{Schaerer2022}
{Schaerer}, D., {Marques-Chaves}, R., {Barrufet}, L., {et~al.} 2022, \aap, 665, L4

\bibitem[{{Schmutz} {et~al.}(1989){Schmutz}, {Hamann}, \& {Wessolowski}}]{Schmutz1989}
{Schmutz}, W., {Hamann}, W.~R., \& {Wessolowski}, U. 1989, \aap, 210, 236

\bibitem[{{Schultz} {et~al.}(2023){Schultz}, {Bildsten}, \& {Jiang}}]{Schultz2023}
{Schultz}, W.~C., {Bildsten}, L., \& {Jiang}, Y.-F. 2023, \apjl, 951, L42

\bibitem[{{Seaton}(1979)}]{Seaton1979}
{Seaton}, M.~J. 1979, \mnras, 187, 73P

\bibitem[{{Shenar} {et~al.}(2014){Shenar}, {Hamann}, \& {Todt}}]{Shenar2014}
{Shenar}, T., {Hamann}, W.~R., \& {Todt}, H. 2014, \aap, 562, A118

\bibitem[{{Sim{\'o}n-D{\'\i}az} {et~al.}(2011){Sim{\'o}n-D{\'\i}az}, {Castro}, {Herrero}, {Puls}, {Garcia}, \& {Sab{\'\i}n-Sanjuli{\'a}n}}]{SimonDiaz2011}
{Sim{\'o}n-D{\'\i}az}, S., {Castro}, N., {Herrero}, A., {et~al.} 2011, in Journal of Physics Conference Series, Vol. 328, Journal of Physics Conference Series, 012021

\bibitem[{{Sim{\'o}n-D{\'\i}az} \& {Herrero}(2007)}]{SimonDiaz2007}
{Sim{\'o}n-D{\'\i}az}, S. \& {Herrero}, A. 2007, \aap, 468, 1063

\bibitem[{{Sim{\'o}n-D{\'\i}az} \& {Herrero}(2014)}]{SimonDiazHerrero2014}
{Sim{\'o}n-D{\'\i}az}, S. \& {Herrero}, A. 2014, \aap, 562, A135

\bibitem[{{Smith} {et~al.}(2002){Smith}, {Norris}, \& {Crowther}}]{Smith2002}
{Smith}, L.~J., {Norris}, R. P.~F., \& {Crowther}, P.~A. 2002, \mnras, 337, 1309

\bibitem[{{Sobolev}(1960)}]{Sobolev1960}
{Sobolev}, V.~V. 1960, {Moving Envelopes of Stars} (Cambridge, MA: Harvard University Press)

\bibitem[{{Steidel} {et~al.}(2014){Steidel}, {Rudie}, {Strom}, {Pettini}, {Reddy}, {Shapley}, {Trainor}, {Erb}, {Turner}, {Konidaris}, {Kulas}, {Mace}, {Matthews}, \& {McLean}}]{Steidel2014}
{Steidel}, C.~C., {Rudie}, G.~C., {Strom}, A.~L., {et~al.} 2014, \apj, 795, 165

\bibitem[{{Stevance} {et~al.}(2023){Stevance}, {Eldridge}, {Stanway}, {Lyman}, {McLeod}, \& {Levan}}]{Stevance2023}
{Stevance}, H.~F., {Eldridge}, J.~J., {Stanway}, E.~R., {et~al.} 2023, Nature Astronomy, 7, 444

\bibitem[{{Sundqvist} {et~al.}(2013){Sundqvist}, {Petit}, {Owocki}, {Wade}, {Puls}, \& {MiMeS Collaboration}}]{Sundqvist2013}
{Sundqvist}, J.~O., {Petit}, V., {Owocki}, S.~P., {et~al.} 2013, \mnras, 433, 2497

\bibitem[{{Sundqvist} \& {Puls}(2018)}]{Sundqvist2018}
{Sundqvist}, J.~O. \& {Puls}, J. 2018, \aap, 619, A59

\bibitem[{{Sundqvist} {et~al.}(2011){Sundqvist}, {Puls}, {Feldmeier}, \& {Owocki}}]{Sundqvist2011}
{Sundqvist}, J.~O., {Puls}, J., {Feldmeier}, A., \& {Owocki}, S.~P. 2011, \aap, 528, A64

\bibitem[{{Tramper} {et~al.}(2014){Tramper}, {Sana}, {de Koter}, {Kaper}, \& {Ram{\'\i}rez-Agudelo}}]{Tramper2014}
{Tramper}, F., {Sana}, H., {de Koter}, A., {Kaper}, L., \& {Ram{\'\i}rez-Agudelo}, O.~H. 2014, \aap, 572, A36

\bibitem[{{Trundle} {et~al.}(2007){Trundle}, {Dufton}, {Hunter}, {Evans}, {Lennon}, {Smartt}, \& {Ryans}}]{Trundle2007}
{Trundle}, C., {Dufton}, P.~L., {Hunter}, I., {et~al.} 2007, \aap, 471, 625

\bibitem[{{Trussler} {et~al.}(2023){Trussler}, {Adams}, {Conselice}, {Ferreira}, {Austin}, {Bhatawdekar}, {Caruana}, {Frye}, {Harvey}, {Lovell}, {Pascale}, {Roper}, {Verma}, {Vijayan}, \& {Wilkins}}]{Trussler2023}
{Trussler}, J. A.~A., {Adams}, N.~J., {Conselice}, C.~J., {et~al.} 2023, \mnras, 523, 3423

\bibitem[{{Vernet} {et~al.}(2011){Vernet}, {Dekker}, {D'Odorico}, {Kaper}, {Kjaergaard}, {Hammer}, {Randich}, {Zerbi}, {Groot}, {Hjorth}, {Guinouard}, {Navarro}, {Adolfse}, {Albers}, {Amans}, {Andersen}, {Andersen}, {Binetruy}, {Bristow}, {Castillo}, {Chemla}, {Christensen}, {Conconi}, {Conzelmann}, {Dam}, {de Caprio}, {de Ugarte Postigo}, {Delabre}, {di Marcantonio}, {Downing}, {Elswijk}, {Finger}, {Fischer}, {Flores}, {Fran{\c{c}}ois}, {Goldoni}, {Guglielmi}, {Haigron}, {Hanenburg}, {Hendriks}, {Horrobin}, {Horville}, {Jessen}, {Kerber}, {Kern}, {Kiekebusch}, {Kleszcz}, {Klougart}, {Kragt}, {Larsen}, {Lizon}, {Lucuix}, {Mainieri}, {Manuputy}, {Martayan}, {Mason}, {Mazzoleni}, {Michaelsen}, {Modigliani}, {Moehler}, {M{\o}ller}, {Norup S{\o}rensen}, {N{\o}rregaard}, {P{\'e}roux}, {Patat}, {Pena}, {Pragt}, {Reinero}, {Rigal}, {Riva}, {Roelfsema}, {Royer}, {Sacco}, {Santin}, {Schoenmaker}, {Spano}, {Sweers}, {Ter Horst}, {Tintori}, {Tromp}, {van Dael}, {van der Vliet}, {Venema}, {Vidali}, {Vinther}, {Vola},
  {Winters}, {Wistisen}, {Wulterkens}, \& {Zacchei}}]{Vernet2011}
{Vernet}, J., {Dekker}, H., {D'Odorico}, S., {et~al.} 2011, \aap, 536, A105

\bibitem[{{Vink} {et~al.}(2001){Vink}, {de Koter}, \& {Lamers}}]{vink2001}
{Vink}, J.~S., {de Koter}, A., \& {Lamers}, H.~J.~G.~L.~M. 2001, \aap, 369, 574

\bibitem[{{Vink} {et~al.}(2023){Vink}, {Mehner}, {Crowther}, {Fullerton}, {Garcia}, {Martins}, {Morrell}, {Oskinova}, {St-Louis}, {ud-Doula}, {Sander}, {Sana}, {Bouret}, {Kub{\'a}tov{\'a}}, {Marchant}, {Martins}, {Wofford}, {van Loon}, {Grace Telford}, {G{\"o}tberg}, {Bowman}, {Erba}, {Kalari}, {Abdul-Masih}, {Alkousa}, {Backs}, {Barbosa}, {Berlanas}, {Bernini-Peron}, {Bestenlehner}, {Blomme}, {Bodensteiner}, {Brands}, {Evans}, {David-Uraz}, {Driessen}, {Dsilva}, {Geen}, {G{\'o}mez-Gonz{\'a}lez}, {Grassitelli}, {Hamann}, {Hawcroft}, {Herrero}, {Higgins}, {John Hillier}, {Ignace}, {Istrate}, {Kaper}, {Kee}, {Kehrig}, {Keszthelyi}, {Klencki}, {de Koter}, {Kuiper}, {Laplace}, {Larkin}, {Lefever}, {Leitherer}, {Lennon}, {Mahy}, {Ma{\'\i}z Apell{\'a}niz}, {Maravelias}, {Marcolino}, {McLeod}, {de Mink}, {Najarro}, {Oey}, {Parsons}, {Pauli}, {Pedersen}, {Prinja}, {Ramachandran}, {Ram{\'\i}rez-Tannus}, {Sabhahit}, {Schootemeijer}, {Reyero Serantes}, {Shenar}, {Stringfellow}, {Sudnik}, {Tramper}, \& {Wang}}]{Vink2023}
{Vink}, J.~S., {Mehner}, A., {Crowther}, P.~A., {et~al.} 2023, \aap, 675, A154

\bibitem[{{Vink} \& {Sander}(2021)}]{Vink2021}
{Vink}, J.~S. \& {Sander}, A. A.~C. 2021, \mnras, 504, 2051

\bibitem[{{{\v{S}}urlan} {et~al.}(2013){{\v{S}}urlan}, {Hamann}, {Aret}, {Kub{\'a}t}, {Oskinova}, \& {Torres}}]{Surlan2013}
{{\v{S}}urlan}, B., {Hamann}, W.~R., {Aret}, A., {et~al.} 2013, \aap, 559, A130

\bibitem[{{Walborn}(1971)}]{Walborn1971}
{Walborn}, N.~R. 1971, \apjs, 23, 257

\bibitem[{{Walborn}(1973)}]{Walborn1973}
{Walborn}, N.~R. 1973, \aj, 78, 1067

\bibitem[{{Walborn} {et~al.}(2002){Walborn}, {Fullerton}, {Crowther}, {Bianchi}, {Hutchings}, {Pellerin}, {Sonneborn}, \& {Willis}}]{Walborn2002}
{Walborn}, N.~R., {Fullerton}, A.~W., {Crowther}, P.~A., {et~al.} 2002, \apjs, 141, 443

\bibitem[{{Walborn} {et~al.}(2010){Walborn}, {Howarth}, {Evans}, {Crowther}, {Moffat}, {St-Louis}, {Fari{\~n}a}, {Bosch}, {Morrell}, {Barb{\'a}}, \& {van Loon}}]{Walborn2010}
{Walborn}, N.~R., {Howarth}, I.~D., {Evans}, C.~J., {et~al.} 2010, \aj, 139, 1283

\bibitem[{{Walborn} {et~al.}(1995){Walborn}, {Lennon}, {Haser}, {Kudritzki}, \& {Voels}}]{Walborn1995}
{Walborn}, N.~R., {Lennon}, D.~J., {Haser}, S.~M., {Kudritzki}, R.-P., \& {Voels}, S.~A. 1995, \pasp, 107, 104

\bibitem[{{Walborn} {et~al.}(2000){Walborn}, {Lennon}, {Heap}, {Lindler}, {Smith}, {Evans}, \& {Parker}}]{Walborn2000}
{Walborn}, N.~R., {Lennon}, D.~J., {Heap}, S.~R., {et~al.} 2000, \pasp, 112, 1243

\end{thebibliography}

\clearpage

\begin{appendix}

\section{Detailed physical treatments within the different expanding atmosphere codes}\label{sec:codedetails}

\subsection{Velocity and density stratification}\label{sec:stratdetails}

\subsubsection{\textsc{Fastwind}}

All \textsc{Fastwind} models in this study employ a $\beta$-velocity law as mentioned in Sect.\,\ref{sec:anacommon}, which is smoothly connected to the hydrostatic regime at $\varv \approx 0.1\,a_\text{s}(T_\text{eff})$, with $a_\text{s}$ denoting the (isothermal) sound speed. The $\beta$-law is implemented in the form of Eq.\,\eqref{eq:betalaw}, with $\beta$ and $\varv_\infty$ given as input parameters. The quantity $R_{0}$ is usually labeled $b$ in their literature and fixed by the connection demand. To account for the radiative acceleration in the solution of the hydrostatic equation, in the very first iterations \textsc{Fastwind} approximates the flux-weighted opacity by the Rosseland opacity, which is further approximated by a Kramer's-like formula, 
\begin{equation}
   \kappa_\text{Ross}(r) \approx \sigma_\text{e} n_\text{e}(r) \left[ 1 + k_\text{c}\,\rho(r)\,T(r)^{-x} \right] 
\end{equation}
\citep{SantolayaRey1997}, for the Rosseland opacity $\kappa_\text{Ross}$ with $n_\text{e}$ denoting the electron number density and $\sigma_\text{e}$ the Thomson cross section. The parameters $k_\text{c}$ and $x$ are obtained from fits to the actually computed $\kappa_\text{Ross}$. These approximations are relaxed in the course of the non-LTE iterations. After an initial phase, the Rosseland opacity is calculated from its actual definition and the radiative acceleration is updated using the current flux-weighted opacities.

\subsubsection{CMFGEN}

For the initial setup of the (quasi-)hydrostatic layers, either a \textsc{TLUSTY} or an old \textsc{CMFGEN} model can be used. To jointly account for the supersonic and hydrostatic parts, CMFGEN implements a velocity law of the form
\begin{equation}
  \label{eq:cmfgen-beta} 
  \varv(r) = \frac{\varv_\text{o}+\left(\varv_\infty - \varv_\text{o}\right) \left( 1 - \frac{R_\ast}{r}\right)^\beta}{ 1 + \frac{\varv_\text{o}}{\varv_\text{core}} \exp{\left(\frac{R_\ast - r}{h_\text{eff}}\right)}},
\end{equation}
as introduced in \citet{Hillier2003} for this single-$\beta$ form. The ``effective scale height'' $h_\text{eff}$ can be updated during the calculations  with the current radiative acceleration to closely fulfill the hydrostatic equation of motion \citep[e.g.,][]{Martins2012}. Other approaches analytically calculate $\varv(r)$ without using the $\beta$-law by means of the $W$-Lambert procedure \citep{MuellerVink2008,GormazMatamala2021}, but it requires a significant effort to determine the extra parameters.

\subsubsection{PoWR}

The PoWR models employed in this work use the default PoWR branch with a pre-described $\beta$-law in the outer part instead of PoWR$^\textsc{hd}$ where the entire velocity field $\varv(r)$ can be obtained from solving the hydrodynamic equation of motion \citep[e.g.,][]{Graefener2005,Sander2017}.
The $\beta$-law implementation in PoWR reads 
\begin{equation}
  \label{eq:powr-beta}
  \varv(r) = p_1 \left( 1 + \frac{R_{20}}{r + p_2} \right)^{\beta}\text{.}
\end{equation}
The two parameters ($p_1$, $p_2$) are determined at the beginning of the model and after updates of the density stratification. The latter happens either when the hydrostatic equation is not sufficiently fulfilled or when the optical depth at the inner boundary deviates notably from the pre-specified value. In the default setting, the two parameters $p_1$ and $p_2$ are adjusted such that $\varv(R_\text{max})$ = $\varv_\infty$ (i.e., $p_1 \approx \varv_\infty$) and one obtains a smooth connection of the velocity gradient $\mathrm{d}\varv/\mathrm{d}r$ when the analytic Eq.\,\eqref{eq:powr-beta} is merged with the numerically solved $\varv(r)$ for the hydrostatic part \citep[cf.\ ][]{Sander2015}. Alternatively, the connection can be forced at a specific fraction of the sound speed, which usually yields a different value for the parameter $p_2$. For large $\beta$-values, the default gradient connection can lead to an unfavorable situation as the gradients for larger $\beta$-laws could enforce a connection too far beneath the sonic point, which can in some cases impact the derived absorption line profiles. In such cases, the forced sonic connection is preferred. In the current work, $\beta$-values around $1$ are employed in the PoWR models and thus the selection of the connection criterion is of no importance. However, a good fit of the LMC targets required the use of a $2\beta$-law, where Eq.\,\eqref{eq:powr-beta} is extended to  
\begin{equation}
  \label{eq:powr-2beta}
  \varv(r) = p_1 \left( 1 + \frac{R_{20}}{r + p_2} \right)^{\beta_1} + p_3 \left( 1 + \frac{R_{20}}{r + p_4} \right)^{\beta_2}\text{.}
\end{equation}
Instead of a single $\beta$-parameter, now two values $\beta_1$ and $\beta_2$ are required with $\beta_2 > \beta_1$. Moreover, a weighting factor $q_{2\beta}$ for the impact to $\varv_\infty$ from the second term needs to be provided, such that $p_1 \approx  \varv_\infty \left( 1- q_{2\beta} \right)$ and $p_3 \approx \varv_\infty q_{2\beta}$. The calculation of $p_2$ is similar to the single-$\beta$ case and $p_4 = R_{20} - R_\text{con}$ follows from the connection radius $R_\text{con}$ determined as described above, as the determination of $R_\text{con}$ uses only the gradient from the $\beta_1$ term.

\subsection{Wind clumping}\label{sec:clumpingdetails}

\subsubsection{\textsc{Fastwind}}\label{sec:clumpingdetailsfw}

To account for wind clumping, \textsc{Fastwind} can use two different approaches. In the standard ``microclumping'' approximation, the wind is composed only of clumps with a void ``inter-clump  medium''. The clumps are further optically thin and a ``clumping factor'' $f_\text{cl} =\langle\rho^2\rangle/\langle\rho\rangle^2$ needs to be defined. Under the above conditions,
 $f_\text{cl} \equiv D$, where $D$ describes the density contrast between the clumps and the mean density. Hence, for a wind with mean density $\langle\rho\rangle$, the clumps as such have a density $D \langle\rho\rangle$. The volume filling factor $f_\text{V}$ can then simply be calculated as $f_\text{V} = f_\text{cl}^{-1}$. For $f_\text{cl} = 1$, a smooth (``unclumped'') wind situation is recovered.
Alternatively, a more extensive effective opacity formalism can be employed to statistically account also for optically thick clumps \citep{Sundqvist2018}. The formalism introduces three additional parameters, $f_\text{ic}$, $f_\text{vel}$, and the ``porosity length'' $h$. For the latter, the default scaling in \textsc{Fastwind} is $h(r) = h_\infty \varv(r) / \varv_\infty$ with $h_\infty = R_{2/3}$ \citep{Sundqvist2018,Brands2022}. This scaling is employed in all models employing the detailed clumping formalism in this work.
With values of $f_\text{ic} > 0$, a non-void inter-clump  medium can be introduced. Similar to $D$, the quantity $f_\text{ic} = \rho_\text{ic}/ \langle\rho\rangle$ describes the ratio of the density to the mean density, but now for the inter-clump  medium. With $f_\text{ic} > 0$, the volume filling factor reads
\begin{equation}
  \label{eq:fv-interclump} 
  f_\text{V} = \frac{\left( 1 - f_\text{ic}\right)^2}{ f_\text{cl} - 2 f_\text{ic} + f_\text{ic}^2}\text{.}
\end{equation}
The mean density is then
\begin{equation}
  \langle\rho\rangle = f_\text{V} \rho_\text{cl} + \left( 1 - f_\text{V} \right) \rho_\text{ic}\text{.} 
\end{equation}

The parameter $f_\text{vel}$ describes a velocity filling factor that can account for the so-called ``velocity porosity'' \citep{Owocki2008}. The value of $f_\text{vel}$ enters in the calculation of the line optical depth for the clumps \citep{Sundqvist2018}, which is proportional to the Sobolev optical depth of the mean wind velocity field divided by the factor $f_\text{vor} := f_\text{vel} / (1 - f_\text{vel})$. For $f_\text{vel} \rightarrow 0$, the line optical depth of the clumps approaches infinity, and the effective opacity is determined by the inter-clump medium alone.

The \textsc{Fastwind}-default clumping stratification is set by the two parameters $\varv_\textrm{cl,start}$ and $\varv_\textrm{cl,max}$. Below $\varv_\textrm{cl,start}$, a smooth atmosphere is assumed. Between $\varv_\textrm{cl,start}$ and $\varv_\textrm{cl,max}$, the clumping factor is increased linearly until the specified value of $f_\text{cl}$ is reached at $\varv_\textrm{cl,max}$. Beyond this point, the constant $f_\text{cl}$ is employed. 
The formalism from \citet{Sundqvist2018} is an effort to describe two wind components without introducing significant computational overhead. Therefore, such models cannot account for a difference in the level populations (and ionization stages) between clumps and inter-clump  medium at a given radius as this would require the calculation of an additional set of non-LTE population numbers and come with a significant computational cost.

\subsubsection{CMFGEN}

Clumping is treated in the ``microclumping'' approximation, assuming a void inter-clump  medium. In all CMFGEN models presented in this work, the volume filling factor $f_\text{V}$ has a velocity-dependent behavior with 
\begin{equation}
\label{eq:cmfgen-clumping}
    f_\text{V}(r) = f_{\text{V},\infty} + (1 - f_{\text{V},\infty})\,e^{-\varv(r)/\varv_\text{cl}},
\end{equation}
as introduced in \citet{Hillier2003}.
The quantity $f_{\text{V},\infty}$ denotes the value at $r \rightarrow \infty$. For a void inter-clump  medium and optically thin clumping, we obtain $D = f_\text{cl}$ and $D_\infty = f_{\text{V},\infty}^{-1}$ (see Sect.\,\ref{sec:clumping}). The free parameter $\varv_{cl}$ is a characteristic velocity varied in a tailored analysis to describe the onset of the clumping. Due to the exponential increase in Eq.\,\eqref{eq:cmfgen-clumping}, this does not imply a completely unclumped wind for $\varv < \varv_\text{cl}$, contrary to the meaning of $\varv_\text{cl}$ in \textsc{Fastwind}.

\subsubsection{PoWR}

The ``microclumping'' approximation is the default setting in PoWR \citep{Hamann1998}, assuming a clumped wind with an overdensity $D \rho$ compared to a smooth wind with the density $\rho$. Similar to CMFGEN, the inter-clump medium is assumed to be void, meaning that the volume filling factor is $f_\text{V} = D^{-1}$. The clumping factor $D \equiv f_\text{cl}$ can be radially dependent with a multitude of choices for the prescription of $D(r)$. For all models in this work, the radius-based prescription 
\begin{equation}
  \label{eq:powrclump}
   D(r) = \begin{cases}
      D_\text{in}  &  \text{for~} r < R_1  \\
     \frac{D_\text{in}}{2} \left[ 1 + \cos \left( x \right) \right] + \frac{D_\infty}{2} \left[ 1 - \cos \left( x \right) \right]   & \text{for~} R_1 < r < R_2 \\
      D_\infty &  \text{for~} r > R_2 \\
   \end{cases}
\end{equation}
is chosen, with 
\begin{equation}
   x = \frac{\pi}{R_2 - R_1} (r-R_1)  
\end{equation}
and $D_{\infty}$ denoting the clumping factor reached at the outer boundary. The inner value is set to $D_\text{in} = 1$. Between the radii $R_1$ and $R_2 \geq R_1$ the clumping factor increases from $D_\text{in}$ to $D_\infty$ in the shape of a cosine. In all models presented in this work, $R_1$ is set to the radius of the sonic point. The outer value differs for each target to improve the spectral fit. The values for $R_2$ are listed as $R_D$ in Tables\,\ref{table:paramsummary-azv377}, \ref{table:paramsummary-sk6950}, and \ref{table:paramsummary-sk66171}. PoWR also has options to account for optically thick clumping or detailed rotational broadening of wind lines in the calculation of the emergent spectrum \citep{Oskinova2007,Shenar2014}, but we do not make use of either feature in the present paper.

\begin{figure}
  \includegraphics[width=\columnwidth]{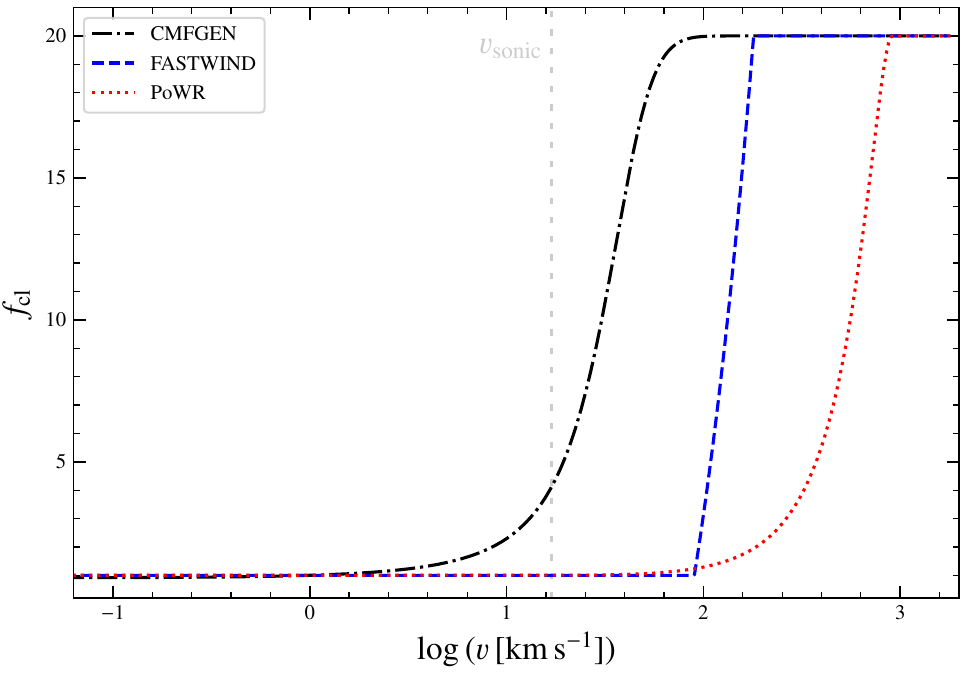}
  \caption{Comparison of the different radial behaviors of the clumping factor $f_\mathrm{cl} \equiv D$, assuming optically thin clumping and input values used for the analysis of Sk\,-66$^{\circ}$ 171. To focus on the clumping stratification, all curves assume the same maximum clumping factor ($f_\mathrm{cl} = 20$) and the same underlying velocity field ($\varv_\infty = 1800\,\mathrm{km}\,\mathrm{s}^{-1}$).}
  \label{fig:clumpstrat}
\end{figure}
With the clumping parametrization used in the PoWR models applied in this work, the maximum clumping factor $D_\infty$ is reached much further out than in \textsc{Fastwind} and CMFGEN models employing clumped winds. Moreover, the increase from the sonic point starts rather slow if described by Eq.\,\eqref{eq:powrclump} due to the differences between radius and velocity space. Both effects are illustrated in Fig.\,\ref{fig:clumpstrat} and have to be considered when interpreting and comparing the resulting values. The PoWR example shown in Fig.\,\ref{fig:clumpstrat} uses $R_D = 3\,R_{20}$, corresponding to a velocity of $\approx 900\,\mathrm{km}\,\mathrm{s}^{-1}$.

\section{Detailed method descriptions}\label{sec:mdetail}

In this section we provide a detailed description of the different approaches, sorted by the different physical aspects. While some individual results are mentioned, a complete overview of all derived values is provided in Tables \ref{table:paramsummary-azv377} to \ref{table:paramsummary-sk66171} with a more broad discussion of the results and their implications given in Sects.\,\ref{sec:results} and \ref{sec:discussion}. 

\subsection{Rotation and macroturbulence}\label{sec:mdetailrot}

\paragraph{F1}\label{sec:f1rot}
To determine the projected rotational and macroturbulent velocities, \texttt{iacob-broad} is employed. Different lines are used for the different stars as metal lines depend on the spectral type and they are faint at low metallicities. \ion{N}{iv} $\lambda$ 5200, 5204, 6380 are used for AzV 377; \ion{O}{iii} $\lambda$ 5592 and \ion{C}{iv} $\lambda$ 5812 for Sk\,-69$^{\circ}$ 50; and \ion{Si}{iv} $\lambda$ 4089, \ion{N}{iii} $\lambda$ 4510, 4514, \ion{O}{iii} 5592 and \ion{C}{iv} $\lambda$ 5801, 5812 for Sk\,-66$^{\circ}$ 171. The obtained projected rotational velocities are, respectively, \vsini= 74$\pm$7, 196$\pm$20 and 97$\pm$10\,km\,s$^{-1}$, where we adopt a 10$\%$ uncertainty when less than four lines are available. \vmac\ is not well constrained, and we adopt estimated values of 60\,km\,s$^{-1}$~ for AzV\,377 and Sk\,-69$^{\circ}$ 50 and 30 km\,s$^{-1}$~ for Sk\,-66$^{\circ}$ 171. 

\paragraph{F2/F3} In the optical-only analyses (F2) the projected rotational velocity $\varv \sin i$ is determined by convolving the spectra within the Kiwi-GA run (cf.\ Appendix~\ref{sec:mdetailwind}). In the optical + UV analyses (F3) the rotational velocity is not fitted, but instead fixed at the best fitting value as found from the optical-only analyses (F2). No explicit macroturbulence is assumed ($\varv_\text{mac} = 0\,\mathrm{km\,s}^{-1}$).  

\paragraph{F4} The rotational velocities are obtained within the spectral fitting process. For this, the synthetic spectra from the calculated model grids are pre-convolved with a set of projected rotational velocities spanning $\varv \sin i = 0$, $20$, $50$, $100$, $150$, $200$, $250$, $300$, and $400\,\mathrm{km\,s}^{-1}$ \citep[see also][]{Bestenlehner2024}. For all targets, a macro-turbulent velocity of $\varv_\text{mac} = 20\,\mathrm{km\,s}^{-1}$ is adopted.

\paragraph{C1} We determine \vsini\ with the Fourier transform analysis from \texttt{iacob-broad}.
For this, we used \ion{He}{i}~4713, \ion{O}{iii} $\lambda$ 5592 and \ion{C}{iv} $\lambda$5801, and computed an overall mean for \vsini. 
We estimate that this approach yields a typical error of $\pm$ 10 \,km\,s$^{-1}$ on \vsini.
Broadening caused by macroturbulence proved uncertain to constrain with this method, and it was derived from a comparison of synthetic spectra with observations.
A reference spectrum is chosen from the C1 initial model grid where \teff\ and \logg\ are close to the expected values for the star's spectral type and luminosity class. The spectrum is convolved with the instrumental resolution (assuming a Gaussian profile for the instruments) and the radial velocity is obtained from direct comparison with the observed spectrum. The error on \vrad\ with such a simple approach is $\pm$ 10 \,km\,s$^{-1}$.
The synthetic spectra are then convolved with the broadening parameters (\vsini\ and \vmac, assuming a radial-tangential profile for the latter), plus instrumental broadening, before comparing them to observations. The shape of line wings gives access to macroturbulence in case \vsini\ is not too large. For Sk\,-69$^{\circ}$\,50 and Sk\,-66$^{\circ}$\,171, we adopted \vmac\ = 80 \,km\,s$^{-1}$ and 20 \,km\,s$^{-1}$, respectively, while for \object{AzV\,377}, we simply assumed \vmac\ = 0\,km\,s$^{-1}$.  

\paragraph{C2} The projected rotational velocity $\varv \sin i$ is determined by convolving the spectra during the fitting process. No explicit macroturbulence is assumed ($\varv_\text{mac} = 0\,\mathrm{km\,s}^{-1}$).  

\paragraph{P1} We employ the \texttt{iacob-broad} to measure the projected rotation velocity $\varv_\text{rot} \sin i$, assuming no macroturbulence. As metal lines are, contrary to helium and hydrogen lines, not significantly affected by pressure broadening, we choose \ion{O}{iv} at $\SI{3397}{\AA}$ and of \ion{N}{iv} at $\SI{3404}{\AA}$, $\SI{3412}{\AA}$, and $\SI{3414}{\AA}$ which are all clearly visible in the optical spectrum of the O5\,V((f)) star. 
The mean value obtained from the different lines is ${\varv_\mathrm{rot} \sin i =\SI{60}{km\,s^{-1}}}$.

\paragraph{P2} We used a combined Fourier transform (FT) and goodness-of-fit (GOF) analysis employing the \texttt{iacob-broad} tool. We applied this method to several absorption lines such as \ion{Si}{iv} $\lambda$ 4089, \ion{He}{i} $\lambda$ 4713, \ion{O}{iii} $\lambda$ 5592, \ion{C}{iv} $\lambda$ 5801, 5812, and  \ion{He}{i} $\lambda$ 5876.
The overall mean  $\varv\sin i$ and macro-turbulent ($\varv_{\mathrm{mac}}$) velocities are given in Tables\,\ref{table:paramsummary-sk66171} and \ref{table:paramsummary-sk6950} . Subsequently, these values, along with instrumental broadening, were used to convolve the model spectra to match the observations.

\subsection{Temperature, surface gravity, and abundances}\label{sec:mdetailtgabu}

Depending on the method, the reddening is either determined before, after, or in parallel to other parameters. 

\paragraph{F1}\label{sec:f1temp} The \textsc{Iacob-Gbat} package \citep{SimonDiaz2011} is used to determine the stellar parameters from the optical \ion{H}{} and \ion{He}{} lines, without re-normalizing the normalized spectrum provided in the \textsc{XshootU} data release \citep{Sana2024}. The package uses a grid of \textsc{Fastwind} models to determine the effective temperature \teff, the surface gravity \logg, the \ion{He}{}/\ion{H}{} number ratio  $y_\text{He}$, the wind strength parameter $Q_\text{ws}$, the $\beta$ exponent (see Eq.~\ref{eq:betalaw}), and the microturbulence velocity $\xi$. The underlying grid for the LMC comprises of nearly $10^5$ model atmospheres at metallicity $Z= 0.008$ calculated with the \textsc{Fastwind} version 10.5 using HTCondor\footnote{http://research.cs.wisc.edu/htcondor/}. The grid covers the parameter ranges listed in Tab.~\ref{tab:gbat_grid}. For the SMC, a dedicated grid centered around the parameters of AzV\,377 at $Z= 0.004$ was used. \textsc{Iacob-Gbat} allows to automatically find the best-fitting model within the grid by looking for the minimum $\chi^2$ obtained by adding the individual $\chi_i^2$ of each individual line, in such a way that lines with a larger uncertainty or photon noise have a smaller weight in the final $\chi^2$.  A full description of the procedure is given in \citet[][Appendix A]{Holgado2018}\footnote{In their appendix A, \citet[][]{Holgado2018} provide relations based on a reduced $\chi^2$, though all previous and current versions of \texttt{iacob\_gbat} apply the standard, non-reduced
quantity.}.

\paragraph{F2/F3} The basic stellar parameters are determined together with the wind parameters, see Appendix~\ref{sec:mdetailwind}.
To compare \textsc{Fastwind} models to the data, the spectra need to be normalized. The UV spectra were normalized using a CMFGEN model from the grids of \citet{Bestenlehner2014} and \citet{Marcolino2024}, with the method as described in \citet{Brands2022}. 
For the fitting process of Sk~-69$^\circ$ 50 and Sk~-66$^\circ$ 171, the following optical lines, sorted by ion and wavelength, are considered: H$\epsilon$, H$\delta$ , H$\gamma$, H$\beta$, H$\alpha$, \ion{He}{i}~$\lambda$4026, \ion{He}{i}~$\lambda$4387, \ion{He}{i}~$\lambda$4471, \ion{He}{i}~$\lambda$4922, \ion{He}{i}~$\lambda$5875, \ion{He}{i}~$\lambda$7065, \ion{He}{ii}~$\lambda$4200, \ion{He}{ii}~$\lambda$4541, \ion{He}{ii}~$\lambda$4686, \ion{He}{ii}~$\lambda$5411, \ion{He}{ii}~$\lambda$6406, \ion{He}{ii}~$\lambda$6527, \ion{He}{ii}~$\lambda$6683, \ion{C}{iii}~$\lambda$4650, \ion{C}{iii}~$\lambda$5695, \ion{C}{iv}~$\lambda$5801, \ion{N}{iii}~$\lambda$4640, \ion{N}{iv}~$\lambda$3480, \ion{N}{iv}~$\lambda$4058, \ion{N}{iv}~$\lambda$6380, \ion{N}{v}~$\lambda$4603, \ion{N}{v}~$\lambda$4620, and \ion{O}{iii}~$\lambda$5592. For the combined optical+UV fit, we additionally include: \ion{P}{v}~$\lambda\lambda$1118-1128, \ion{C}{iv}~$\lambda$1169, \ion{C}{iii}~$\lambda$1176, \ion{N}{v}~$\lambda$1240, \ion{O}{iv}~$\lambda$1340, and \ion{Si}{iv}~$\lambda\lambda$1394-1402, and \ion{C}{iv}~$\lambda$1550 for both Sk~-69$^\circ$ 50 and Sk~-66$^\circ$ 171. For Sk~-66$^\circ$ 171, we further include \ion{N}{iv}~$\lambda$1718 and \ion{N}{iii}~$\lambda$1751. 
For AzV\,377 the same lines are used as for Sk\,-66$^\circ$ 171, excluding \ion{He}{ii}~$\lambda$6406, \ion{N}{iv}~$\lambda$4058, \ion{N}{iv}~$\lambda$6380, and the \ion{N}{v} lines, but including the \ion{He}{i}~$\lambda$4143, \ion{He}{i}~$\lambda$4713, \ion{He}{i}~$\lambda$5015, \ion{He}{ii}~$\lambda$1640, \ion{N}{iii}~$\lambda$4515, and \ion{Si}{iii}~$\lambda$1113.

\paragraph{F4} The spectroscopic analysis pipeline minimizes the $\chi^2$ and returns a 4D ($T_{\rm eff}$, $\log g$, $\dot{M}$, and $y_\text{He}$) probability distribution function. The 3 variations of CNO abundances provide only a rough estimate as they are not treated as independent parameters. The methodology is described in \citet{bestenlehner2022} and \citet{Bestenlehner2024}. Uncertainties of the stellar parameters have been derived by defining $1\,\sigma$ confidence intervals. The resulting uncertainties are typically larger compared to the other methods presented in this work due to correlations between all four stellar parameters. 

Without weighting specific spectral lines, the $\chi^2$ is dominated by Balmer lines, which are the strongest and broadest lines in early-type stars. Aiming to be able to eventually analyze the entire ULLYSES sample from mid-B dwarfs to early-O supergiants, we chose an approach that is not tailored to a specific parameter regime. In the wavelength range of the Balmer lines we remove every second wavelength point, thereby decreasing the wavelength resolution by a factor of two. In regions containing metal lines we add points to artificially increase the wavelength resolution by a factor of two. Helium lines are left untouched. In addition, we mask interstellar lines and bands. For mid- and late-O stars, the temperature determination is dominated by the ionization balance of \ion{He}{i} and \ion{}{ii}. For early-O stars and for B stars, the nitrogen and silicon lines respectively contribute enough to the total $\chi^2$ to obtain reliable temperatures. The surface gravity is derived on the basis of the Balmer line wings. The \ion{He}{}/\ion{H}{} number ratio $y_\text{He}$ is determined from the H and He lines, except for Sk\,-69$^\circ$ 50 where grid issues prevented a determination of this quantity.

\paragraph{C1} $T_\text{eff}$ and $\log g$ are constrained using a $\chi^{2}$ analysis on the initial model grid, relying on the optical spectra only. For AzV\,377, we re-normalized the spectrum by eye to optimize the use of the entire wavelength range, in particular close to the Balmer jump. For the LMC stars, we applied no re-normalization at this step. The $\chi^{2}$ calculations is limited to a list of lines which includes all the hydrogen and helium lines as quoted above for models F2/F3, but we did not use the \mbox{CNO}-lines listed therein. We first determine \teff\ and \logg\ from using the grids of models. We stress that since \ion{He}{i} lines are predominant in this list, we gave more weight to the \ion{He}{ii} in order to avoid a systematic bias of the results towards low \teff. This weight was set to 2 in this work.

In a second step after the luminosity determination (see below), $y_\text{He}$ and \mbox{CNO} abundances are determined from the analysis of UV and optical (normalized) spectra. Starting from a model with the determined $T_\text{eff}$, $\log g$, we computed new sets of models varying the abundances and microturbulence $\xi$. 
We then apply the same type of $\chi^{2}$ analysis as before, but this time for a set of He, C, N, or O lines. Essentially, we computed new atmosphere models for typically five to seven different abundances and used values of 5, 10, 15 and 20 $\mathrm{km}\,\mathrm{s}^{-1}$ for $\xi$ (applied in the line absorption profiles in the CMF radiative transfer as well as in calculation of the synthetic spectra). We refer to \citet{Martins2024} for more details on the methodology as well as a discussion of the dispersion of the results on the abundances. 
This approach does not propagate uncertainties in the fundamental parameters into the errors on the abundance measurements \citep[see also ][]{martins12b, martins15}.

\paragraph{C2} Utilizing an earlier set of calculations previously used in \citet{Maryeva2014,Gvaramadze2018,Gvaramadze2019} we ran a series of custom CMFGEN models for the two LMC stars. The initial assignment of $T_\text{eff}$, $L$, and $\log g$ is based on the spectral types of the sample stars and the O-star calibration by \citet{Martins2005}. We then proceed with a more detailed parameter determination. To measure $T_\text{eff}$, we used the \ion{Si}{iii}-\ion{Si}{iv}, \ion{N}{ii}-\ion{N}{iii}-\ion{N}{iv}, \ion{C}{iii}-\ion{C}{iv}, and \ion{He}{i}-\ion{He}{ii} lines. In the initial models for estimating $T_\text{eff}$ and $L$ (see below), we assume $y_\text{He} = 0.11$ for Sk\,-69$^{\circ}$\,50 and $y_\text{He} = 0.14$ for Sk\,-66$^{\circ}$\,171. We further assume increased abundances of nitrogen, and decreased carbon and oxygen abundances in these initial models, while for all other elements (Al, Si, P, S, Fe), subsolar abundances scaled to the metallicity of LMC were taken. After the determination of $T_\text{eff}$ and $L$, we compared the locations of the stars in the H-R diagram to Geneva evolutionary tracks from \citet{Eggenberger2021} for a better estimate of H, He, C, N, and O. Finally, these abundances are then slightly adjusted to better match the observed spectra. The resulting models show a considerably deficiency in the \ion{C}{iii}~4647-50-51 feature, which could not be resolved without affecting the fit quality of other lines. While it is possible to decrease the intensity of this triplet by decreasing the carbon abundances, this would also decrease the otherwise sufficient intensities of the \ion{C}{iii}~5696 and \ion{C}{iv}~5801,5812 lines. 

\paragraph{P1} To estimate the effective temperature and the surface gravity of the star, we use the ratio between the \ion{He}{i} and \ion{He}{ii} lines. The \ion{He}{i} lines play a crucial role in setting the lower limit of the temperature. In our fitting procedure, we excluded \ion{He}{i} lines with lower level 1s2p $^1$P as these are known to be unreliable \citep{Najarro+2006}. To determine the surface gravity of the star, the wings of $\mathrm{H\beta}$, $\mathrm{H\gamma}$, and $\mathrm{H\delta}$ are used, as they are almost insensitive to the wind. The best fit is achieved with an effective temperature of ${T_\mathrm{eff}=44.6\pm2.0}$\,kK and a surface gravity of ${\log(g/\si{cm\,s^{-2}})=4.0\pm0.1}$. Notably, temperature and surface gravity are linked to each other and obtaining the best fit is an iterative process. For the microturbulent velocity in the hydrostatic integration, a value of $10\,\mathrm{km}\,\mathrm{s}^{-1}$ is chosen, similar to what is used as a minimum in the formal integral. Compared to a model that does not include turbulence in the hydrostatic equation, we expect a higher $\log g$ of about $0.04$ according to Eq.\,\eqref{eq:loggturb}.

In the optical and UV spectra of \object{AzV\,377}, one can see several metal lines associated with CNO elements, yielding valuable information on their surface abundances. For carbon, we use the \ion{C}{IV} $\lambda1169$ and the \ion{C}{III} $\lambda1175$ lines in the UV and the \ion{C}{iV} $\lambda\lambda5801, 5812$ doublet in the optical. We find that these lines are best matched when lowering the carbon abundance to ${X_\mathrm{C}=\num{15e-5}\pm \num{5e-5}}$. For nitrogen, we see strong absorption lines of \ion{N}{IV} at $\SI{3404}{\AA}$, $\SI{3412}{\AA}$, and $\SI{3414}{\AA}$ as well as weak emission of \ion{N}{IV} at $\SI{4058}{\AA}$. These lines can only be matched when using an increased nitrogen abundance of ${X_\mathrm{N}=\num{80e-5}\pm\num{15e-5}}$. To determine the oxygen abundance, the \ion{O}{IV} $\lambda\lambda1339,1343$ doublet and the \ion{O}{III} $\lambda\lambda 1409, 1410, 1411, 1412$ multiplet in the UV and the \ion{O}{IV} $\lambda\lambda3381, 3386, 3390, 3397, 3410$ multiplet in the optical are used, yielding ${X_\mathrm{O}=50^{+20}_{-10}\times\num{e-5}}$.

\paragraph{P2} The stellar temperature is primarily constrained by the helium ionization balance. Since \ion{He}{i} singlet lines are more susceptible to model details \citep{Najarro+2006}, we give them less weight during the fitting process. In addition, we used the \ion{Si}{iii}-\ion{Si}{iv}, \ion{C}{iii}-\ion{C}{iv}, and \ion{N}{iii}-\ion{N}{iv} line ratios to determine the temperature. After getting a constraint on the temperature, we measured the surface gravity using the pressure-broadened wings of the Balmer lines. The main diagnostic lines are H$\delta$, H$\gamma$, and H$\beta$, since they are less affected by wind emission.
For Sk\,-66$^{\circ}$\,171, the combined optical and UV spectra are best reproduced using a model with $T_\text{eff} = 29.3$\,kK and $\log g = 3.08$. 
The \ion{N}{iii} emission lines in the optical mark a notable exception, where the temperature is too low to replicate them. On the other hand, higher temperature models reproducing the \ion{N}{iii} emission lines would spoil the overall spectral fit, especially the strength of \ion{He}{i} and \ion{He}{ii} lines. Moreover, the \ion{C}{iv\,$\lambda$1169} and \ion{C}{iii\,$\lambda$1176} lines in the UV are temperature-sensitive, and we combine them with other optical diagnostics to obtain the best fit. The combined UV and optical spectra of Sk\,-69$^{\circ}$\,50 are found to be best reproduced by $T_\text{eff} = 33.4\,$kK and $\log g = 3.42$. The microturbulent velocity of $\xi = 14\,\mathrm{km}/\mathrm{s}^{-1}$ included in the hydrostatic equation leads to an increased $\log g$ of about $0.1\,$dex following Eq.\,\eqref{eq:loggturb}. The microturbulence entering the hydrostatic equation is slightly lower than our eventually preferred value for the minimum microturbulence in the formal integral ($18$ and $20\,\mathrm{km}\,\mathrm{s}^{-1}$).

We adjusted the CNO abundances in the models to reproduce the observed strength of their respective absorption lines. For Sk\,-69$^{\circ}$\,50, we increased the nitrogen mass fraction in the model by a factor of 60 times the baseline LMC values. The best-fit model reproduces \ion{N}{iv} and \ion{N}{iii} absorption lines in the optical and UV ranges while under-predicting the \ion{N}{iii} emission. To match the strength of the \ion{N} {iii\,$\lambda$4097}  absorption line in Sk\,-66$^{\circ}$\,171, we increased the nitrogen mass fraction in the model by a factor of ten. The He line strength for both stars is found to be better predicted by models with a modestly higher He enrichment. For the remaining elements we adopted typical LMC abundance values derived from OB stars \citep{Trundle2007} if available, or were otherwise adopted as half-solar abundances.

\subsection{Luminosity and reddening}\label{sec:mdetaillumi}

\paragraph{F1} We derive the stellar radius and the resulting luminosity by comparing the model magnitudes from the best fit with the $B, V, J, H$ and $K$ observed magnitudes listed in Tab.~\ref{tab:sample}. To this end, we also need to determine the extinction. We have used the extinction law by \cite{JMaiz2014} and characterized it by the R$_{\rm 5495}$ index. Ideally, all photometric bands will give the same radius for the correct extinction. In practice, we take the extinction where the dispersion for the different radii values is lowest. We note that this dispersion is the major contribution to the error budget for the stellar radius. We get a best value of $R_{\rm 5495}= 3.8$ and $4.0$ for Sk\,-69$^{\circ}$ 50 and Sk\,-66$^{\circ}$ 171, whereas for \object{AzV 377} the best value is at $R_{\rm 5495}= 2.5$. This points to a different extinction nature for both cases, although the differences with the canonical value $R_{5495}= 3.1$ are small. For AzV\,377, the $K$-band was not included in the luminosity determination due to an inaccuracy in the $K$-magnitude distributed among the team for this star which was only corrected after the F1 calculations were already completed.

\paragraph{F2/F3} To derive the luminosity, Kiwi-GA requires an anchor magnitude, for which we use the $K_\text{s}$-band motivated by its generally low extinction. We derive the absolute $K_\text{s}$-magnitude by fitting the \citet{Fitzpatrick1999} extinction law through the $U$, $B$, $V$, $J$, $H$, and $K_\text{s}$ magnitudes listed in \citet{Vink2023}, adopting the CMFGEN model employed for the normalization as the intrinsic model.  With the obtained $A_{K_\text{s}}$ value and the adopted LMC distance given in Sect.\,\ref{sec:sample}, we derive $M_{K_\text{s}} = -5.92$\,mag (Sk\,-66$^\circ$ 171), $M_{K_\text{s}} = -4.89$\,mag (Sk\,-69$^\circ$ 50), and $M_{K_\text{s}} = -3.67$\,mag (AzV\,377). For F2 and F3, the flux-calibrated UV spectra do not enter the luminosity determinations.

\paragraph{F4} The photometric data from Table\,\ref{tab:sample} is utilized to determine the reddening and the luminosity $L$. We fit the model SED with the {\sc limfit} routine\footnote{\url{https://lmfit.github.io/lmfit-py/index.html}} to the optical to near-IR photometry by applying the extinction law by \cite{JMaiz2014}, similar to the procedure performed in \citet{Bestenlehner2022a}. The uncertainties from the SED fit are used to estimate the uncertainties on the photometric data. Together with the uncertainties on the stellar parameters, these are propagated into the error of $L$.

\paragraph{C1} Once $T_\text{eff}$, $\log g$ have been derived from the analysis of the optical spectrum, the stellar luminosity is constrained by comparing the SED of the corresponding model to the ``observational'' SED built from the flux-calibrated \fuse\ + \hst\ spectra in the UV, as well as optical and NIR photometry. For the latter, we employed the photometric data listed in Table \ref{tab:sample} and converted the magnitudes to flux, using the \textsc{synphot} Python packages available on \textsc{astropy}. The bolometric corrections were derived from \citet{mp06} for the previously derived effective temperature. 
In the determination of the luminosity, the interstellar extinction has to be derived as a side-product. For this, the synthetic spectra were reddened, using the \citet{Fitzpatrick2019} extinction law for the Milky Way foreground reddening, and \citet{Gordon2003} for the internal reddening in the Magellanic Clouds. The distances given in Sect.\,\ref{sec:sample} are used to scale the synthetic flux for the sample stars, enabling proper comparisons to the observations. 

\paragraph{C2} At first, the luminosity $L$ is roughly estimated by comparing the spectral energy distribution (SED) of the model spectrum with the photometric measurements over the whole spectral range. Beside the photometric data given in Table~\ref{tab:sample}, we also used data from the \textit{XMM-Newton} Serendipitous Ultraviolet Source Survey \citep[XMM-SUSS;][]{Page2012}, \textit{Spitzer} SAGE infrared photometry \citep{Bonanos+2009}, and the Wide-field Infrared Survey Explorer (WISE)  \citep{Cutri2013}.
  For a more accurate determination of the luminosities, the synthetic magnitudes of the stars are then calculated in the $U$, $B$, and $V$ filters and compared with observations. In order to obtain the magnitudes for the model spectra, we first rescale the fluxes to the distance of the LMC. The resulting fluxes are then corrected for the interstellar extinction with the IDL procedure \textsc{fm-unred}. The LMC2 option is selected, employing average LMC reddening parameters from \citet{Misselt1999}. Afterwards, the calculated spectra are convolved with the transmission curves of the standard $U$, $B$, and $V$ filters, and the corresponding zero points are applied using the {\sc pysysp} Python package \citep{Casagrande2014}.
  
\paragraph{P1} The luminosity is constrained by fitting the synthetic model flux to the photometry from Table\,\ref{tab:sample}, yielding a luminosity of ${\log\,(L/L_\odot)=5.3\pm0.1}$. The reddening is modeled as a combination of the color excess arising from the Galactic foreground with ${E_\mathrm{B-V,\,Gal}=\SI{0.03}{mag}}$ and the local SMC reddening with ${E_\mathrm{B-V,\,SMC}=0.028}$\,mag, using the reddening laws by \citet{Seaton1979} and \citet{Bouchet+1985}, respectively.

\paragraph{P2} The luminosity and color excess $E_\mathrm{B-V}$ are derived by fitting the model SED to the photometry from Table\,\ref{tab:sample} and the flux-calibrated UV spectra from ULLYSES. We provide the total extinction including the contribution from the Galactic foreground (assuming $E_{\rm B-V} $ = 0.04\,mag), adopting the reddening law from \cite{Seaton1979} for the Milky Way part, and the reddening law described in \cite{Howarth1983} with $R_{V}=3.2$ for the LMC contribution.

\subsection{Wind parameters, including\ clumping and additional X-rays}\label{sec:mdetailwind}

Some of the methods separate the determination of the stellar parameters from the wind parameters, while others vary both at the same time.

\paragraph{F1} The optical spectrum for the studied targets does not allow to derive the terminal wind velocities. We thus rely on Eq.\,(5) from \citet{Hawcroft2023} for \object{AzV 377} -- which consequently has a large error -- and the terminal velocity values by Brands et al.\ (in prep., part of the XShootU series) for Sk\,-69$^{\circ}$ 50 and Sk\,-66$^{\circ}$ 171. To determine the mass-loss rates, Eq.\,\ref{eq:qws} is evaluated. For the dwarf star \object{AzV\,377} only an upper limit for $Q_\text{ws}$ and thus $\dot{M}$ can be derived, while the two LMC stars with stronger winds enabled to constrain $Q_\text{ws}$ and thus $\dot{M}$ (with the additional information abut $\varv_\infty$). All models are unclumped,
and no additional X-rays were added to any of the models.

\paragraph{F2/F3} Once the input for a Kiwi-GA run is prepared, the normalized spectrum, the luminosity anchor, and the line selection are fed to the algorithm, and the best fit including uncertainty values are derived. We do this twice for each star: once with only the optical line selection, and once with the combined optical and UV line selection. For the optical-only run, we have to make assumptions about the wind clumping due to the lack of more diagnostics. We adopt fixed ``macroclumping'' parameters, with clumping factor $f_\textrm{cl} = 10$, inter-clump  density contrast $f_\text{ic} = 0.1$, velocity-porosity $f_\text{ic} = 0.5$, and onset velocity of clumping $\varv_\text{cl,start} = 0.05\varv_\infty$ (for details about these clumping parameters, see Appendix\,\ref{sec:clumpingdetailsfw} and \citealt{Sundqvist2018}). The terminal velocity is also fixed in F2, to a value estimated from the position of the blue edge of C~\textsc{iv}~$\lambda\lambda$1548-52.  We further fix the wind acceleration parameter $\beta = 1.0$, and the micro turbulence velocity to $15$~km~s$^{-1}$. The parameters we fit are: $T_\textrm{eff}$, $\log g$, $\varv_\text{rot} \sin i$, He, C, N and O abundance, and the mass-loss rate $\dot{M}$. For the optical + UV run, we adopt the He abundance and the $\varv_\text{rot} \sin i$ to the best fit values from the optical-only run, and then fit the following free parameters: $T_\mathrm{eff}$, $\log g$, C, N, and O abundance, $\dot{M}$, $\beta$, terminal velocity $\varv_\infty$, wind turbulence velocity $\varv_\textrm{windturb}$, $f_\textrm{cl}$, inter-clump density contrast $f_\textrm{ic}$, velocity filling factor $f_\textrm{vel}$, and clumping onset velocity $\varv_\textrm{cl}$.
We assume an LMC metallicity of $Z = 0.5\,Z_\odot$. Furthermore, we include X-rays following the prescription of \citet{Carneiro2016}. The X-rays are described by a radius-dependent shock temperature,
\begin{equation}
   T_\text{S}(r) = \frac{3}{16} \frac{\mu m_\text{H}}{k_\text{B}} u(r)^2, 
\end{equation}
with a jump velocity, 
\begin{equation}
   u(r) = u_\infty \left[ \frac{\varv(r)}{\varv_\infty} \right]^{\gamma_\text{x}}
,\end{equation}
that has two free parameters $u_\infty$ and $\gamma_\text{x}$. We fix $\gamma_\text{x} = 0.75$ \citep{Brands2022} and assume $u_\infty = 0.3 \varv_\infty$ for the maximum jump velocity $u_\infty$. The necessary X-ray volume filling fraction is chosen such that the X-ray output luminosity is approximately $10^{-7}$ times the stellar luminosity (for details, see Brands et al., in prep., part of the XShootU series). 

\paragraph{F4} As we apply this method only to the optical spectra, no value for $\varv_\infty$ can be measured. As mentioned for F1, $Q_\text{ws}$ can be estimated from the emission line profiles of H${\alpha}$, and, in the case of mid-early O stars, also from \ion{He}{ii}~$\lambda$4686, but $\varv_\infty$ needs to be known to derive the mass-loss rate $\dot{M}$. For F4, no assumptions for $\varv_\infty$ were made and thus no mass-loss rates were constrained. For \object{AzV\,377}, only an upper limit for $Q_\text{ws}$ can be determined (cf.\ Table\,\ref{table:paramsummary-azv377}). For the two LMC targets, we find best fits using models with $Q_\text{ws} = -12.3$. All models are unclumped ($f_\textrm{cl} = 1$) and no additional X-rays were added to any of the models. 

\paragraph{C1} The initial grid models do neither include clumping nor X-rays, and their wind mass-loss rates are computed using the \cite{vink2001} formula for
the corresponding metallicity, adopted to be 0.2\,\zsun\ and 0.5\,\zsun\ for SMC and LMC, respectively. Moreover, $\varv_\infty = 3.0\,\varv_\text{esc,eff}$ is assumed to be consistent with \citet{Martins2021} and references therein. The wind acceleration parameter is further fixed to $\beta = 1.0$ 
For the LMC stars, we then refined the wind parameters by calculating dedicated models beyond the initial assumptions of the LMC grid.
The wind terminal velocities were determined from the UV P\,Cygni profiles (\ion{C}{iv} resonance profiles are saturated for both Sk\,-69$^{\circ}$\,50, and Sk\,-66$^{\circ}$\,171). All the models computed later on adopted the measured values of $\varv_\infty$.
The mass-loss rate, the $\beta$ exponent of the velocity law, and the volume filling factor for the wind clumping were constrained from the UV and optical spectra. For the models with clumping, we adopted $\varv_\text{cl} = 30\, \mathrm{km}\,\mathrm{s}^{-1}$ \citep[see, e.g.,][]{Bouret2005}. We started with $\log L_{\rm X}/L_{\rm bol}$ = -7 but further tuned this quantity to improve the strength of \mbox{N~{\sc iv}} $\lambda\lambda$1238--1242. For the SMC star \object{AzV\,377}, no wind parameters were determined.

\paragraph{C2} The terminal velocity $\varv_{\infty}$ is estimated using spectral lines with P\,Cygni profiles in the UV. Intensities of emission lines were used for refining the mass-loss rate $\dot{M}$. For the clumping parameters, we adopted $f_{\text{V},\infty}$=0.2 and 0.3 as well as $\varv_\text{cl}$=20~km\,s$^{-1}$ and $10$~km\,s$^{-1}$ for Sk\,-69$^{\circ}$\,50 and Sk\,-66$^{\circ}$\,171, respectively. We further included additional X-rays in the models for both stars to reproduce the \ion{N}{V} $\lambda 1239,1243$ doublet. In a last step, the microturbulent velocity $\xi$ was also adjusted to better match the spectra. 

\paragraph{P1} For AzV\,377 no optical wind diagnostics are available. In the UV, \ion{N}{V} $\lambda\lambda1239, 1243$ doublet has a strong pronounced P\,Cygni profile, while the \ion{C}{IV} $\lambda\lambda1548, 1551$ doublet only shows a weak absorption trough. After measuring the terminal velocity from these lines, we find that this odd morphology can only be explained with a mass-loss rate of $\log\,(\dot{M} [M_\odot\,\mathrm{yr}^{-1}])=\num{-7.8\pm0.1}$. The density contrast $D$ is assumed to be $10$, which corresponds to the clumping factor $f_\text{cl}$ in this case as we assume a void inter-clump  medium. No clumping is assumed below the sonic point. No additional X-rays are included in the models as AzV\,377 has a high enough temperature to intrinsically show \ion{N}{V}.

\paragraph{P2} After fixing $T_\ast$ and log\,$g_\ast$ (see above), we adjust the wind parameters and recalculate the models accordingly. We measure the terminal velocities ($\varv_\infty$) from the blue edge of the absorption trough of the \ion{C}{iv\,$\lambda\lambda$1548--1551} P-Cygni line. The $\beta$ parameter of the velocity law is varied such that the synthetic spectrum can reproduce the profile shapes of the UV resonance lines and the optical wind emission lines, such as H$\alpha$ and \ion{He}{ii\,$\lambda4686$}. To get the best fit, we had to assume a double-$\beta$ law. For Sk\,-69$^{\circ}$\,50 we use $\beta_1=0.8$ and $\beta_2=1.5$ with $q_{2\beta} = 0.80$. In the case of Sk\,-66$^{\circ}$\,171, higher $\beta$ values ($\beta_1=1.1$ and $\beta_2=1.8$ with $q_{2\beta} = 0.85$) provide a better fit to the observed spectra. By consistently fitting the wind lines in the UV and optical, we infer the mass-loss rate and the clumping parameters. The primary diagnostic lines used are the UV resonance doublets such as \ion{C}{iv\,$\lambda\lambda$1548--1551},  \ion{Si}{iv\,$\lambda\lambda$1393--1403}, and \ion{N}{v}\,$\lambda\lambda$1238--1242 along with H$\alpha$ emission. Since the models for both stars are found to be too cool to reproduce the \ion{N}{v\,$\lambda\lambda$1238--1242} line profile, we incorporated an additional X-ray field in the models \citep{Baum1992}. For  Sk\,-69$^{\circ}$\,50, we used $T_\text{X} = 1\,$MK and an X-ray onset radius of $R_\text{X,min} = 1.1\,R_\ast$, while for  Sk\,-66$^{\circ}$\,171, we used $T_\text{X} = 3\,$MK and $R_\text{X,min} = 1.2\,R_\ast$. The X-ray filling factor was set to $0.1$ for both stars.

\section{Best-fit result overview and atomic data coverage}

\subsection{Overview of best-fit values}

In Tables~\ref{table:paramsummary-azv377}, \ref{table:paramsummary-sk6950}, and \ref{table:paramsummary-sk66171}, we provide an overview of the best-fitting parameters determined with each of the methods. Abundances are given in different formats for easier comparison. Depending on the method, not all parameters were determined. Fixed or assumed values are denoted in italics.

\subsection{Wide-range spectral comparison}

\begin{figure*}
       \includegraphics[width=18cm]{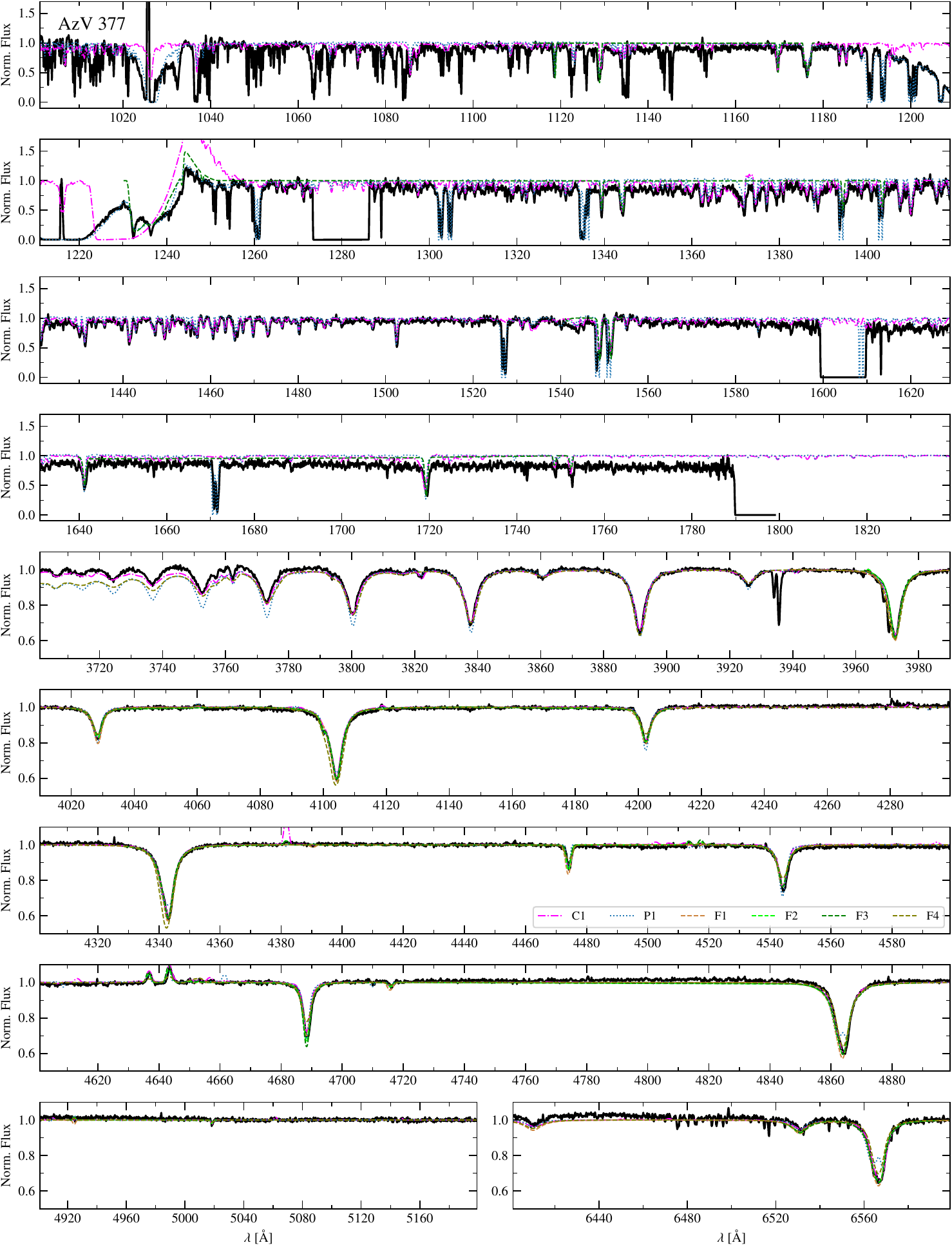}%
       \caption{Comparison of the different model spectra (colored lines, see legend) with the normalized observations of AzV\,377 (thick black line) across a wide range of UV and optical wavelengths.}%
       \label{fig:master_AV377}%
     \end{figure*}

\begin{figure*}
       \includegraphics[width=18cm]{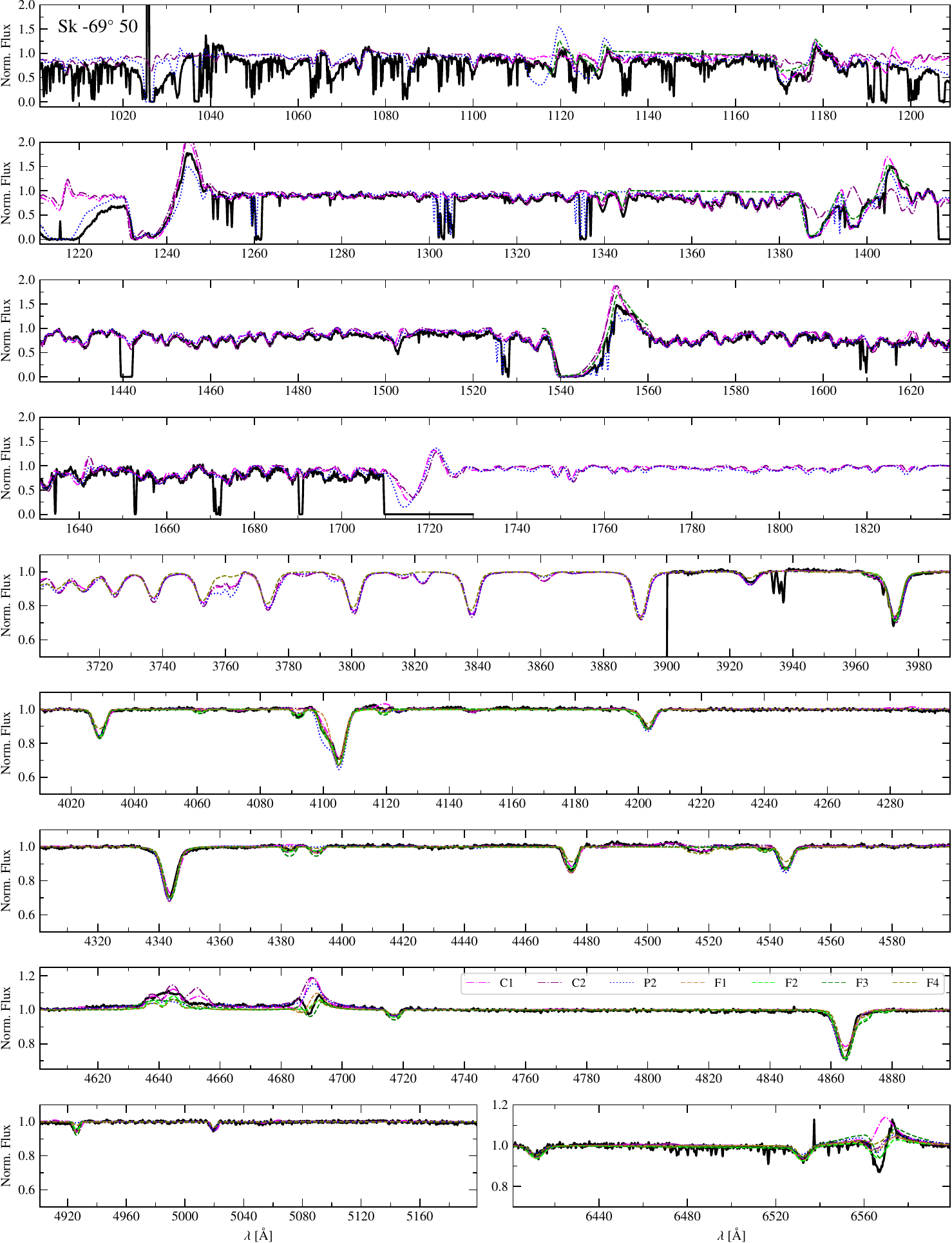}%
       \caption{Comparison of the different model spectra (colored lines, see legend) with the normalized observations of  Sk\,-69$^{\circ}$\,50 (thick black line) across a wide range of UV and optical wavelengths.}%
       \label{fig:master_SK6950}%
     \end{figure*}

\begin{figure*}
       \includegraphics[width=18cm]{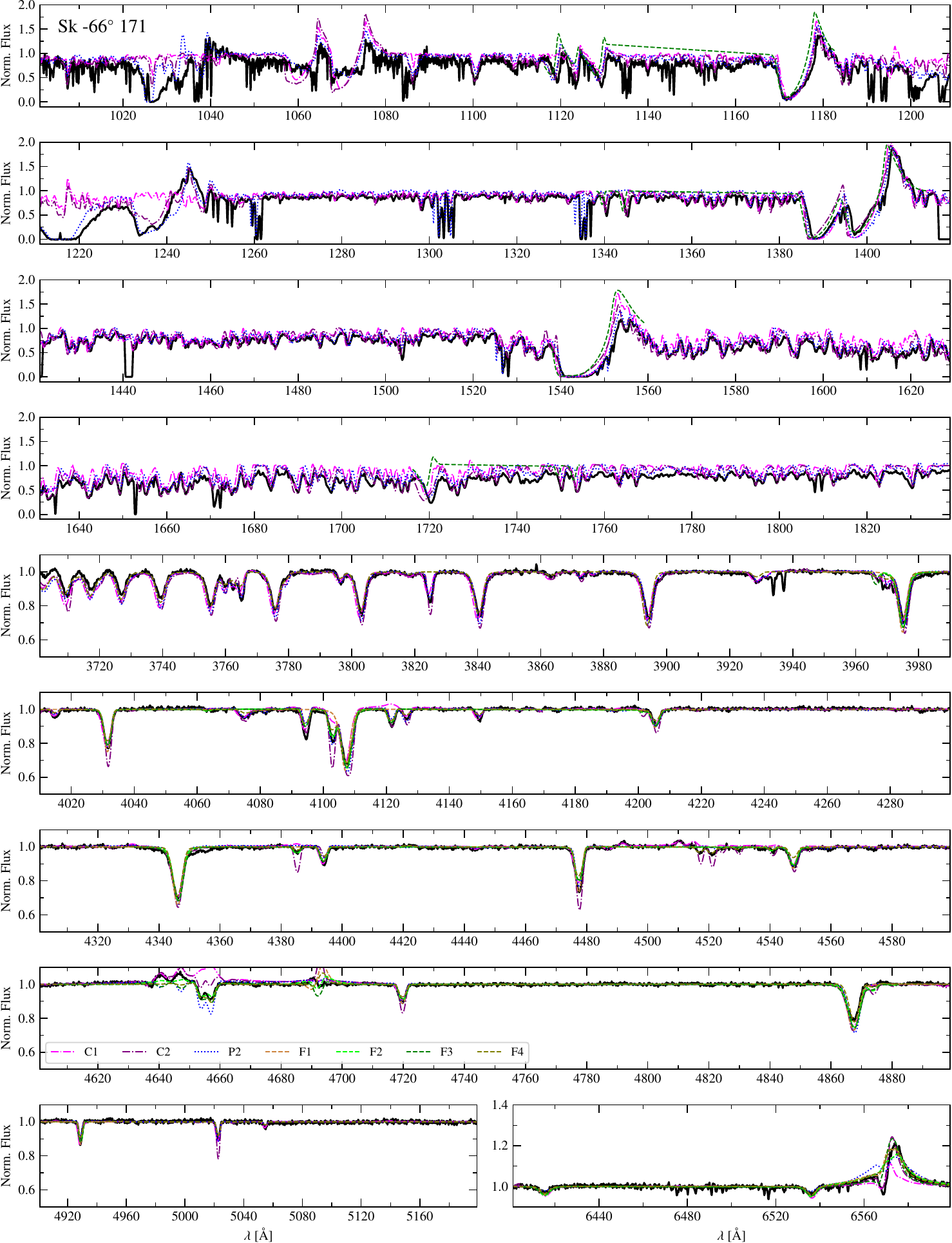}%
       \caption{Comparison of the different model spectra (colored lines, see legend) with the normalized observations of  Sk\,-66$^{\circ}$\,171 (thick black line) across a wide range of UV and optical wavelengths.}%
       \label{fig:master_SK66717}%
     \end{figure*}

In addition to the discussions about individual lines in Sects.\,\ref{sec:speclines} and \ref{sec:speclinediff}, we provide comparison plots in a large wavelength range in Figs.\,\ref{fig:master_AV377}, \ref{fig:master_SK6950}, and \ref{fig:master_SK66717}. Each of these figures shows the normalized observations compared to all the model spectra. For clarity, we show only one normalization of the observations, while some methods performed individual (re-)normalization of the data in some wavelength regimes. This is especially apparent in the UV range, where normalization is tricky due to the forest of iron lines and commonly done with the help of a continuum from an atmosphere model itself. As the resulting normalization is then model-dependent in such cases, this can lead to an apparent disagreement between the normalized observation shown in our figures and in the continuum level of methods using a (slightly) different normalization.

\subsection{Ion coverage}

Table~\ref{tab:ioncoverage} provides an overview of all the ions taken into account by each of the methods. The handling of the atomic data varies considerably between the different model atmosphere codes. Below, we give information for each of the codes (denoted by their associated methods) to better interpret the values given Table~\ref{tab:ioncoverage}.

\paragraph{F1/F2/F3/F4} The number of levels and lines provided in Table \ref{tab:ioncoverage} for the methods F1 to F4 refer to the explicit and background elements in \textsc{Fastwind} (see Sect.~\ref{sec:fastwind}) as used in the specific models. Model F1 has been calculated with only H and He as explicit elements. The models for F2 and F3 treat H, He, C, N, O, Si, P as explicit elements, and the models for F4 do this for H, He, C, N, O, Si. 

Most of the denoted levels are packed with respect to fine-structure, and are depacked when calculating the radiative transfer \citep[for details, see][]{Puls2005, Puls2020}. In addition to the displayed levels and lines,
transitions from other elements are considered as well, covering almost all atoms from H to Zn (except those with a negligible abundance). Moreover, the complete line list used for calculating the SEDs 
comprises not only the quoted explicit lines (when depacked), but also the multitude of additional lines with the ground-state or a metastable level as the lower level and a high-lying upper-level, where
the source function is approximated by a two-level atom approach.

\paragraph{C1/C2} In \textsc{CMFGEN}, there is a distinction between the levels taken explicitly into account for the non-LTE solution of the statistical equations and the levels in the radiative transfer. To keep the number of statistical equations manageable, levels are groups into superlevels with only each superlevel entering the rate equations. In the radiative transfer however, the levels are instead accounted for explicitly (i.e., no opacity sampling or redistribution). The grouping into superlevels can be changed by the user. Consequently, Table \ref{tab:ioncoverage} lists two columns for the superlevels and the ``total levels'' which refer to the number of levels accounted for in the radiative transfer. A more in-depth description of the superlevel approach is provided in \citet{Hillier1998}.

\paragraph{P1/P2} \textsc{PoWR} differentiates between ``normal'' and iron-group elements. For normal elements, recommended sets of atomic data are available where the fine-structure of the higher levels is usually packed into single levels. The user then usually just requests which elements and ions should be taken into account. Each of these levels is then treated in full non-LTE, i.e., explicitly enters the rate equations and their transition are taken into account for the radiative transfer (unless they are radiatively forbidden). As this can lead to very large sets of equations, the user further has the option to limit the total number of levels for a given ion. 

For iron and other elements of the ``iron group'' (Sc to Ni), a different approach is used in order to ensure that full blanketing effect is taken into account while also the forest of UV lines can be reproduced sufficiently. Grouped by energy bands, and in more recent years also by even and odd parity, the large list of iron levels is reduced to a set of superlevels. The allowed transitions between the superlevels are grouped into ``superlines'' for which a wavelength-dependent cross-section is calculated. To account for different Doppler velocities (and thus profile broadenings) in the radiative transfer, different pre-calculated sets are available. In the actual model calculation, only the population numbers for the superlevels are calculated, taking into account their different nature compared to the ``normal'' levels. In the radiative transfer, the pre-calculated cross-sections ensure that all transitions are taken into account at their correct wavelengths. The \textsc{PoWR} superlevel concept (without parity splitting) is described in more detail in \citet{Graefener2002}.

\begin{table*}%
        \caption{Best-fit parameters for the O5 V((f)) star AzV\,377 derived with the different methods. Adopted values are in italics.}
        \label{table:paramsummary-azv377}
        \centering
  \begin{tabular}{lccccccc}
    \hline\hline
    \rule{0cm}{2.8ex}
    AzV\,377     &  F1  &   F2  &  F3  &  F4  &  C1  &  P1       \\ 
    \hline
    \rule{0cm}{2.8ex}%
    \rule{0cm}{2.8ex}$T_\text{eff}$ (kK)    
      &  $44.0\pm0.5$  &  $46.00^{+1.00}_{-1.25}$  & $47.25^{+2.25}_{-0.25}$ &  $44.7^{+3.0}_{-1.0}$  &  $44.0\pm1.5$   &  $44.1^{+2}_{-2}$  \\
    \rule{0cm}{2.8ex}$\log g$ (cm\,s$^{-2}$)  
      &  $3.99\pm0.05$ &  $4.04^{+0.08}_{-0.08}$ &  $4.08^{+0.14}_{-0.06}$ &  $4.00^{+0.13}_{-0.10}$  &  $4.1\pm0.1$   &  $4.0^{+0.1}_{-0.1}$ \\
    \rule{0cm}{2.8ex}$\log L$ ($L_\odot$)   
      &  $5.34\pm0.04$ &  $5.36^{+0.03}_{-0.04}$ &  $5.39^{+0.06}_{-0.03}$ &  $5.34^{+0.14}_{-0.08}$ &   $5.30\pm0.05$   &  $5.3\pm0.1$  \\
    \rule{0cm}{2.8ex}$R_{2/3}$ ($R_\odot$)  
      &  $8.1\pm0.5$  &  $7.62^{+0.26}_{-0.26}$ &  $7.46^{+0.24}_{-0.29}$ &  $7.8^{+0.6}_{-0.9}$  &  $8.67\pm0.80$   &  $7.8^{+0.9}_{-0.8}$  \\
    \rule{0cm}{2.8ex}$M_\mathrm{spec}$ ($M_\odot$)  
      &  $23\pm3$  &    $29^{+9}_{-3}$ &  $24^{+8}_{-3}$ &  $22^{+9}_{-4}$  &  $34\pm11$   &  $21.9^{+5.7}_{-4.5}$  \\
                       
    \rule{0cm}{2.8ex}$\beta_1$ 
      &  $\leq 1.5$  &  \textit{1.0} &  $0.5^{+0.35}_{-0.05}$ &  \textit{1.0}  &  $1.0$   &  $0.8$  \\
    \rule{0cm}{2.8ex}$\beta_2$ 
      &  --  &   --  &  -- &  --  &  --  &  --  \\
    \rule{0cm}{2.8ex}$\log \dot{M}$ ($M_\odot \mathrm{yr}^{-1}$)  
      &  $\leq-7.38$  &   $-7.17^{+0.20}_{-0.33}$  &  $-8.10^{+0.28}_{-0.22}$ &  --  &  --   & $-7.8^{+0.1}_{-0.1}$ \\
    \rule{0cm}{2.8ex}$\log (\dot{M}\sqrt{f_\text{cl}})$ ($M_\odot \mathrm{yr}^{-1}$)  
      &  $\leq-7.38$  &   $-6.67^{+0.18}_{-0.30}$  &  $-7.49^{+0.35}_{-0.48}$ &  --  &  --   & $-7.3^{+0.1}_{-0.1}$ \\
    \rule{0cm}{2.8ex}$\varv_{\infty}$ (km\,s$^{-1}$)   
      &  $2356\pm615$ &   \textit{3500}  &  $1975^{+150}_{-125}$ &  --  &  --   &  $1800\pm200$  \\
     \rule{0cm}{2.8ex}$\varv \sin i$ (km\,s$^{-1}$)   
      &  $74\pm7$ &   $75^{+20}_{-15}$  &  \textit{75} &  $\sim 100$  &  $80\pm10$   &  $60\pm10$  \\
         
    \rule{0cm}{2.8ex}$\varv_{\mathrm{rad}}$ (km\,s$^{-1}$)     
      &  $-187$ &   $\sim175$  &  $\sim175$ &  $\sim 195$  &  $190\pm10$   &  $178\pm20$  \\
    \rule{0cm}{2.8ex}$\varv_{\mathrm{mac}}$ (km\,s$^{-1}$)  
      &  $60$ &  \textit{0}  &  \textit{0}  &  \textit{20}  &  \textit{0}   &  \textit{0}  \\
    \rule{0cm}{2.8ex}$\xi$ (km\,s$^{-1}$) 
      &  $5$  &   \textit{15}  &  \textit{15} &  10  &  $10$   &  $10$  \\
    \rule{0cm}{2.8ex}$\varv_\text{windturb}$ (km\,s$^{-1}$) 
      &  --  &   \textit{350}  &  $39.5^{+197.5}_{-39.5}$ &  --  &  $350$   &  $180$  \\
               
        \rule{0cm}{2.8ex}$X_{\rm H}$ (mass fr.) 
      &  0.59  &   $0.59^{+0.06}_{-0.03}$  &  \textit{0.59} &  0.66  &  $0.69\pm0.02$   &  $0.74^{+0.05}_{-0.05}$  \\
        \rule{0cm}{2.8ex}$X_{\rm He}$ (mass fr.) 
      &  0.37  &   $0.40^{+0.03}_{-0.06}$  &  \textit{0.40} &  0.34  &  $0.30\pm0.02$  &  $0.26^{+0.05}_{-0.05}$  \\
        \rule{0cm}{2.8ex}$X_{\rm C}/10^{-4}$ (mass fr.)
      &  \textit{4.5}  &  $0.5^{+3.1}_{-0.3}$   &  $2.0^{+6.8}_{-1.2}$ &  --  &  $8.3\pm3.5$   &  $1.5^{+0.5}_{-0.5}$  \\
        \rule{0cm}{2.8ex}$X_{\rm N}/10^{-4}$ (mass fr.)
      &  \textit{1.3}  &   $7.4^{+9.0}_{-3.0}$  &  $10.5^{+15.5}_{-5.5}$ &  --  &  $19.4\pm7.0$   &  $8.0^{+1.5}_{-1.5}$ \\
        \rule{0cm}{2.8ex}$X_{\rm O}/10^{-3}$ (mass fr.) 
      &  \textit{1.1}  &   $0.8^{+3.4}_{-0.7}$  &  $0.7^{+5.2}_{-0.5}$  &  --  &  $0.55\pm0.2$   &  $0.5^{+0.2}_{-0.1}$ \\
        \rule{0cm}{2.8ex}$y_\text{He}$  
      &  $0.15\pm 0.02$  &   $0.17^{+0.02}_{-0.04}$ & \textit{0.17} &  $\sim0.13$  &  $0.11\pm0.01$ &  $0.088$  \\
        \rule{0cm}{2.8ex}$\epsilon_{\rm C}$ 
      &  \textit{7.7}  &  $6.80^{+0.85}_{-0.55}$ &  $7.45^{+0.60}_{-0.35}$ &  --  &  $7.0\pm0.20$  &  $7.22^{+0.15}_{-0.17}$  \\
        \rule{0cm}{2.8ex}$\epsilon_{\rm N}$ 
      &  \textit{7.2}  &  $7.95^{+0.30}_{-0.20}$ &  $8.10^{+0.35}_{-0.30}$ &  --  &  $8.30\pm0.20$   &  $7.89^{+0.07}_{-0.10}$  \\
        \rule{0cm}{2.8ex}$\epsilon_{\rm O}$ 
      &  \textit{8.0}  &  $7.9^{+0.7}_{-1.7}$    &  $7.85^{+0.90}_{-0.50}$ &  --  &  $7.70\pm0.20$   &  $7.63^{+0.14}_{-0.11}$  \\
          
        \rule{0cm}{2.8ex}$E_{B-V}$ (mag)   
      &  0.09  &   0.07  &  0.07 &  0.08  &  $0.10\pm0.02$   &  $0.058$ \\
        \rule{0cm}{2.8ex}$\log \left(L_\text{X} / L\right)$ 
      &  --  & $-7.43^{+0.10}_{-0.18} $  &  $-8.07^{+0.26}_{-0.11}$ &  --  &  --   &  --  \\
        \rule{0cm}{2.8ex}$\log\,Q_{\mathrm{H}}$ (s$^{-1}$)     
      &  --  &   $49.14^{+0.04}_{-0.05}$  &  $49.18^{+0.08}_{-0.03}$ &  $49.1$  &  $49.25$   &  $49.08$  \\
        \rule{0cm}{2.8ex}$\log\,Q_{\mathrm{He\,I}}$ (s$^{-1}$)  
      &  --  &   $48.53^{+0.05}_{-0.08}$  &  $48.58^{+0.11}_{-0.03}$ &  $48.4$  &  $48.75$   &  $48.42$  \\
        \rule{0cm}{2.8ex}$\log\,Q_{\mathrm{He\,II}}$ (s$^{-1}$) 
          &  --   &   $43.79^{+0.17}_{-0.29}$  &  $43.65^{+0.36}_{-0.25}$  &  --  &  $44.08$ & $43.38$\vspace{1.5ex}\\
 
    \rule{0cm}{2.8ex}$\log Q_\text{ws}$ 
      &  $\leq -13.8$ &   --  &  $-13.8$ &  $<-13.0$  &  --   &  $-13.5$  \\
    \rule{0cm}{2.8ex}$D_\infty$ or $f_\text{cl}$ or $f^{-1}_\text{V}$\tablefootmark{($\dagger$)}  
       & \textit{1}  &   \textit{10}  &  $16^{+17}_{-15}$  &  \textit{1}  &  \textit{1}  & $10$  \\
    \rule{0cm}{2.8ex}$\varv_\text{cl}$ (km\,s$^{-1}$)  
       & --  &   --  &  --  &  --  &  --   &    --   \\   
    \rule{0cm}{2.8ex}$\varv_\text{cl,start}$ (km\,s$^{-1}$)  
       & --  &   --  &  $158^{+356}_{-138}$  &  --  &  --   &    --   \\   
    \rule{0cm}{2.8ex}$R_D$ ($R_{20}$)  
       & --  &   --  &  --  &  --  &  --   &  $10$   \\   
    \rule{0cm}{2.8ex}$\log f_\text{ic}$  
       & --  &   --  &  $-1.20^{+0.65}_{-0.80}$  &  --  &  --   &    --   \\   
    \rule{0cm}{2.8ex}$f_\text{vel}$  
       & --  &   --  &  $0.82^{+0.06}_{-0.22}$  &  --  &  --   &    --   \\   
    \rule{0cm}{2.8ex}$T_{20}$ (kK)  & --  &   --  &  --  &  --  &  $44.2$  &   $44.5^{+2}_{-2}$          \\
    \rule{0cm}{2.8ex}$R_{20}$ ($R_\odot$)  &  --  &  --  &  --  &  -- & $8.61$  &   $26.2^{+2}_{-3}$  \\ 
    \rule{0cm}{2.8ex}$T_{100}$ (kK) & --  &   --  &  --  &  --  &  $44.3$  & --  &  -- \\\hline
  \end{tabular}
  \tablefoot{
    \tablefoottext{$\dagger$}{For optically thin clumping with no inter-clump  medium: $D_\infty = f_\text{cl} = f^{-1}_\text{V}$ (cf.\ Sect.\,\ref{sec:clumping})}
  }
\end{table*}%

\begin{table*}%
        \caption{Best-fit parameters for the O7(n)(f)p star Sk\,-69$^{\circ}$\,50 derived with the different methods. Adopted values are in italics.}
        \label{table:paramsummary-sk6950}
        \centering
 \begin{tabular}{lccccccc}
    \hline\hline
    \rule{0cm}{2.8ex}
    Sk\,-69$^{\circ}$\,50    &  F1  &   F2  &  F3\tablefootmark{($\ddagger$)}  &  F4  &  C1  &  C2  &  P2       \\ 
    \hline
    \rule{0cm}{2.8ex}%
    \rule{0cm}{2.8ex}$T_\text{eff}$ (kK)    
      &  $34.2\pm0.5$  &  $36.25^{+0.25}_{-1.50}$ &  $34.50^{+0.25}_{-0.25}$ &  $34.4^{+1.8}_{-11.4}$  &  $35.0\pm0.2$ &  $34.0\pm0.2$   &  $33.4^{+1}_{-1}$  \\
    \rule{0cm}{2.8ex}$\log g$ (cm\,s$^{-2}$)  
      &  $3.23\pm0.06$  &  $3.38^{+0.05}_{-0.12}$ &  $3.25^{+0.03}_{-0.05}$  &  $3.3^{+0.2}_{-0.9}$ &  $3.4^{+0.1}_{-0.1}$  &  $3.25$         &  $3.42^{+0.1}_{-0.1}$  \\
    \rule{0cm}{2.8ex}$\log L$ ($L_\odot$)   
      &  $5.43\pm0.02$  &  $5.54^{+0.02}_{-0.05}$ &  $5.48^{+0.02}_{-0.02}$  &  $5.42^{+0.12}_{-0.41}$ &  $5.45\pm0.05$ &  $5.41\pm0.03$  &  $5.35^{+0.1}_{-0.1}$  \\
    \rule{0cm}{2.8ex}$R_{2/3}$ ($R_\odot$)  
      &  $14.8\pm0.3$   & $15.0^{+0.5}_{-0.4}$ &  $15.5^{+0.4}_{-0.4}$  &  $14.4^{+1.5}_{-6.8}$  &  $14.2^{+0.1}_{-0.1}$ & $14.7$         &  $14.1^{+1}_{-1}$  \\
    \rule{0cm}{2.8ex}$M_\mathrm{spec}$ ($M_\odot$)  
      &  $17\pm2$       &   $19.5^{+1.6}_{-3.6}$  &  $21.2^{+1.2}_{-1.5}$ &  $15^{+13}_{-11}$  &  $18.5^{+2.8}_{-2.8}$  &  $14.0$         &  $19.2^{+5}_{-4}$  \\
                       
    \rule{0cm}{2.8ex}$\beta_1$ 
      &  $1.0\pm0.1$  &  \textit{1.0}  &  $0.95^{+0.10}_{-0.05}$ &  \textit{1.0}  &  1.2  &  $1.35$   &  $0.8$  \\
    \rule{0cm}{2.8ex}$\beta_2$ 
      &  --  &  --  & -- &  --  &  --  &  --  &  $1.5$  \\
    \rule{0cm}{2.8ex}$\log \dot{M}$ ($M_\odot \mathrm{yr}^{-1}$)  
      &  $-5.64^{+0.15}_{-0.16}$  &   $-6.11^{+0.05}_{-0.05}$  &  $-6.16^{+0.03}_{-0.13}$ &  --  &  $-6.26^{+0.08}_{-0.08}$  &  $-5.92^{+0.06}_{-0.06}$   &  $-5.8^{+0.1}_{-0.1}$   \\
    \rule{0cm}{2.8ex}$\log (\dot{M}\sqrt{f_\text{cl}})$ ($M_\odot \mathrm{yr}^{-1}$)  
      &  $-5.64^{+0.15}_{-0.16}$  & $-5.61^{+0.05}_{-0.05}$ &  $-5.39^{+0.09}_{-0.11}$  &  --  &  --  &  $-5.57^{+0.06}_{-0.06}$   &  $-5.45^{+0.16}_{-0.16}$   \\
    \rule{0cm}{2.8ex}$\varv_{\infty}$ (km\,s$^{-1}$)   
      &  \textit{1925} &  \textit{1876}  &  $1925^{+25}_{-150}$  &  --  &  $1805^{+100}_{-100}$  &  $2100$   &  $1800^{+200}_{-200}$  \\
     \rule{0cm}{2.8ex}$\varv \sin i$ (km\,s$^{-1}$)   
      &  $196\pm20$  &  $185^{+30}_{-10}$  &  \textit{185} &  $\sim$$200$  &  $197$  &  200   &  $180^{+20}_{-20}$  \\
         
    \rule{0cm}{2.8ex}$\varv_{\mathrm{rad}}$ (km\,s$^{-1}$)     
      &  $-230$  &   $-214^{+16}_{-16}$  &  $-216^{+16}_{-16}$ &  $\sim$$250$  &  $-235$  &  $\sim$210   &  $-228^{+20}_{-20}$ \\
    \rule{0cm}{2.8ex}$\varv_{\mathrm{mac}}$ (km\,s$^{-1}$)  
      &  $60$  &   \textit{0}  &  \textit{0} &  \textit{20}  &  $80$  &  \textit{0}   &  $70$  \\
    \rule{0cm}{2.8ex}$\xi$ (km\,s$^{-1}$) 
      &  $20$  &   \textit{15}  &  \textit{15} &  10  &  $15$  &  $20$   &  $20$  \\
    \rule{0cm}{2.8ex}$\varv_\text{windturb}$ (km\,s$^{-1}$) 
      &  --  &   \textit{187}  &  $260^{+77}_{-29}$ &  --  &  180  &  --  &  $180$  \\
               
        \rule{0cm}{2.8ex}$X_{\rm H}$ (mass fr.) 
      &  0.62  & $0.63^{+0.05}_{-0.05}$ & \textit{0.63} &  $<0.74$  &  $0.63$  &  $0.654$ &  $0.65$ \\
        \rule{0cm}{2.8ex}$X_{\rm He}$ (mass fr.) 
      &  0.37  & $0.36^{+0.05}_{-0.05}$ & \textit{0.36} &  $>0.26$  &  $0.36$  &  $0.34$  &  $0.34^{+0.1}_{-0.05}$  \\
        \rule{0cm}{2.8ex}$X_{\rm C}/10^{-4}$ (mass fr.)
      &  \textit{3.7}  &   $1.9^{+4.0}_{-1.5}$  &  $3.1^{+11.8}_{-0.5}$ &  --  &  $4.5^{+2}_{-0.5}$ &  $4^{+2}_{-2}$  &  $5^{+1}_{-1}$ \\
        \rule{0cm}{2.8ex}$X_{\rm N}/10^{-4}$ (mass fr.)
      &  \textit{1.2}  &   $40^{+69}_{-17}$  &  $80^{+17}_{-33}$ &  --  &  $40^{+9}_{-10}$ &  $12^{+4}_{-4}$   &  $50^{+10}_{-10}$  \\
        \rule{0cm}{2.8ex}$X_{\rm O}/10^{-3}$ (mass fr.) 
      &  \textit{2.2}  &   $0.04^{+3.47}_{-0.01}$  &  $1.3^{+11.2}_{-0.7}$ &  --  &  $1.9^{+1.}_{-0.5}$  &  $1.6^{+0.6}_{-0.6}$   &  $3^{+0.5}_{-0.5}$  \\
        \rule{0cm}{2.8ex}$y_\text{He}$  
      &  $0.15\pm0.02$  &   $0.14\pm0.03$  &  \textit{0.14}  &  $>0.09$  &  $0.14\pm0.02$  &  $0.13$   &  $0.13$  \\
        \rule{0cm}{2.8ex}$\epsilon_{\rm C}$ 
      &  \textit{7.7}  &   $7.40^{+0.45}_{-0.65}$  &  $7.60^{+0.65}_{-0.05}$ &  --  & $7.77$  &  $7.71$   &  $7.81$  \\
        \rule{0cm}{2.8ex}$\epsilon_{\rm N}$ 
      &  \textit{7.1}  &   $8.65^{+0.40}_{-0.20}$  &  $8.95^{+0.05}_{-0.20}$ &  --  & $8.65$  &  $8.24$   &  $8.82$  \\
        \rule{0cm}{2.8ex}$\epsilon_{\rm O}$ 
      &  \textit{8.3}  &   $6.60^{+1.90}_{-0.05}$  &  $8.10^{+0.95}_{-0.30}$ &  --  & $8.27$  &  $8.18$   &  $8.46$  \\
          
        \rule{0cm}{2.8ex}$E_{B-V}$ (mag)   
      &  0.07  &   0.20  &  0.20 &  0.1  &  0.11  &  $0.095$   &  $0.095$  \\
        \rule{0cm}{2.8ex}$\log \left(L_\text{X} / L\right)$ 
      &  --  &   $-7.16$  &  $-7.04$ &  --  &  $-6.4$  &  $-5.87$   &  $-8.1$ \\
        \rule{0cm}{2.8ex}$\log\,Q_{\mathrm{H}}$ (s$^{-1}$)     
      &  --  &   $49.21^{+0.03}_{-0.06}$  &  $49.06^{+0.03}_{-0.02}$ &  $49.1$  & $49.55$   &   --   &$48.93$  \\
        \rule{0cm}{2.8ex}$\log\,Q_{\mathrm{He\,I}}$ (s$^{-1}$)  
      &  --  &   $48.22^{+0.04}_{-0.11}$  &  $47.91^{+0.04}_{-0.02}$ &  $48.1$  &  $48.55$  &  --   &  $47.73$\\
        \rule{0cm}{2.8ex}$\log\,Q_{\mathrm{He\,II}}$ (s$^{-1}$) 
          &  --   &   $41.40^{+0.07}_{-0.15}$  &  $41.13^{+0.09}_{-0.04}$  &  --  &  $39.50$  &  --   &$40.46$\vspace{1.5ex}\\
 
    \rule{0cm}{2.8ex}$\log Q_\text{ws}$ 
      &  $-12.1\pm0.1$  &   $-12.3$  &  $-12.1\pm0.1$ &  -12.3  &  -12.2  &  -12.3   &  $-12.1$  \\
    \rule{0cm}{2.8ex}$D_\infty$ or $f_\text{cl}$ or $f^{-1}_\text{V}$\tablefootmark{($\dagger$)} 
       & \textit{1}  &   \textit{10}  &  $35^{+8}_{-4}$  &  \textit{1}  &  $20^{+5}_{-7}$  &  $5$  &  $5^{+3}_{-3}$  \\
    \rule{0cm}{2.8ex}$\varv_\text{cl}$ (km\,s$^{-1}$)  
       & --  &   --  &  --  &  --  &  30  &  $20$  &    --   \\   
    \rule{0cm}{2.8ex}$\varv_\text{cl,start}$ (km\,s$^{-1}$)  
       & --  &   \textit{94}   &  $146^{+73}_{-37}$  &  --  &  --  &  --  &    --   \\   
    \rule{0cm}{2.8ex}$R_D$ ($R_{20}$)  
       & --  &   --  &  --  &  --  &  --  &  --  &   $5$   \\   
    \rule{0cm}{2.8ex}$\log f_\text{ic}$  
       & --  &   \textit{-1}  &  $-0.64^{+0.10}_{-0.04}$  &  --  &  --  &  --  &    --   \\   
    \rule{0cm}{2.8ex}$f_\text{vel}$  
       & --  &   \textit{0.5}  &  $\cdots$   &  --  &  --  &  --  &    --   \\   
    \rule{0cm}{2.8ex}$T_{20}$ (kK)         &  --  &  --  &  --  &  --  &  --  &  $35.9\pm0.2$  &  $34^{+1}_{-1}$    \\
    \rule{0cm}{2.8ex}$R_{20}$ ($R_\odot$)  &  --  &  --  &  --  &  --  &  --  &  13.2  &  $13.7^{+1}_{-1}$  \\
    \rule{0cm}{2.8ex}$T_{100}$ (kK)        &  --  &  --  &  --  &  --  &  --  &  $36.0\pm0.2$  &   --  \\\hline
  \end{tabular}
  \tablefoot{
    \tablefoottext{$\dagger$}{For optically thin clumping with no inter-clump  medium: $D_\infty = f_\text{cl} = f^{-1}_\text{V}$ (cf.\ Sect.\,\ref{sec:clumping})}
    \tablefoottext{$\ddagger$}{Three dots ($\cdots$) indicate that a quantity is included as a free parameter, but cannot be constrained. For an unconstrained $f_\mathrm{cl}$, $\log (\dot{M}\sqrt{f_\text{cl}})$ and $\log Q_\text{ws}$ are computed assuming $f_\mathrm{cl} = 15$.}  
  }
\end{table*}%

\begin{table*}
  \caption{Best-fit parameters for the O9 Ia star Sk\,-66$^{\circ}$\,171 derived with the different methods. Adopted values are in italics.}
  \label{table:paramsummary-sk66171}
  \centering
  \begin{tabular}{lccccccc}
    \hline\hline
    \rule{0cm}{2.8ex}
    Sk\,-66$^{\circ}$\,171     &  F1  &   F2  &  F3\tablefootmark{($\ddagger$)} &  F4  &  C1  &  C2  &  P2       \\ 
    \hline
    \rule{0cm}{2.8ex}%
    \rule{0cm}{2.8ex}$T_\text{eff}$ (kK)    
        &  $31.1\pm0.5$  &  $31.25\pm1.00$  &  $32.25^{+0.75}_{-1.75}$  &  $29.1^{+2.0}_{-0.7}$  &  $30.0\pm1.0$  &  $30.3\pm0.2$  & $29.3^{+1}_{-1}$ \\
    \rule{0cm}{2.8ex}$\log g$ (cm\,s$^{-2}$)  
        &  $3.11\pm0.05$  & $3.05^{+0.05}_{-0.13}$ & $3.20^{+0.48}_{-0.13}$  &  $2.9^{+0.2}_{-0.1}$  &  $3.0\pm0.1$  &  $3.15$  & $3.08^{+0.15}_{-0.10}$  \\
    \rule{0cm}{2.8ex}$\log L$ ($L_\odot$)   
        &  $5.73\pm0.01$  &  $5.75^{+0.04}_{-0.04}$  &  $5.80^{+0.03}_{-0.07}$  &  $5.64^{+0.14}_{-0.08}$  &  $5.71\pm0.02$  &  $5.72\pm0.02$  &   $5.7^{+0.1}_{-0.1}$  \\
    \rule{0cm}{2.8ex}$R_{2/3}$ ($R_\odot$)  
        &  $25.3\pm0.3$  &  $25.9^{+1.0}_{-0.9}$  &  $25.8^{+1.1}_{-0.9}$  &  $22.1^{+6.3}_{-2.7}$  &  $26.5\pm1.7$  &  $26.5$  &   $27.2^{+2}_{-3}$     \\
    \rule{0cm}{2.8ex}$M_\mathrm{spec}$ ($M_\odot$)  
        &  $31\pm1$  &  $27.4^{+3.0}_{-5.7}$  &  $40.4^{+67.9}_{-7.4}$  &  $14^{+9}_{-6}$  &  $25.6\pm3.2$  &  $36.2$  &   $33.2^{+9}_{-7}$   \\
                       
    \rule{0cm}{2.8ex}$\beta_1$ 
        &  $\geq 1.1$ & \textit{1.0} &  $1.6^{+0.7}_{-0.3}$  &  \textit{1.0}  &  1.3  &  $2.0$  &   $1.1$    \\
    \rule{0cm}{2.8ex}$\beta_2$ 
        &  --  &  --  &  --  &  --  &  --  &  --  &    $1.8 $                \\
    \rule{0cm}{2.8ex}$\log \dot{M}$ ($M_\odot \mathrm{yr}^{-1}$)  
        &  $-5.27\pm0.10$  &  $-5.72^{+0.05}_{-0.08}$  &  $-6.19^{+0.55}_{-0.08}$  &  --  & $-5.96^{+0.5}_{-0.04}$  &  $-5.68^{+0.03}_{-0.03}$     & $-5.8^{+0.1}_{-0.1}$   \\
    \rule{0cm}{2.8ex}$\log (\dot{M}\sqrt{f_\text{cl}})$ ($M_\odot \mathrm{yr}^{-1}$)  
        &  $-5.27\pm0.10$  &  $-5.22^{+0.05}_{-0.08}$  &  $-5.60$  &  --  &  -5.37  &  $-5.42^{+0.03}_{-0.03}$     & $-5.15^{+0.15}_{-0.15}$   \\
    \rule{0cm}{2.8ex}$\varv_{\infty}$ (km\,s$^{-1}$)   
        &  \textit{1850}  &  \textit{1876}  &  $1850^{+75}_{-100}$ &  --  &  $1700\pm100$   &  $2000$   &  $1800^{+200}_{-200}$    \\  
     \rule{0cm}{2.8ex}$\varv \sin i$ (km\,s$^{-1}$)   
        &  $97\pm10$  &  $125^{+10}_{-25}$  &  \textit{125}  &  $\sim$$100$  &  $98$  &  $90$   &   $98\pm10$       \\
         
    \rule{0cm}{2.8ex}$\varv_{\mathrm{rad}}$ (km\,s$^{-1}$)     
        &  $412$  &   $407\pm16$ &  $408\pm2$ &  $\sim$$420$  &  $410$  &  $\sim$450   &  $410^{+20}_{-20}$        \\
    \rule{0cm}{2.8ex}$\varv_{\mathrm{mac}}$ (km\,s$^{-1}$)  
      &  30  &   \textit{0}  &  \textit{0} &  20  &  $20$  &  \textit{0}   &    $10$ \\
    \rule{0cm}{2.8ex}$\xi$ (km\,s$^{-1}$) 
      &  $\geq$ 15  &   \textit{15}  &  \textit{15} &  10  &  $15$  &  $25\pm5$   &    $18$  \\
    \rule{0cm}{2.8ex}$\varv_\text{windturb}$ (km\,s$^{-1}$) 
      &  --  &  \textit{187}  &  $315^{+37}_{-102}$ &  --  &  250  &  --   &  $180$  \\

                \rule{0cm}{2.8ex}$X_{\rm H}$ (mass fr.) 
       &  0.71  &   $0.68^{+0.04}_{-0.1}$  &  \textit{0.68} &  0.74  &  $0.62$  &  $0.61$   &   $0.6^{+0.05}_{-0.1}$           \\
                \rule{0cm}{2.8ex}$X_{\rm He}$ (mass fr.) 
       &  0.28  &   $0.31^{+0.1}_{-0.04}$  &  \textit{0.31} &  0.26  &  $0.37$  &  $0.38$   &   $0.39^{+0.1}_{-0.05}$           \\
                \rule{0cm}{2.8ex}$X_{\rm C}/10^{-4}$ (mass fr.)
       &  \textit{4.3}  &   $7^{+37}_{-2}$  &  $15^{+55}_{-8}$ &  --  &  $10\pm1$  &  $9\pm1$  &  $8\pm1$               \\
                \rule{0cm}{2.8ex}$X_{\rm N}/10^{-4}$ (mass fr.)
       &  \textit{1.3} &   $14^{+12}_{-10}$  &  $6.9^{+7.7}_{-6.3}$ &  --  &  $4.5^{+2.9}_{-1.0}$  &  $13.8\pm4.6$   &   $8\pm2$    \\
                \rule{0cm}{2.8ex}$X_{\rm O}/10^{-3}$ (mass fr.) 
       &  \textit{2.5}  &   $1.0^{+6.4}_{-0.6}$  &  $5.6^{+2.8}_{-5.0}$ &  --  &  $2.9^{+1.0}_{-0.5}$  &  $1.39\pm0.35$  &   $3.0\pm0.5$   \\
                \rule{0cm}{2.8ex}$y_\text{He}$  
       &  $0.10\pm0.02$  & $0.11^{+0.06}_{-0.02}$ & \textit{0.11} &  $0.09$  &  $0.15\pm0.01$  &  $0.155$ &   $0.164$   \\
                \rule{0cm}{2.8ex}$\epsilon_{\rm C}$ 
       &  \textit{7.7}  &  $7.95^{+0.75}_{-0.05}$  &  $8.25^{+0.65}_{-0.30}$ &  --  &  8.13  &  $8.04$  &  $8.1$  \\
                \rule{0cm}{2.8ex}$\epsilon_{\rm N}$ 
       &  \textit{7.1}  &  $8.15^{+0.25}_{-0.55}$  &  $7.85^{+0.30}_{-1.0}$ &  --  &  7.71  &  $8.2$  &  $8.0$  \\
                \rule{0cm}{2.8ex}$\epsilon_{\rm O}$ 
       &  \textit{8.3}  &  $7.95^{+0.85}_{-0.35}$  &  $8.70^{+0.15}_{-0.95}$ &  --  &  8.46  &  $8.15$  &  $8.5$  \\
          
                \rule{0cm}{2.8ex}$E_{B-V}$ (mag)   &  0.07  &   0.18  &  0.18 &  0.08  &  0.12  &  $0.11$  &  $0.09$                 \\
                \rule{0cm}{2.8ex}$\log \left(L_\text{X} / L\right)$ &  --  &   $-6.88$  &  $-6.91$ &  --  &  -6.1  &  $-7.25$  & -6.0   \\
                \rule{0cm}{2.8ex}$\log\,Q_{\mathrm{H}}$ (s$^{-1}$)   &  --  &   $49.30^{+0.07}_{-0.04}$  &  $49.31^{+0.05}_{-0.14}$  &  $49.1$  &  49.71  &  --  &  $49.12$\\
                \rule{0cm}{2.8ex}$\log\,Q_{\mathrm{He\,I}}$ (s$^{-1}$)   & --  &   $47.97^{+0.13}_{-0.09}$  &  $47.90^{+0.13}_{-0.43}$  &  $47.6$  &  48.3  &  --  &  $47.15$ \\
                \rule{0cm}{2.8ex}$\log\,Q_{\mathrm{He\,II}}$ (s$^{-1}$)   & --  &   $41.58^{+0.09}_{-0.14}$  &  $41.64^{+0.07}_{-0.15}$  &  --  &  38.76  &  --  &  $42.22$\vspace{1.5ex}\\
 
     \rule{0cm}{2.8ex}$\log Q_\text{ws}$ 
        &  $-12.3\pm0.1$ &  $-12.2$ &  $-12.6$  &  $-12.3$  &  $-12.4$  &  $-12.5$  &  $-12.2$  \\
   \rule{0cm}{2.8ex}$D_\infty$ or $f_\text{cl}$ or $f^{-1}_\text{V}$\tablefootmark{($\dagger$)}
       & \textit{1}  &   \textit{10}  &  $\cdots$  &  \textit{1}  &  14  &  $3.33$  &  $20^{+10}_{-10}$   \\
    \rule{0cm}{2.8ex}$\varv_\text{cl}$ (km\,s$^{-1}$)  
       & --  &   --  &  --  &  --  &  30  &  $10$  &    --   \\   
    \rule{0cm}{2.8ex}$\varv_\text{cl,start}$ (km\,s$^{-1}$)  
       & --  &   \textit{94}  &  $\cdots$  &  --  &  --  &  --  &    --   \\   
    \rule{0cm}{2.8ex}$R_D$ ($R_{20}$)  
       & --  &   --  &  --  &  --  &  --  &  --  &    $3$   \\   
    \rule{0cm}{2.8ex}$\log f_\text{ic}$  
       & --  &   \textit{-1}  &  $\cdots$  &  --  &  --  &  --  &    --   \\   
    \rule{0cm}{2.8ex}$f_\text{vel}$  
       & --  &   \textit{0.5}  &  $\cdots$  &  --  &  --  &  --  &    --   \\   
    \rule{0cm}{2.8ex}$T_{20}$ (kK)  & --  &   --  &  --  &  --  &  --  &  $30.8\pm0.2$  &   $30^{+2}_{-1}$            \\
    \rule{0cm}{2.8ex}$R_{20}$ ($R_\odot$)  &  --  &  --  &  --  &  --  &  --  & 25.6  &   $26.2^{+2}_{-3}$       \\
    \rule{0cm}{2.8ex}$T_{100}$ (kK) & --  &   --  &  --  &  --  &  $\cdots$  &  $31.1\pm0.2$  &   --  \\\hline
  \end{tabular}
  \tablefoot{
    \tablefoottext{$\dagger$}{For optically thin clumping with no inter-clump  medium: $D_\infty = f_\text{cl} = f^{-1}_\text{V}$ (cf.\ Sect.\,\ref{sec:clumping})}
    \tablefoottext{$\ddagger$}{Three dots ($\cdots$) indicate that a quantity is included as a free parameter, but cannot be constrained. For an unconstrained $f_\mathrm{cl}$, $\log (\dot{M}\sqrt{f_\text{cl}})$ and $\log Q_\text{ws}$ are computed assuming $f_\mathrm{cl} = 15$.}  
  }%
\end{table*}%

\onecolumn%
\begin{longtable}{l|cccc|cc|cc|cc|cc}%
  \caption{\label{tab:ioncoverage}Ions, levels, and line transitions covered in the different methods.}\\
    \hline\hline\rule{0cm}{2.0ex}
                  &  \multicolumn{4}{c|}{F1, F2, F3, F4\tablefootmark{(F)}}  &  \multicolumn{2}{c|}{C1}   & \multicolumn{2}{c|}{C2} &  \multicolumn{2}{c|}{P1}   & \multicolumn{2}{c}{P2}\\                    
     Ion           &  Levels & Lines &  Levels & Lines  &   Super- & Total    &  Super- &  Total    &   Levels & Lines  &  Levels & Lines\\
                   &  (expl) & (expl)&  (bg)   & (bg) &  levels  &  Levels   & levels  &  Levels  &   &  &   &  \\\hline%
  \endfirsthead%
  \caption{Continued.}\\%
  \hline\hline\rule{0cm}{2.0ex}
                  &  \multicolumn{4}{c|}{F1, F2, F3, F4\tablefootmark{(F)}}  &  \multicolumn{2}{c|}{C1}   & \multicolumn{2}{c|}{C2} &  \multicolumn{2}{c|}{P1}   & \multicolumn{2}{c}{P2}\\                    
     Ion           &  Levels & Lines &  Levels & Lines  &   Super- & Total    &  Super- &  Total    &   Levels & Lines   &  Levels & Lines\\
                   &  (expl) & (expl)&  (bg)   & (bg) &  levels  &  Levels   & levels  &  Levels  &   &    &    &   \\\hline%
  \endhead%
  \endfoot%
   \ion{H}{I}      &  20 &  190 &  --  &  --   &  30   &   30   &   20  & 30    &  22   &  231   &   22   &  231   \\    
   \ion{H}{II}     &   1 &  --  &  --  &  --   &   1   &   --   &    1  & --    &   1   &   --   &    1   &  --    \\    
   \ion{He}{I}     &  49 &  341 &  --  &  --   &  131  &   237  &   69  & 69    &  35   &  595   &   35   &  595   \\    
   \ion{He}{II}    &  20 &  190 &  --  &  --   &  30   &   30   &   22  & 30    &  26   &  325   &   26   &  325   \\    
   \ion{He}{III}   &   1 &  --  &  --  &  --   &   1   &   --   &    1  & --    &   1   &   --   &    1   &  --    \\    
   \ion{C}{I}      &  -- &  --  &  22  &  98   &  --   &   --   &   --  & --    &   --  &   --   &   15   &  105   \\    
   \ion{C}{II}     &  67 &  422 &  36  &  284  &  --   &   --   &  40   & 92    &   --  &   --   &   32   &  496   \\    
   \ion{C}{III}    &  70 &  644 &  50  &  520  &  99   &   243  &  99   & 243   &  40   &  780   &   40   &  780   \\    
   \ion{C}{IV}     &  50 &  141 &  27  &  103  &  64   &   64   &  64   & 64    &  25   &  300   &   25   &  300   \\    
   \ion{C}{V}      &   1 &   -- &   5  &   8   &  --   &   --   &   --  & --    &  29   &  406   &    5   &   10   \\    
   \ion{C}{VI}     &  -- &  --  &   1  &  --   &  --   &   --   &   --  & --    &  15   &  105   &   15   &  105   \\    
   \ion{N}{I}      &  -- &  --  &  28  &  70   &  --   &   --   &   --  & --    &  --   &   --   &   10   &   45   \\    
   \ion{N}{II}     &  50 &  129 &  21  &  64   &  --   &   --   &  45   & 85    &  --   &   --   &   38   &  703   \\    
   \ion{N}{III}    &  41 &  359 &  40  &  356  &  57   &   287  &  41   & 82    &  56   &  1540  &   36   &  630   \\    
   \ion{N}{IV}     &  50 &  520 &  50  &  520  &  44   &   70   &  44   & 76    &  38   &  703   &   38   &  703   \\    
   \ion{N}{V}      &  27 &  104 &  27  &  104  &  41   &   49   &  41   & 49    &  20   &  190   &   20   &  190   \\    
   \ion{N}{VI}     &   1 &   -- &   5  &  6    &  --   &   --   &   --  & --    &  14   &   91   &   14   &   91   \\    
   \ion{O}{I}      &  -- &  --  &  48  &  534  &  --   &   --   &   --  & --    &  --   &   --   &   13   &   78   \\    
   \ion{O}{II}     &  50 &  595 &  50  &  595  &  --   &   --   &  54   & 123   &  37   &  666   &   37   &  666   \\    
   \ion{O}{III}    &  50 &  554 &  50  &  554  &  36   &   104  &  88   & 170   &  33   &  528   &   33   &  528   \\    
   \ion{O}{IV}     &  44 &  435 &  44  &  435  &  30   &   64   &  38   & 78    &  25   &  300   &   25   &  300   \\    
   \ion{O}{V}      &  50 &  524 &  50  &  524  &  32   &   56   &  32   & 56    &  36   &  630   &   36   &  630   \\    
   \ion{O}{VI}     &  27 &  102 &  27  &  102  &  --   &   --   &  25   & 31    &  16   &  120   &   16   &  120   \\    
   \ion{O}{VII}    &   1 &   -- &   1  &   --  &  --   &   --   &   --  & --    &  15   &  105   &   15   &  105   \\    
   \ion{Ne}{I}     &  -- &  --  &  41  &  127  &  --   &   --   &   --  & --    &  --   &   --   &    8   &   28   \\        
   \ion{Ne}{II}    &  -- &  --  &  50  &  592  &  14   &   48   &  --   &  --   &  --   &   --   &    1   &    0   \\        
   \ion{Ne}{III}   &  -- &  --  &  38  &  319  &  23   &   71   &  --   &  --   &  --   &   --   &   18   &  153   \\        
   \ion{Ne}{IV}    &  -- &  --  &  50  &  577  &  17   &   52   &  --   &  --   &  --   &   --   &   35   &  595   \\        
   \ion{Ne}{V}     &  -- &  --  &  50  &  534  &  --   &   --   &   --  &  --   &  --   &   --   &   54   & 1431   \\        
   \ion{Ne}{VI}    &  -- &  --  &  50  &  343  &  --   &   --   &   --  &  --   &  --   &   --   &   49   & 1176   \\        
   \ion{Ne}{VII}   &  -- &  --  &  1   &   0   &  --   &   --   &   --  &  --   &  --   &   --   &    1   &    0   \\    
   \ion{Mg}{I}     &  -- &  --  &  21  &  74   &  --   &   --   &   --  & --    &  --   &   --   &    1   &   --   \\    
   \ion{Mg}{II}    &  -- &  --  &  26  &  104  &  36   &   44   &   --  & --    &  32   &  496   &   12   &   66   \\    
   \ion{Mg}{III}   &  -- &  --  &  50  &  529  &  --   &   --   &   --  & --    &  43   &  903   &   10   &   45   \\    
   \ion{Mg}{IV}    &  -- &  --  &  50  &  589  &  --   &   --   &   --  & --    &  17   &  136   &    1   &   --   \\    
   \ion{Mg}{V}     &  -- &  --  &  50  &  547  &  --   &   --   &   --  & --    &   1   &   --   &   --   &   --   \\
   \ion{Al}{II}    &  -- &  --  &  18  &  57   &  --   &   --   &  38   & 58    &  --   &   --   &   10   &   45   \\    
   \ion{Al}{III}   &  -- &  --  &  27  &  97   &  --   &   --   &  17   & 45    &  --   &   --   &   10   &   45   \\    
   \ion{Al}{IV}    &  -- &  --  &  50  &  529  &  --   &   --   &   --  & --    &  --   &   --   &   10   &   45   \\    
   \ion{Al}{V}     &  -- &  --  &  50  &  588  &  --   &   --   &   --  & --    &  --   &   --   &   10   &   45   \\    
   \ion{Si}{II}    &  34 &  174 &  40  &  293  &  --   &   --   &  52   & 80    &  --   &   --   &    1   &    0   \\    
   \ion{Si}{III}   &  28 &  57  &  50  &  480  &  50   &   50   &  33   & 33    &  24   &  276   &   24   &  276   \\    
   \ion{Si}{IV}    &  18 &  52  &  25  &  90   &  66   &   66   &   22  & 33    &  23   &  253   &   23   &  253   \\    
   \ion{Si}{V}     &   1 &   0  &  50  &  531  &  --   &   --   &   --  & --    &  25   & 1326   &    1   &    0   \\        
   \ion{Si}{VI}    &  -- &  --  &  50  &  596  &  --   &   --   &   --  & --    &  10   &   45   &   --   &   --   \\        
   \ion{Si}{VII}   &  -- &  --  &   1  &   0   &  --   &   --   &   --  & --    &   1   &    0   &   --   &   --   \\        
   \ion{P}{IV}     &  19 &  27  &  19  &  27   &  --   &   --   &  30   & 90    &  12   &   66   &   12   &   66   \\        
   \ion{P}{V}      &  25 &  90  &  25  &  90   &  --   &   --   &  16   & 62    &  11   &   55   &   11   &   55   \\        
   \ion{P}{VI}     &  14 &  41  &  14  &  41   &  --   &   --   &   --  & --    &   1   &    0   &    1   &    0   \\        
   \ion{S}{III}    &  -- &  --  &  14  &  32   &  39   &   78   &   24  & 44    &  24   &   27   &   23   &  253   \\        
   \ion{S}{IV}     &  -- &  --  &  13  &  22   &  40   &   108  &  51   & 142   &  25   &  300   &   11   &   55   \\        
   \ion{S}{V}      &  -- &  --  &  44  &  404  &  37   &   144  &  31   & 98    &  20   &  190   &   10   &   45   \\        
   \ion{S}{VI}     &  -- &  --  &  18  &  59   &  --   &   --   &   --  & --    &  22   &  231   &    1   &    0   \\        
   \ion{S}{VII}    &  -- &  --  &  14  &  39   &  --   &   --   &   --  & --    &  15   &  102   &   --   &   --   \\        
   \ion{Ar}{III}   &  -- &  --  &  13  &  21   &  24   &   138  &   --  &  --   &  --   &   --   &   --   &   --  \\
   \ion{Ar}{IV}    &  -- &  --  &  11  &  22   &  30   &   102  &   --  &  --   &  --   &   --   &   --   &   --  \\
   \ion{Ar}{V}     &  -- &  --  &  40  &  328  &  14   &   29   &   --  &  --   &  --   &   --   &   --   &   --  \\
   \ion{Ca}{III}   &  -- &  --  &  15  &  43   &  29   &   88   &   --  &  --   &  --   &   --   &   --   &   --  \\
   \ion{Ca}{IV}    &  -- &  --  &  50  &  176  &  19   &   72   &   --  &  --   &  --   &   --   &   --   &   --  \\
   \ion{Fe}{I}     &  -- &  --  &  11  &  6    &  --   &   --   &   --  &  --   &  --   &   --   &    1   &   --   \\        
   \ion{Fe}{II}    &  -- &  --  &  50  &  405  &  --   &   --   &   --  &  --   &  --   &   --   &    3\tablefootmark{(S)} &    2\tablefootmark{(S)}   \\                                                
   \ion{Fe}{III}   &  -- &  --  &  50  &  246  &  65   &  607   &  104  & 1433  &   1   &   --   &   13\tablefootmark{(S)}  &   40\tablefootmark{(S)}   \\                                                   
   \ion{Fe}{IV}    &  -- &  --  &  45  &  253  & 100   & 1000   &   50  & 1000  &  18\tablefootmark{(S)}   &   77\tablefootmark{(S)}   &   18\tablefootmark{(S)}   &   77\tablefootmark{(S)}   \\                                    
   \ion{Fe}{V}     &  -- &  --  &  50  &  451  & 139   & 1000   &   61  &  300  &  22\tablefootmark{(S)}   &  107\tablefootmark{(S)}   &   22\tablefootmark{(S)}   &  107\tablefootmark{(S)}   \\                                   
   \ion{Fe}{VI}    &  -- &  --  &  50  &  452  &  59   & 1000   &   57  &  439  &  29\tablefootmark{(S)}   &  194\tablefootmark{(S)}   &   29\tablefootmark{(S)}   &  194\tablefootmark{(S)}   \\                                   
   \ion{Fe}{VII}   &  -- &  --  &  22  &  91   &  --   &   --   &   --  &   --  &  19\tablefootmark{(S)}   &   87\tablefootmark{(S)}   &   19\tablefootmark{(S)}   &   87\tablefootmark{(S)}   \\                                    
   \ion{Fe}{VIII}  &  -- &  --  &  42  &  300  &  --   &   --   &   --  &   --  &  14\tablefootmark{(S)}   &   49\tablefootmark{(S)}   &   14\tablefootmark{(S)}   &   49\tablefootmark{(S)}   \\                                   
   \ion{Fe}{IX}    &  -- &  --  &   1  &  0    &  --   &   --   &   --  &   --  &   1   &   --   &   15\tablefootmark{(S)}   &   56\tablefootmark{(S)}   \\                                                   
   \ion{Fe}{X}     &  -- &  --  &  --  &  --   &  --   &   --   &   --  &   --  &   --  &   --   &    1    &   --   \\    
   \ion{Ni}{III}   &  -- &  --  &  40  &  281  &  24   &  150   &   --  &  --   & \multicolumn{2}{c|}{\tablefootmark{(G)}} & \multicolumn{2}{c}{\tablefootmark{(G)}}    \\
   \ion{Ni}{IV}    &  -- &  --  &  50  &  528  &  36   &  200   &   --  &  --   & \multicolumn{2}{c|}{\tablefootmark{(G)}} & \multicolumn{2}{c}{\tablefootmark{(G)}}   \\
   \ion{Ni}{V}     &  -- &  --  &  41  &  70   &  46   &  183   &   --  &  --   & \multicolumn{2}{c|}{\tablefootmark{(G)}} & \multicolumn{2}{c}{\tablefootmark{(G)}}    \\
   \ion{Ni}{VI}    &  -- &  --  &  45  &  253  &  40   &  182   &   --  &  --   & \multicolumn{2}{c|}{\tablefootmark{(G)}} & \multicolumn{2}{c}{\tablefootmark{(G)}}    \\\hline 
  \multicolumn{13}{p{0.92\columnwidth}}{%
   \tablefoot{%
    \tablefoottext{F}{For \textsc{Fastwind} models (F1-F4), there are two sets of levels/lines with explicit (``expl'') and background (``bg'') elements, see Sect.\,\ref{sec:fastwind}.}
    \tablefoottext{G}{Ni levels and lines are included in the Fe superlevels and superlines which contain the whole iron group \citep[see][for details]{Graefener2002}.}
    \tablefoottext{S}{Numbers listed for Fe in PoWR-based methods denote superlevels and superlines: Superlevels contain a set of levels (of the same parity) within an energy band. Superlines describe a combined treatment of all transitions between two superlevels \citep[see][for more details]{Graefener2002}.}
    }
  }
  \\%
\end{longtable}

\end{appendix}

\end{document}